\documentclass[12pt]{article}
\usepackage{graphicx}
\usepackage{cancel}
\usepackage{authblk}
\usepackage{amsmath}
\usepackage{amsthm}
\usepackage{amssymb}
\usepackage{bm,upgreek}
\usepackage{makecell}
\usepackage[sort&compress,numbers]{natbib}
\usepackage[colorlinks=true,linkcolor=black, citecolor=blue, urlcolor=blue]{hyperref}

\graphicspath{{figures/}} 

\begin{document}

\title{Identification of the length scale parameter of simplified strain gradient elasticity from standard Mode I fracture tests}

\author[1,2]{Yury Solyaev}
\author[2]{Kirill Shelkov}
\author[2]{Pavel Polyakov}

\affil[1]{Institute of Applied Mechanics of Russian Academy of Sciences, Moscow, Russia}
\affil[2]{Moscow Aviation Institute, Moscow, Russia}

\setcounter{Maxaffil}{0}
\renewcommand\Affilfont{\itshape\small}

\date{\today}

\maketitle

\begin{abstract}
Recently, it was shown that additional material constants of strain gradient elasticity (SGE) can be identified for brittle and quasi-brittle materials based on the analysis of experimental data on the crack size effect. In the present paper, we perform precise numerical simulations within SGE and derive regression relations for processing experimental data from standard fracture mechanics tests under pure Mode I loading conditions (CCT, SENT, SENB). We consider the simplified SGE, whose constitutive relations contain a single length scale parameter $l$ in addition to the classical elastic constants. We show that, for brittle materials, this parameter can be explicitly identified as $l \approx 0.362 (K_{Ic}/\sigma_{ult})^2$. This identification ensures that the fracture loads predicted by classical linear elastic fracture mechanics (LEFM) and by the simplified SGE coincide for relatively long Mode I cracks. However, within the simplified SGE, these fracture loads are evaluated from the nonsingular stress field using the maximum principal stress criterion. For quasi-brittle materials, we derive regression relations that describe the non-classical size effect on strength. This effect is usually treated within nonlinear fracture mechanics but can be naturally captured by SGE. Examples of identification of the length scale parameter $l$ based on the established relations and the experimental data for chopped fiber composites and for porous and dense quasi-brittle ceramics are presented.
\end{abstract}

%\begin{keyword}
%Strain gradient elasticity \sep Crack problems \sep Regression analysis \sep Length scale parameter \sep Identification \sep Size effect
%\end{keyword}
%
%\author[label1,label2]{Yury Solyaev\corref{cor1}}
%\ead{yos@iam.ras.ru}
%\author[label1]{Kirill Shelkov}
%\author[label2]{Pavel Polyakov}
%\address[label1]{Institute of Applied Mechanics of Russian Academy of Sciences, Moscow, Russia}
%\address[label2]{Moscow Aviation Institute, Moscow, Russia}

\section{Introduction}
\label{intro}

Strain gradient elasticity (SGE) was developed by Toupin and Mindlin~\cite{toupin1962elastic, mindlin1964microstructure} under the assumption that the strain energy density of a material depends not only on the strain but also on the gradient of strain. Three equivalent formulations of the constitutive equations of isotropic SGE have been established in terms of strain gradient, second gradient of displacement, and symmetric/antisymmetric parts of the second gradient of displacement~\cite{mindlin1964microstructure}. These formulations contain five additional material constants apart from the classical Lam\'e parameters. Later, applied and simplified gradient theories were developed as particular cases of the Toupin-Mindlin theory, reducing the number of independent constants to three~\cite{hutchinson1997strain, lam2003experiments}, two~\cite{lazar2022reduced, lurie2023variant} and one~\cite{aifantis1992role, barchiesi2021granular, lurie2021dilatation, eremeyev2022strong}.

In the present study, we use the so-called simplified strain gradient elasticity theory, which employs the Aifantis form of constitutive equations with a single length scale parameter~\cite{aifantis1992role,askes2011gradient}. Although this theory should be regarded as an approximation for real materials (which may require a more general formulation of the constitutive relations \cite{dell2009generalized}), it is widely used in different applications due to its simplicity. Within this theory, it has been shown explicitly that a wide class of classical singular problems can be regularised~\cite{gutkin1999dislocations,gourgiotis2009plane,askes2011gradient,lazar2006dislocations,lazar2017non,solyaev2022elastic,dell2024deformation,gourgiotis2026contact}. In applications to fracture mechanics problems, it has been shown that regular strain and stress fields around the tip of cracks can be evaluated and the smooth cusp-like crack opening can be described even within the linear elastic formulation of SGE. These results have been obtained based on asymptotic analysis~\cite{gourgiotis2009plane, aravas2009plane,sciarra2013asymptotic,solyaev2024higher}, analytical methods~\cite{gourgiotis2009plane} and numerical methods~\cite{akarapu2006numerical,aravas2009plane,papanicolopulos2010crack,vasiliev2021failure,solyaev2025enriched} for isotropic homogeneous materials as well as for orthotropic, flexoelectric and inhomogeneous materials~\cite{profant2026asymptotic,profant2024discussion,kotoul2018asymptotic}.

Recently, it has been suggested that the failure analysis of pre-cracked bodies can be performed within SGE solutions involving failure criteria methods for regularised stresses as an alternative approach to classical linear elastic fracture mechanics \cite{askes2015understanding, vasiliev2019estimation,vasiliev2021new, barchiesi2021computation, vasiliev2021failure, sessa2024implicit}. Moreover, it has been shown that SGE solutions capture the non-classical size effect and transition from long to short crack regimes in quasi-brittle materials~\cite{askes2015understanding,vasiliev2021new}. This result was also obtained by using the J-integral-based fracture criterion, which was applied to plane strain problems with finite size cracks in Ref.~\cite{gourgiotis2009plane}. Additional confirmation that the predictions of linear elastic fracture mechanics (LEFM) and SGE for the fracture loads of bodies containing penny-shaped cracks coincide for relatively large cracks was provided in Ref.~\cite{solyaev2025energy} based on the Eshelby equivalent inclusion method. 

Thus, an advantage of SGE is the ability to describe the strength of bodies with cracks of arbitrary length within a linear elastic formulation, using either criteria formulated in terms of regular stresses or energy criteria based on the J-integral \cite{askes2015understanding,vasiliev2021new}. Another advantage is the ability to describe in a unified manner the fracture loads for any type of stress concentrator -- including cracks, sharp notches, sharp edges, etc \cite{bagni2016gradient, vasiliev2021new, razavi2023length, solyaev2025evaluation} -- while using a single set of material characteristics: elastic properties, strength characteristics (e.g., ultimate strength), and a set of non-classical length scale parameters that enter the constitutive relations of SGE.

Identification of these additional parameters can be performed based on the analysis of the crack size effect on fracture load in standard fracture mechanics tests under static or fatigue loading conditions \cite{askes2015understanding, vasiliev2019estimation, vasiliev2021new}. Such processing of experimental data within the framework of the simplified SGE has been performed for PMMA~\cite{vasiliev2021new}, for some ceramics and metals~\cite{askes2015understanding,bagni2016gradient,vasiliev2021new,razavi2023length}. It has been shown that for many materials, the length scale parameter of the simplified SGE is of the same order of magnitude as the standard length scale parameter of fracture mechanics $L_{LEFM} = K_{Ic}^2/(\pi\sigma_{ult}^2)$ (defining the fracture process zone size), although for quasi-brittle materials it can be either larger or smaller than this value~\cite{askes2015understanding,vasiliev2021new}. We can also note alternative approaches for identifying the additional constants of SGE that have been well developed to date for various materials, composites and metamaterials, based on pull-out and indentation tests \cite{rezaei2024procedure,mokios2012gradient}, homogenisation methods \cite{gitman2005representative,bacca2013mindlin,ganghoffer2021variational,placidi2026revised}, based on analysis of deformation patterns \cite{dell2016large}, and based on the analysis of dispersion relations in dynamics \cite{rosi2018validity,solyaev2023second} and analysis parameters of interatomic interaction (for ideal crystals) \cite{shodja2018toupin,lazar2022mathematical}, see also Refs. \cite{muller2020experimental, fedele2024review, forest2020strain} for review.

In the present study, we perform numerical simulations for three types of standard fracture mechanics tests on specimens with Mode I cracks. The calculations are carried out using the developed enriched finite element method (FEM)~\cite{solyaev2025enriched}. This is a $C^1$-continuous FEM which was initially developed in Refs.~\cite{Zervos2001, papanicolopulos2010crack}. Enrichment was applied to the elements placed around the crack tip. These elements incorporate additional shape functions that correspond to asymptotic solutions of plane strain crack problems of SGE~\cite{gourgiotis2009plane,aravas2009plane}. 
Based on the obtained solutions, we are able to accurately determine the stress and strain concentrations at the crack tip (which become regular within SGE), as well as the values of the amplitude factors appearing in the asymptotic solutions. Using these amplitude factors, the J-integral is subsequently calculated based on known relations~\cite{aravas2009plane,sciarra2013asymptotic}. 

We investigate the dependence of the maximum hoop stress concentration, the amplitude coefficients, and the J-integral values on the crack length for standard fracture mechanics tests: centre-cracked tension (CCT), single-edge notched tension (SENT), and single-edge notched bending (SENB).
Assuming the validity of the stress-based approach for evaluating the fracture loads of the specimens \cite{askes2015understanding,vasiliev2021new}, we propose two methods that can be used for identification of the length scale parameter of simplified SGE based on these Mode I tests. 

First, for relatively long cracks, we find an explicit linear relation between the SGE length scale parameter and the corresponding LEFM characteristic length. A similar relation has been established previously based on the so-called implicit gradient method \cite{askes2015understanding}. In the present study, we refine the value of the proportionality coefficient between the length scale parameters of SGE and LEFM based on the performed precise numerical analysis with rigorous satisfaction of SGE boundary conditions (for discussion see \cite{lazar2015non}). The established simple relation can be used for a rapid estimate of the SGE
length scale parameter for brittle materials. We also show that astablished relation is valid for the central as well as for the edge Mode I cracks. 
%
%Based on the analysis of the performed numerical simulations, we also developed the the regression relations that can be used to calculate the failure loads for given geometry of specimens, length of crack and the value of material length scale parameter of SGE. Inversely, the length scale parameter of SGE can be identified based on these regression relations if the experimental data on the fracture loads are known.
%%to process experimental data for the validation of the considered gradient theory and for the identification of its length scale parameter. 
%
%Moreover, we show that for relatively long cracks, an explicit relationship can be established between the length scale parameter of SGE and that of classical linear elastic fracture mechanics. Moreover, by performing detailed numerical simulations, we refine the proportionality coefficient between these parameters compared to earlier estimates~\cite{}. 

Secondly, for relatively short cracks, we demonstrate the ability of SGE to describe the non-classical size effect on strength and propose convenient regression relations that can be used to interpret the corresponding experiments. Examples of identification of SGE length scale parameter based on these relations are presented for random chopped fiber composite and for quasi-brittle ceramics.

%Based on the analysis of experimental data, it is also shown that the value of the identified length scale parameter correlates with the characteristic size of inhomogeneities (mean width of profile elements) on the fracture surface of the specimens.

The remainder of the paper is organised as follows. In Section~2, we provide a brief description of SGE formulation. Its numerical implementation within the considered enriched $C^1$-continuous FEM is presented in Section~3. In Section~4, we demonstrate the convergence of the considered numerical scheme and choose its optimal parameters (the number of integration points and the number and size of enriched elements around the crack tip). We show the correspondence of the obtained numerical solutions for the J-integral to the semi-analytical solution obtained within SGE in Ref.~\cite{gourgiotis2009plane} and to the classical solution (for relatively long cracks). In Section~5, we propose two approaches for the identification of the length scale parameter $l$ of SGE based on Mode I crack tests. For purely brittle materials, we obtain an explicit definition of $l$ in terms of $K_{Ic}$ (Section 5.1). For quasi-brittle materials, we propose appropriate regression relations for the dependence of fracture loads on the crack length that allow the identification of $l$ (Section 5.2). Examples of identification are presented in Section~6. The analysis was performed using experimental data available in published studies, as well as data obtained from our own experiments conducted with highly porous SiO$_2$ ceramic samples.

%%%%%%%%%%%%%%%%%%%%%%%%%%%%
\section{Simplified strain gradient elasticity}

In the formulation of SGE, the strain energy density depends not only on the strains ($\varepsilon_{ij}$) themselves but also on their first gradients ($\kappa_{ijk}=\varepsilon_{ij,k}$)~\cite{mindlin1964microstructure}:

\begin{equation}\label{eq:sge_energy}
	w(\varepsilon_{ij}, \kappa_{ijk}) =
	\frac{1}{2} C_{ijkl} \varepsilon_{ij} \varepsilon_{kl}
	+ \frac{1}{2} A_{ijklmn} \kappa_{ijk} \kappa_{lmn}
\end{equation}
where $C_{ijkl} = \lambda\delta_{ij}\delta_{kl} +
\mu(\delta_{ik}\delta_{jl}+\delta_{il}\delta_{jk})$ is the classical
fourth-order elasticity tensor, and $A_{ijklmn}$ is the sixth-order tensor
containing additional material constants.

Within the simplified SGE~\cite{aifantis1992role}, $A_{ijklmn}$ takes the form
\begin{equation}\label{eq:simple_A}
	A_{ijklmn} = l^{2} C_{ijlm} \delta_{kn},
\end{equation}
where $l$ is the sole additional material constant that has the
dimension of length and that is called the "length scale parameter". The classical theory is recovered when $l = 0$.

The constitutive relations are:
\begin{align}
	\tau_{ij} &= \frac{\partial w}{\partial \varepsilon_{ij}}
	= C_{ijkl} \varepsilon_{kl}
	= \lambda \delta_{ij} \varepsilon_{kk} + 2\mu \varepsilon_{ij},
	\label{eq:stress} \\
	\mu_{ijk} &= \frac{\partial w}{\partial \kappa_{ijk}}
	= A_{ijklmn} \kappa_{lmn}
	= l^{2} C_{ijlm} \kappa_{lmk}
	= l^{2} \tau_{ij,k},
	\label{eq:double_stress}
\end{align}
so that the double stress $\mu_{ijk}$ in the simplified theory is simply
the gradient of the classical stress scaled by $l^{2}$.

In the framework of the simplified SGE, the boundary-value problem is formulated as follows:

\begin{equation}\label{eq:BVP}
	\begin{cases}
		\sigma_{ij,j} + b_i = 0,              & x_i \in \Omega,   \\
		t_i = \bar{t}_i \ \ \text{or}\ \ u_i = \bar{u}_i,           & x_i \in \partial\Omega, \\
		m_i = \bar{m}_i \ \ \text{or}\ \ u_{i,j} n_j = \bar{g}_i,  & x_i \in \partial\Omega, \\
		s_i = \bar{s}_i \ \ \text{or}\ \ u_i = \bar{u}_i^e,         & x_i \in \partial\partial\Omega,
	\end{cases}
\end{equation}
where $\sigma_{ij} = \tau_{ij} - \mu_{ijk,k}$ is the total stress tensor, $\bar{u}_i$
and $\bar{g}_i$ are the prescribed displacements and normal gradients of displacements on the body boundary $\partial\Omega$, $\bar{u}_i^e$ is the prescribed displacement on the body edges $\partial\partial\Omega$, $\bar{t}_i^e$, $\bar{m}_i^e$ is surface traction and surface double traction and $\bar{s}_i^e$ is the prescribed edge traction. The traction vectors are defined as follows:
\begin{gather}
	t_i=\sigma_{ij}n_j+D_j\left(\mu_{ijk}n_k\right)+\left(D_ln_l\right)\mu_{ijk}n_jn_k,
	\label{eq:traction}\\
	m_i=\mu_{ijk}n_jn_k,
	\label{eq:double_traction}\\
	s_i=\left[\mu_{ijk}c_jn_k\right].
	\label{eq:edge_traction}
\end{gather}
where $D_i = (...)_{,i} - n_i (...)_{,k} n_k$ is the surface gradient operator, $n_j$ is the outward unit normal vector, and $c_j$ is the co-normal vector, which is tangent to the surface and normal to the given edge.

In this work, we mainly analyze stress fields $\tau_{ij}$ \eqref{eq:stress}, which, for most classical singular problems of elasticity, become regular in SGE \cite{,gourgiotis2009plane,askes2011gradient,lazar2006dislocations,lazar2017non,solyaev2022elastic,dell2024deformation,gourgiotis2026contact}. These stresses are related to strains via Hooke's law \eqref{eq:stress}, and the strains are likewise regular in these class of problems. We also note that the equilibrium equations \eqref{eq:BVP} and boundary conditions \eqref{eq:traction} in SGE include the contributions of the double stress gradient $\mu_{ijk,l}$. This is precisely what gives rise to cohesive traction effect on crack faces in SGE solutions, analogous to the Barenblatt-Dugdale models (see discussion in \cite{aravas2009plane, sciarra2013asymptotic, solyaev2024higher}). 

In typical engineering problems, tractions $t_i=\bar t_i$ together with zero double tractions $m_i=0$ are prescribed on the boundaries within SGE \cite{andreaus2016numerical,shekarchizadeh2022benchmark}. Conditions for edge tractions can be used in the problems with line forces \cite{dell2024deformation}. Alternatively, displacement conditions are specified in constraint zones. On symmetry surfaces, in addition to the classical conditions on normal displacements, conditions of zero normal gradients of tangential displacement (i.e. shear) should be imposed \cite{papanicolopulos2010crack}. For a more thorough discussion of the formulation and solution procedures in SGE, the reader is referred to Refs. \cite{georgiadis2006energy, dell2017higher, eremeyev2022strong, gao2007variational}.

%%%%%%%%%%%%%%%%%%%%%%%%%%%%
\section{Numerical implementation}

\subsection{FEM formulation}
The variation of total strain energy in body $\Omega$ within SGE is given by:

\begin{equation}\label{eq:var_energy}
	\delta W = \int_{\Omega} \bigl(\tau_{ij}\delta\varepsilon_{ij}
	+ \mu_{ijk}\delta\kappa_{ijk}\bigr)\,dv.
\end{equation}

The variation of external work~\cite{papanicolopulos2010crack,solyaev2025enriched} reads
\begin{equation}\label{eq:var_ext}
	\delta W^{\text{ext}} = \int_{\Omega} b_i \delta u_i \,dv
	+ \int_{\partial\Omega} \bar{t}_i \delta u_i \,ds
	+ \int_{\partial\Omega} \bar{m}_i n_k \delta u_{i,k} \,dv
	+ \int_{\partial\partial\Omega} \bar{s}_i \delta u_i \,dl,
\end{equation}

Thus, the weak form of SGE problems is given by relation $\delta W = \delta W^{\text{ext}}$. It can be written in matrix form as follows \cite{papanicolopulos2010crack, solyaev2025enriched}:
%\begin{multline}\label{eq:variational}
%	\int_{\Omega} \bigl(\tau_{ij}\delta\varepsilon_{ij}
%	+ \mu_{ijk}\delta\varepsilon_{ij,k}\bigr) dv
%	= \int_{\Omega} b_i \delta u_i \,dv
%	+ \int_{\partial\Omega} \bar{t}_i \delta u_i \,ds \\
%	+ \int_{\partial\Omega} \bar{m}_i n_k \delta u_{i,k} \,dv
%	+ \int_{\partial\partial\Omega} \bar{s}_i \delta u_i \,dl,
%\end{multline}
%and it 

\begin{multline}\label{eq:variationalWF}
\Big(\int_{\Omega} \bigl( \mathbf{B}_1^{T}\mathbf{C}\mathbf{B}_1
+ \mathbf{B}_2^{T}\mathbf{A}\mathbf{B}_2 \bigr)\,d\Omega\Big)\widehat{\textbf u}=\int_{\Omega} \mathbf{N}^{T}\mathbf{b}\,d\Omega
+ \int_{\partial\Omega} \mathbf{N}^{T}\mathbf{t}\,ds\\
+ \int_{\partial\Omega} \overline{\mathbf{N}}^{T}\mathbf{m}\,ds
+ \int_{\partial\partial\Omega} \mathbf{N}^{T}\mathbf{s}\,dl
\end{multline}
where $\widehat{\mathbf{u}}$ is the vector of nodal degrees of freedom that is used for the interpolation of displacement field ($\mathbf{u} = \mathbf{N}\widehat{\mathbf{u}}$). $\mathbf{N}$ is the matrix of shape functions. 
$\mathbf{B}_1$ and $\mathbf{B}_2$ are the derivative matrices containing first and second derivatives of the shape functions, respectively, that are used to define strain $\boldsymbol{\varepsilon} = \mathbf{B}_1\widehat{\mathbf{u}}$ and the strain gradient $\boldsymbol{\kappa} = \mathbf{B}_2\widehat{\mathbf{u}}$. $\mathbf{C}$ and $\mathbf{A}$ are the constitutive matrices that are used to define the stress $\boldsymbol{\tau} = \mathbf{C}\boldsymbol{\varepsilon}$ and the double stress $\boldsymbol{\mu} = \mathbf{A}\boldsymbol{\kappa}$.
The superscript "$^T$" denotes matrix transpose. Explicit forms of the matrices $\mathbf{N}$, $\mathbf{B}_1$, $\mathbf{B}_2$, $\mathbf{C}$ and $\mathbf{A}$ are presented in Appendix A.

For the problems without body forces, edge tractions and double tractions, the weak form \eqref{eq:variationalWF} provides the following discrete system:

\begin{equation}
\begin{gathered}
\mathbf{K}\widehat{\mathbf{u}} = \mathbf{f},\\
\mathbf{K} = \int_{\Omega} \bigl( \mathbf{B}_1^{T}\mathbf{C}\mathbf{B}_1
+ \mathbf{B}_2^{T}\mathbf{A}\mathbf{B}_2 \bigr)\,d\Omega,\qquad
\mathbf{f} = \int_{\partial\Omega} \mathbf{N}^{T}\mathbf{t}\,ds
\label{eq:fem}
\end{gathered}
\end{equation}

\begin{figure}[t!]
	\centering
	
	(a)\includegraphics[width=0.5\textwidth]{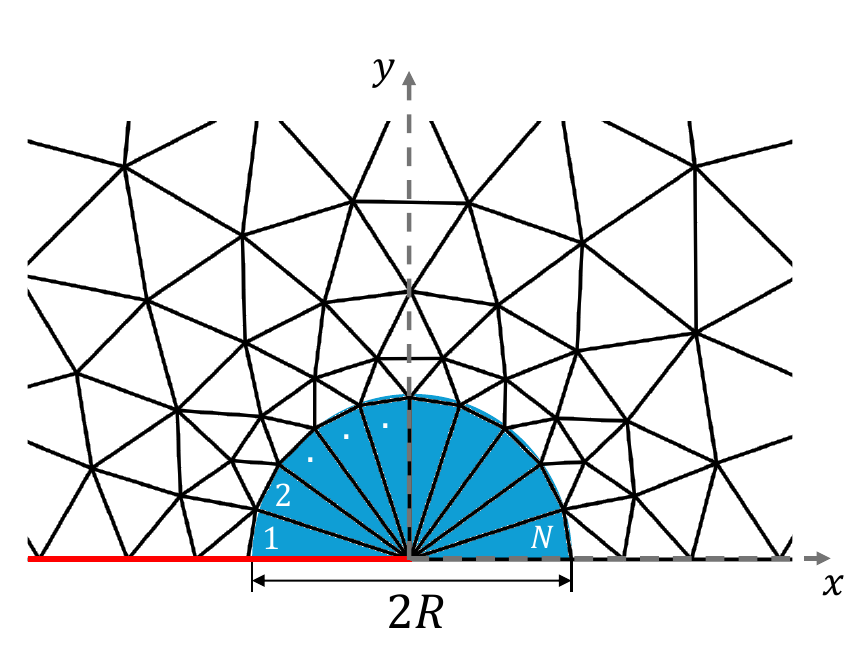}
	
	\vspace{10pt}
	
	(b)\includegraphics[width=0.85\textwidth]{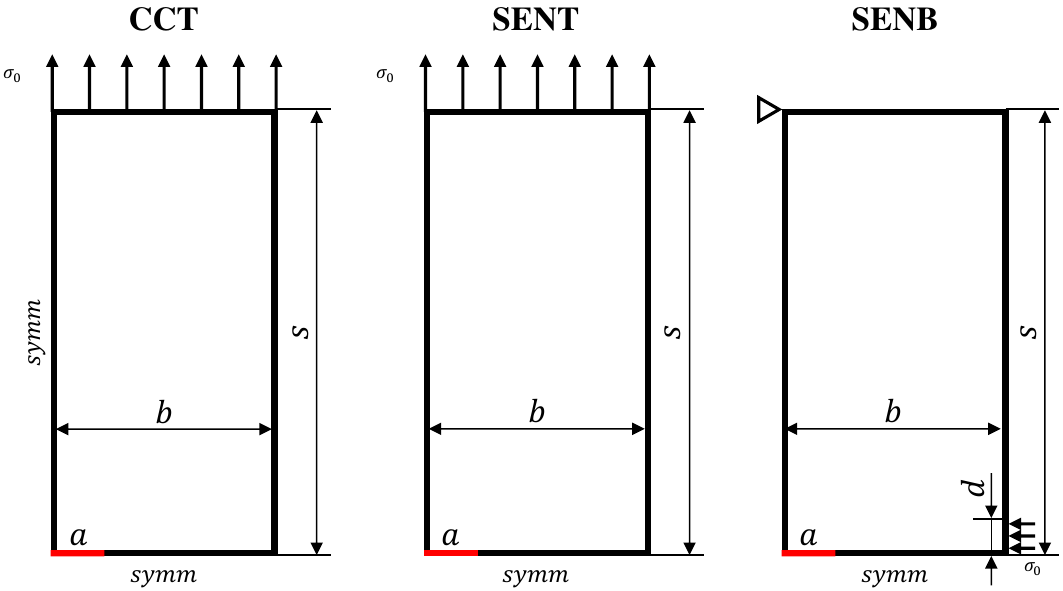}
	
	\caption{(a): Finite element mesh near the crack tip (enriched elements are shown in blue), (b): Schematic representation of the specimen geometries and boundary conditions for the CCT, SENT, and SENB models. The crack face is highlighted in red.}
	\label{fig:model}
\end{figure}

In the considered numerical calculations, the domains were discretized with the three-node triangular elements providing $C^{1}$-continuity of the approximation at the element nodes. $C^{1}$-continuity is required since the second-order derivatives of shape functions should be calculated in \eqref{eq:fem} (in matrix $\textbf B_2$).
The fifth-order polynomials of Bell triangle \cite{dasgupta1990higher} are used as the shape functions following the FEM formulation presented in Refs. \cite{Zervos2001, papanicolopulos2010crack}. 
The enriched elements (placed around the crack tip, Fig. \ref{fig:model}a) contain additional shape functions related to the asymptotic solution for SGE crack problems \cite{solyaev2025enriched}. 
 
Note that the vector of nodal variables $\widehat{\mathbf{u}}$ in the considered method includes the nodal displacements as well as their first and second derivatives within the considered FEM formulation.
Thus, the elements without enrichment have 36 degrees of freedom (12 per node), including displacements and their first and second derivatives. In the general case, the enriched elements have 40 degrees of freedom, including four additional amplitude factors $K_n$ ($n=1...4$) that arise in SGE asymptotic solutions for the plane strain cracks under mixed-mode loading conditions \cite{gourgiotis2009plane, aravas2009plane, solyaev2025enriched}. In the present study, we consider only the Mode I loading so that the number of additional amplitude factors was reduced to two ($K_1, K_2$) and the number of degrees of freedom in the enriched elements was 38. The applied enrichment procedure was similar to those used in classical singular finite elements \cite{benzley1974representation}, with the difference that the asymptotic solution of the SGE is used as the enrichment functions preserving $C^{1}$-continuity of interpolation, as described in~\cite{solyaev2025enriched}.

The J-integral within the simplified SGE for the Mode I cracks can then be evaluated using the obtained amplitude factors $K_1$ and $K_2$~\cite{aravas2009plane,solyaev2025enriched}:
\begin{equation}
J = \frac{1+\eta}{8\mu}\,\pi l^{2}\Bigl((3K_{1}+K_{2})^{2}+8K_{2}^{2}(\eta+2)\Bigr), \label{eq:ji} 
\end{equation}
where $\eta = 3 - 4\nu$ is the Kolosov constant and $\mu$ is a Lam\'e constant.

The method under consideration was implemented in Abaqus. A detailed description of the finite element formulations for the standard shape functions and enrichment functions (including the case of mixed-mode loading with four additional amplitude factors), the integration scheme and Abaqus implementation using a UEL subroutine, can be found in Refs.~\cite{solyaev2025enriched,shelkov2026implementation}.

\subsection{Loading cases and boundary conditions}
Three types of boundary value problems were considered in the numerical simulations according to standard fracture mechanics tests: centre-cracked tension (CCT), single-edge notched tension (SENT), and single-edge notched bending (SENB) (see Fig.~\ref{fig:model}b). The specimen dimensions are defined by the length $s$ and width $b$, with crack length $a$.
The calculations exploited symmetry with respect to the $x$-axis, which coincides with the crack extension line. The corresponding symmetry conditions within SGE are defined by \cite{papanicolopulos2009three, solyaev2025enriched}:
\begin{equation}
 -a\leq x\leq 0, \,\,y=0:\quad v=0, \quad u_{,y}=0
\end{equation}

These conditions correspond to the classical requirements for the absence of normal displacements on the plane of symmetry, as well as to additional generalized symmetry conditions of SGE that explicitly require the normal derivatives of tangential displacements (shear) to vanish on the symmetry plane. In the FEM implementation, at nodes lying on the symmetry plane ($y=0$), the corresponding degrees of freedom ($v, u_{,y}$) in the vector $\widehat{\mathbf{u}}$ were set to zero. In addition, to ensure that the symmetry conditions are satisfied between the nodes, it is further necessary, within the framework of a $C^1$-FEM, to explicitly set to zero the corresponding first and second tangential derivatives of displacements, which are also included in $\widehat{\mathbf{u}}$ (these are $v_{,x}, v_{,xx}, u_{,xy}$)\cite{papanicolopulos2009three,solyaev2025enriched}.

For the CCT specimen, symmetry about the $y$-axis was additionally taken into account: 
\begin{equation}
  x=0:\quad u=0, \quad v_{,x}=0
\end{equation}
that were prescribed explicitly for the corresponding components of $\widehat{\mathbf{u}}$ together with additional conditions for the tangential derivatives ($u_{,y}=0$, $u_{,yy}=0$, $v_{,xy}=0$) needed for $C^1$-FEM.

With regard to the applied loading, a uniform traction $\bar t_2=\sigma_0$ was applied at the top edge for the CCT and SENT specimens. For the SENB specimen, the load was applied as a traction $\bar t_1=\sigma_0 = P/(2hd)$ distributed over a small area of size $d=0.01b$, yielding a resultant total bending force $P$ applied in the three-point bending experiment. On all other faces, including the crack faces, the traction-free ($\bar t_i=0$) and double-traction-free ($\bar m_i=0$) boundary conditions were imposed. 

The number $N$ and the size $R$ of the enriched elements placed around the crack tip (Fig. 1a) were selected in accordance with the results of convergence analysis presented in the next section.

%%%%%%%%%%%%%%%%%%%%%%%%%%%%

\section{Convergence analysis}
\label{sec:convergence}

To ensure the accuracy and reliability of the numerical solutions obtained with the enriched $C^1$-FEM formulation described in Section 3, a series of convergence studies was performed using a CCT specimen with a crack-to-width ratio $a/b=0.01$ (approximating a crack in an infinite plate). Two aspects of the finite-element scheme were investigated: the number of integration points required for the enriched elements, and the mesh refinement near the crack tip. The last one is defined by number of elements $N$ and their size $R$. In addition, the convergence of the numerical solution to the known semi-analytical solution \cite{gourgiotis2009plane} for the calculated $J$-integral was assessed.

\subsection{Integration order}

\begin{figure}[b!]
\centering
(a)\includegraphics[width=0.28\textwidth]{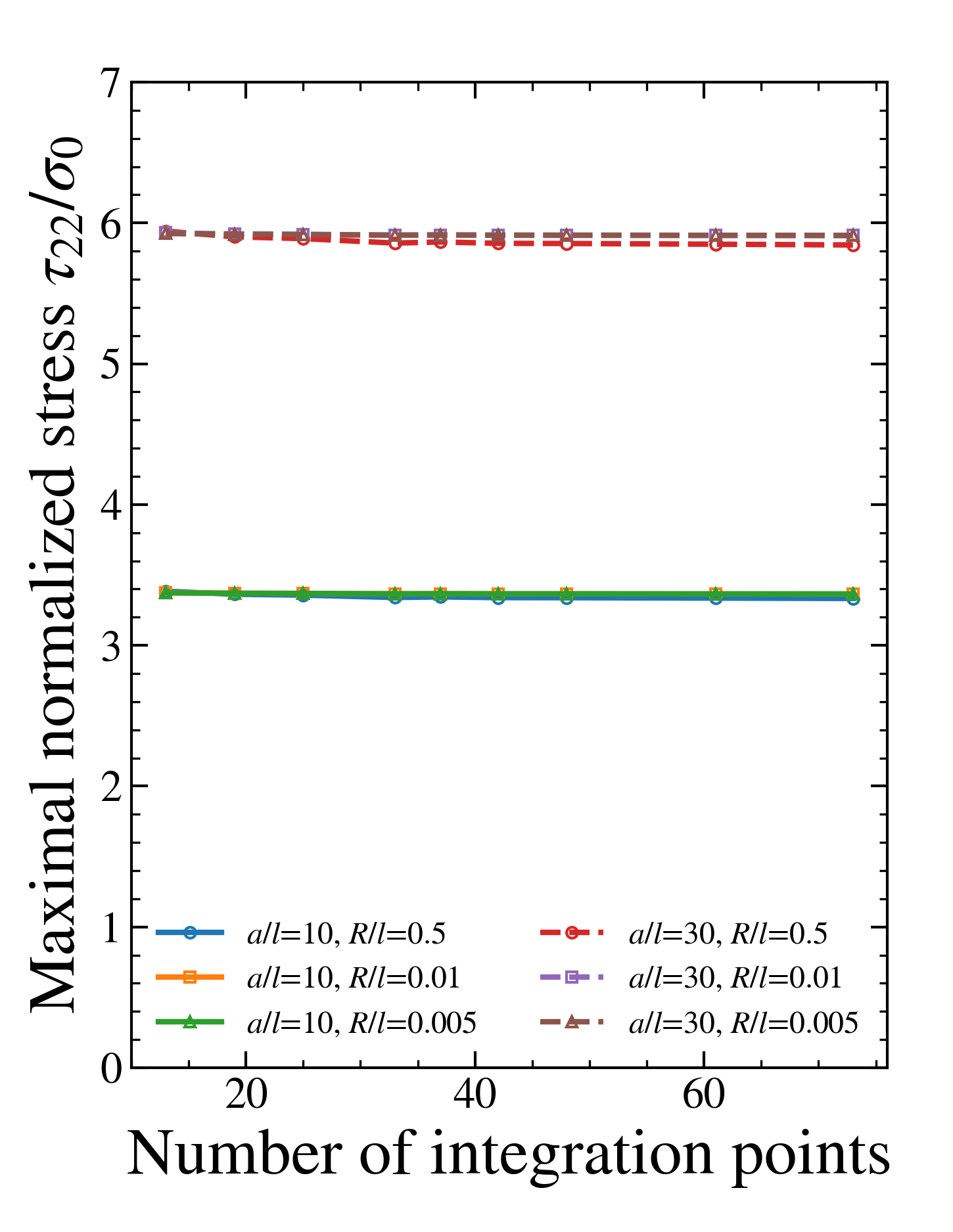}
(b)\includegraphics[width=0.28\textwidth]{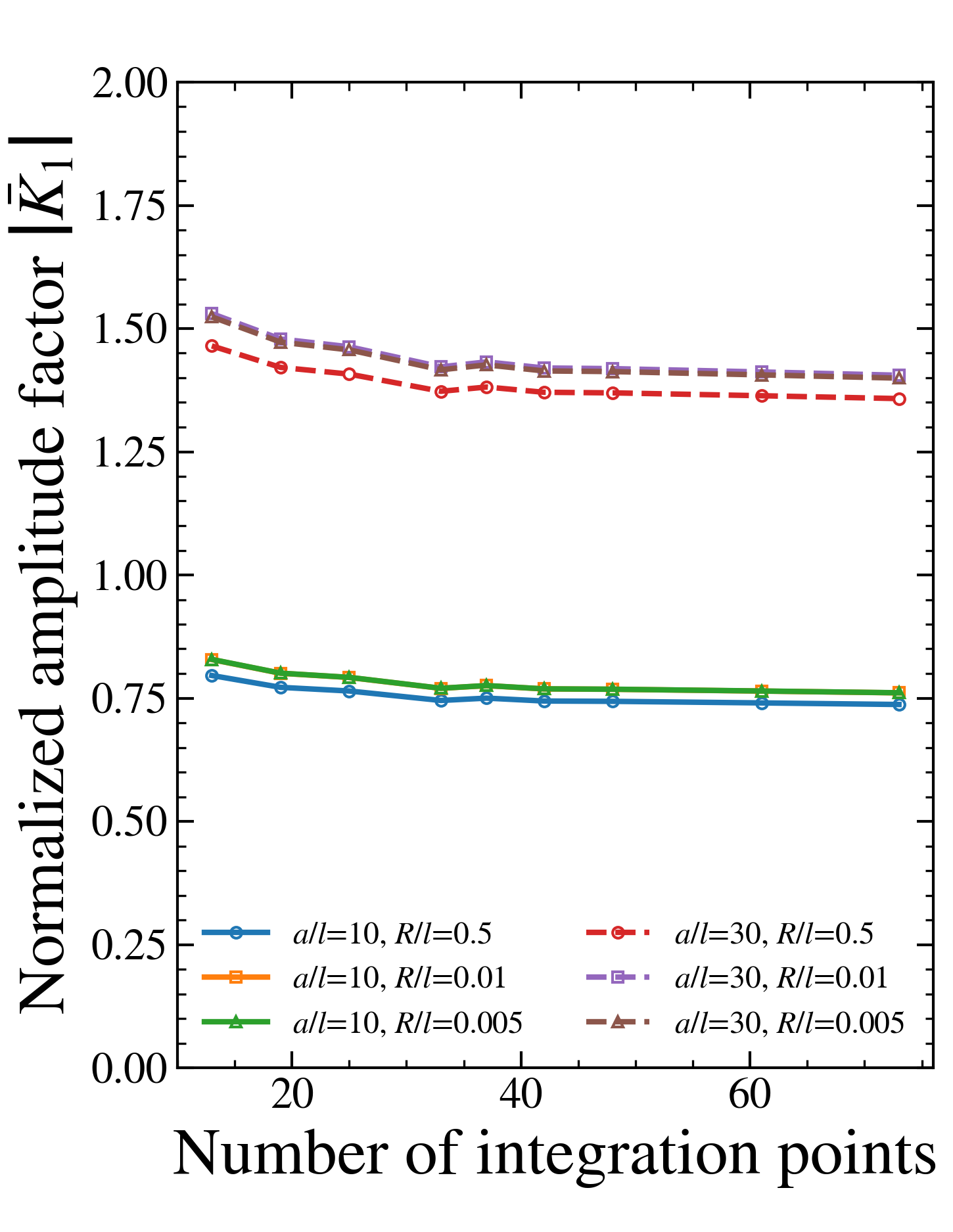}
(c)\includegraphics[width=0.28\textwidth]{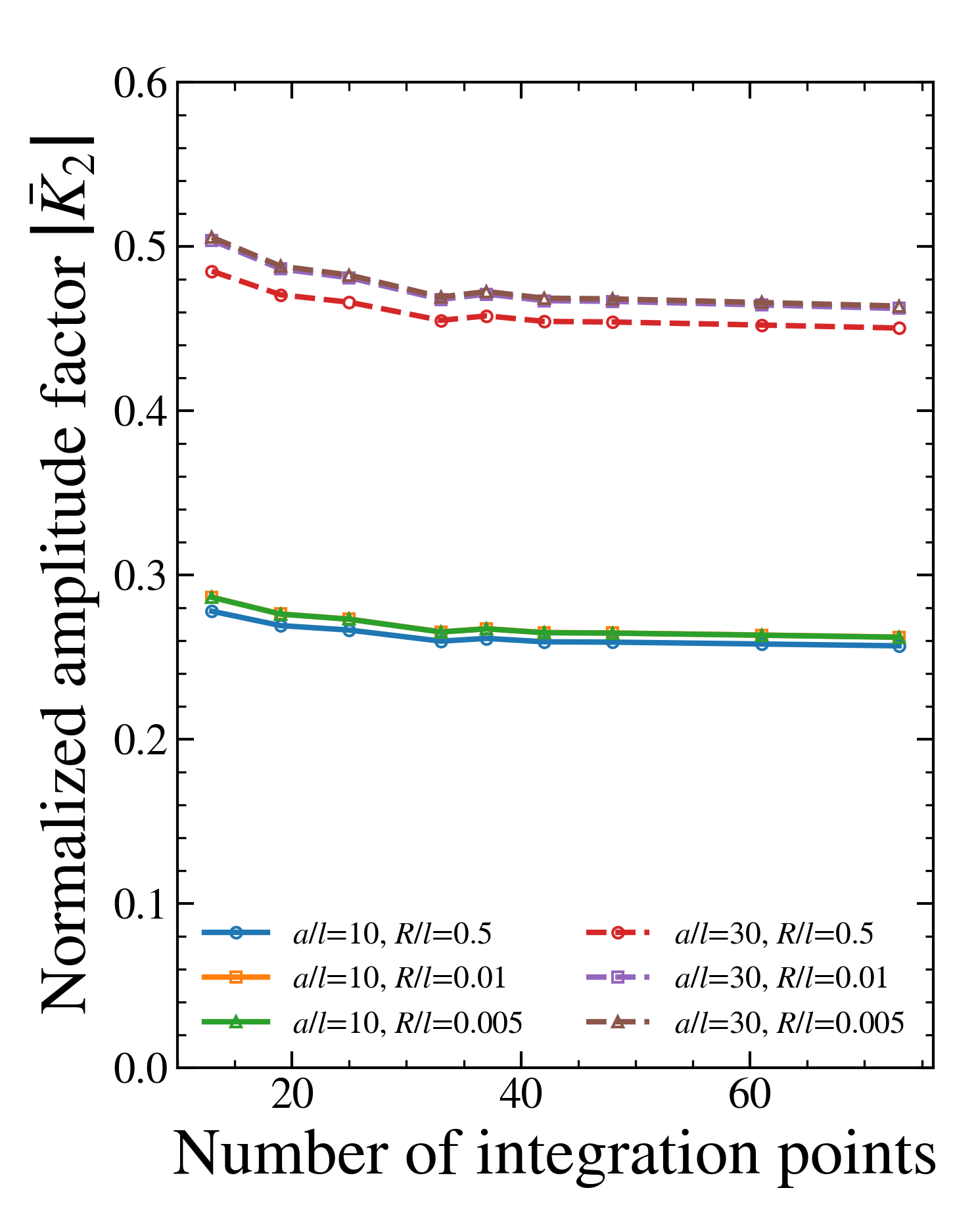}
\caption{Convergence of the enriched $C^1$-FEM solution with respect to the number of integration points: (a) regularised stress $\tau_{22}$, (b) first amplitude factor $K_1$, (c) second amplitude factor $K_2$}
\label{fig:inter}
\end{figure}

The enriched elements incorporate additional shape functions derived from the asymptotic SGE crack-tip solutions~\cite{solyaev2025enriched}. Unlike the standard three-node triangular elements of $C^1$-continuous FEM, which employ a reduced Gauss quadrature with 13 integration points~\cite{zervos2001finite}, the enriched elements in SGE require a higher integration order to accurately evaluate the stiffness matrix contributions associated with the steep gradients near the crack tip.

A convergence study was carried out by examining the dependence of the computed maximum regularised stresses at the crack tip $\tau_{22}$ \eqref{eq:stress} and the two amplitude factors $K_1$, $K_2$ on the number of Gauss integration points. The Gauss integration schemes were taken from \cite{dunavant1985high}. The results are presented in Fig.~\ref{fig:inter}. It was found that 42 integration points per enriched element are sufficient to obtain converged values for all quantities of interest, while the standard 13-point rule \cite{zervos2001finite} can be used in all other non-enriched elements. This result refines the previous finding reported in~\cite{solyaev2025enriched}, where 13 integration points were considered sufficient for enriched elements. However, that study~\cite{solyaev2025enriched} only examined the convergence of regularised stresses, and not that of the amplitude factors. As shown in Fig.~\ref{fig:inter}(b,c), a larger number of integration points is required for $K_1$, $K_2$ to converge. In this figure and in the following, the values of amplitude factors are normalised with respect to the dimensional group: $(\widetilde K_i=K_i\sigma_0/\sqrt{l}$ ($i=1,2$).

\subsection{Mesh refinement}

A second convergence study was conducted to assess the influence of the size and number of enriched elements arranged around the crack tip. Several mesh configurations were examined, each characterized by the relative size of the smallest element $R/l$ and the number of radial layers of elements $N$ emanating from the crack tip (see Fig. 1a).

\begin{figure}[htbp]
\centering
(a)\includegraphics[width=0.28\textwidth]{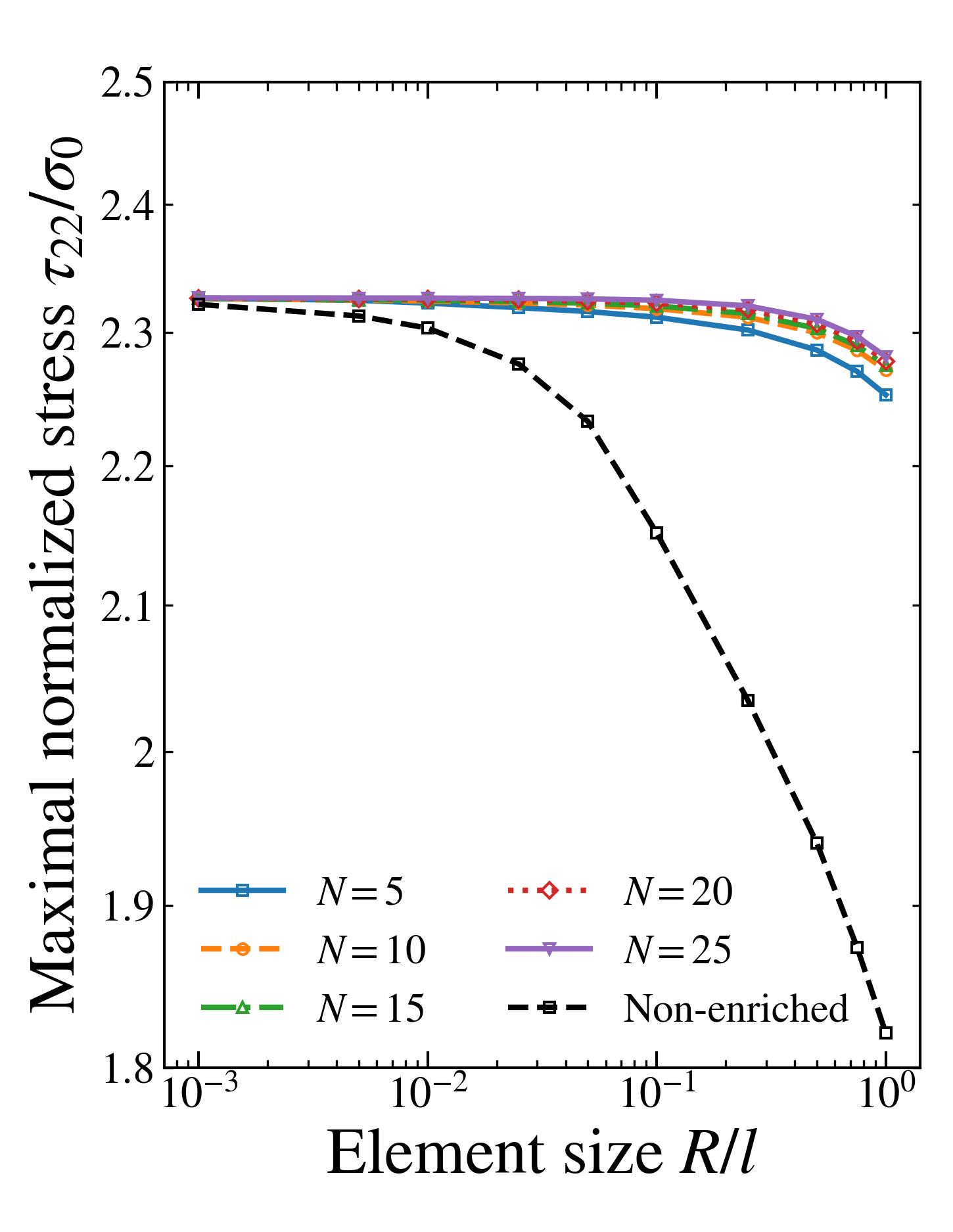}
(b)\includegraphics[width=0.28\textwidth]{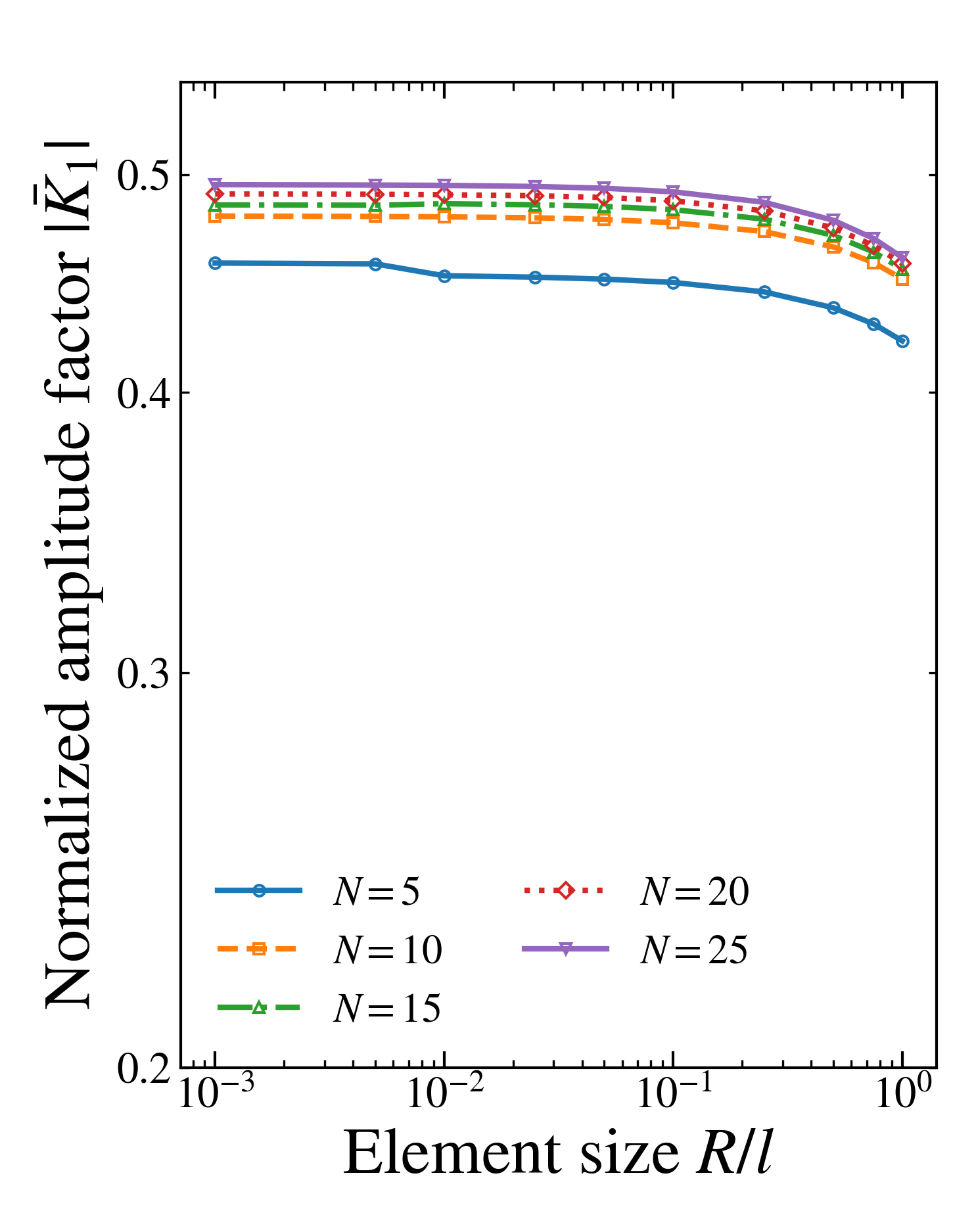}
(c)\includegraphics[width=0.28\textwidth]{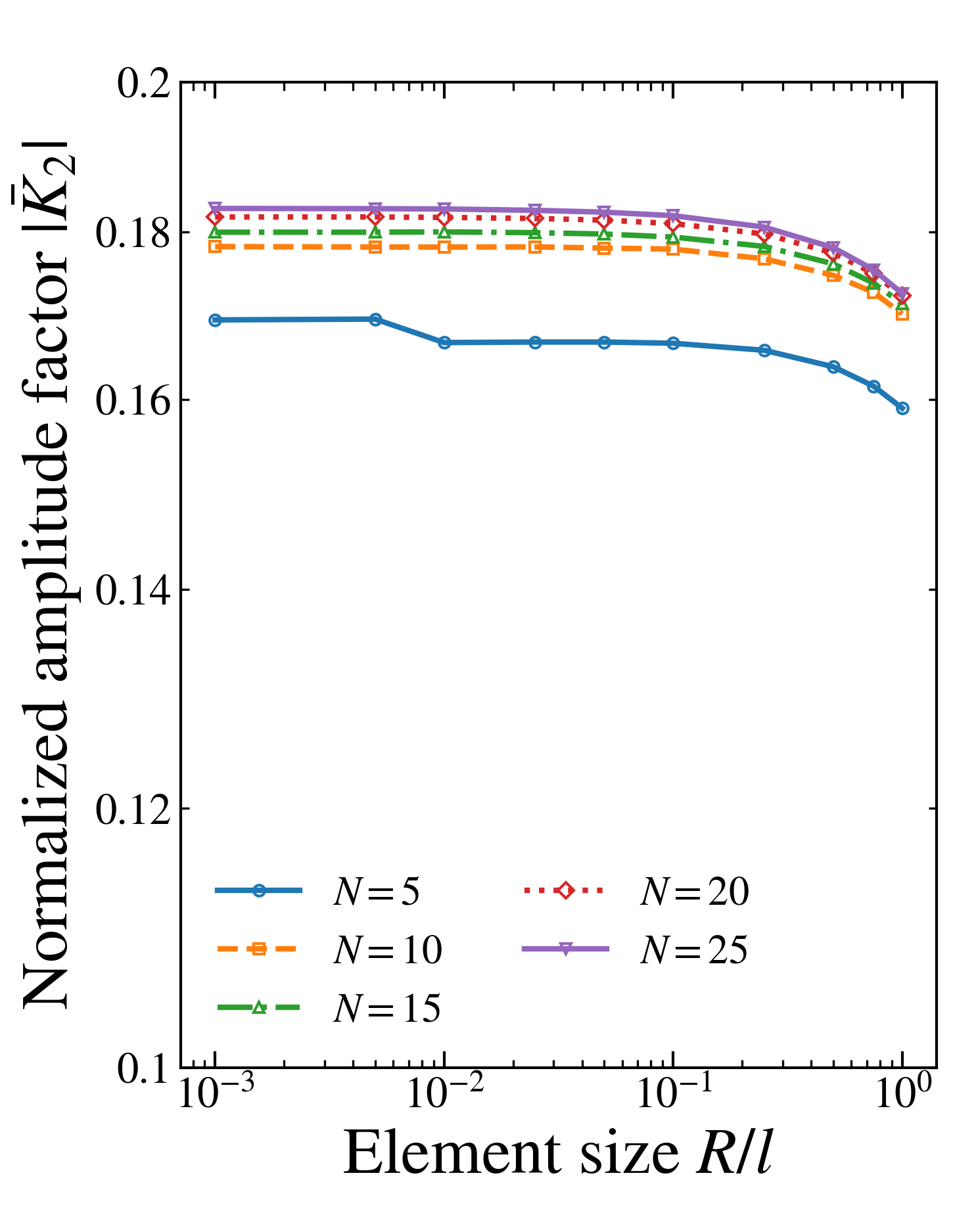}
\caption{Convergence of the enriched $C^1$-FEM solution with respect to the element size at the crack tip for different numbers of elements around the crack tip: (a) stress $\tau_{22}$, (b) first amplitude factor $K_1$, (c) second amplitude factor $K_2$}
\label{fig:conv}
\end{figure}

The convergence of maximum stress $\tau_{22}$ and amplitude factors $K_1$, $K_2$ is shown in Fig.~\ref{fig:conv} for different mesh densities. 
%The results demonstrate that a sufficient number of elements around the crack tip is required to capture the regularised stress fields characteristic of the SGE solution. Based on this study, recommendations for the optimal mesh configuration can be established, balancing accuracy and computational efficiency. 
The results show that reliable stress estimates can be obtained using $R/l=0.1$ and $N=5...10$. However, the computed amplitude factors depend strongly on $N$, and at least $N=15$ is required for their accurate determination.

\subsection{Validation of calculated J-integral values}

\begin{figure}[t]
\centering
(a)\includegraphics[width=0.45\textwidth]{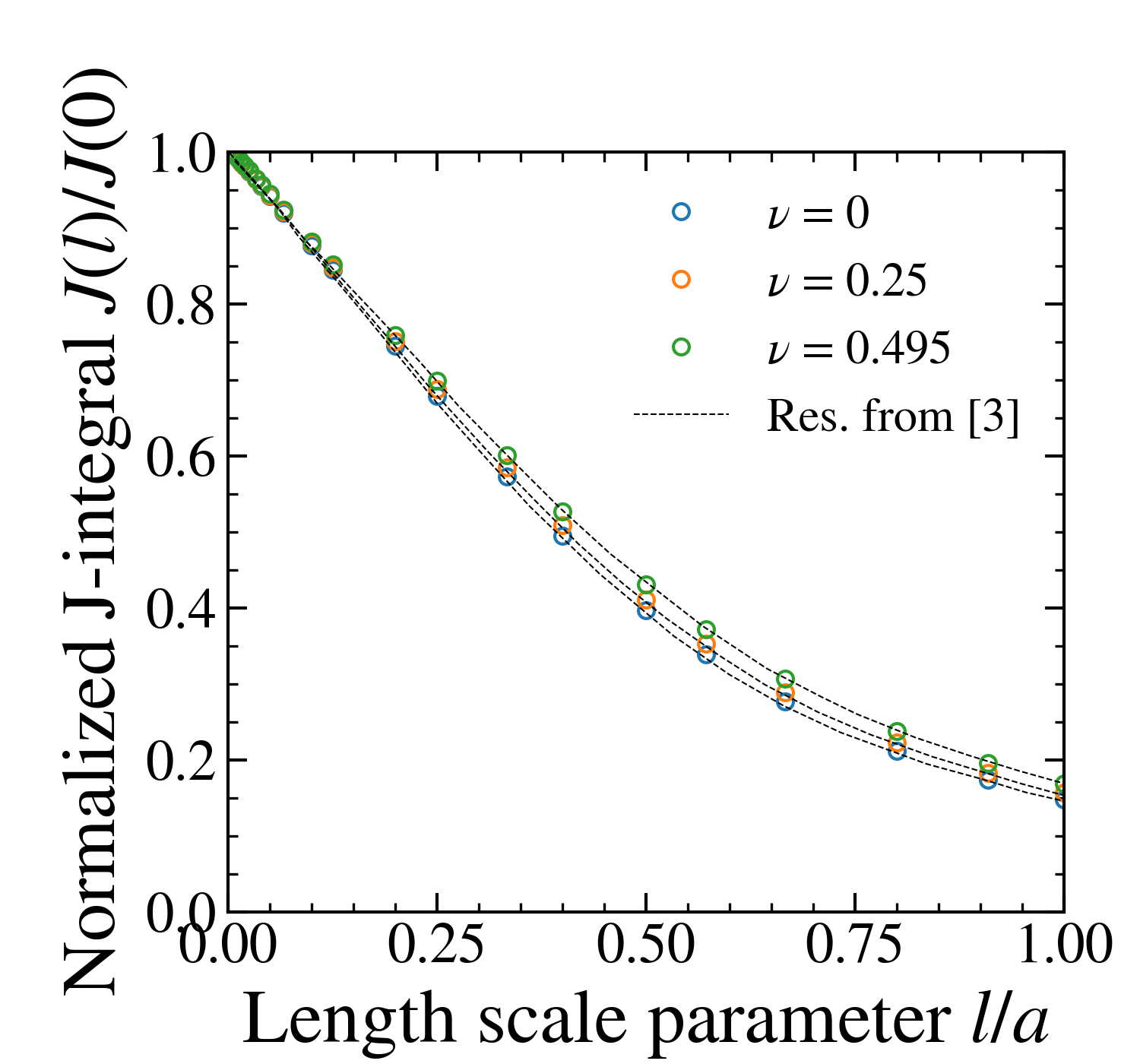}
(b)\includegraphics[width=0.45\textwidth]{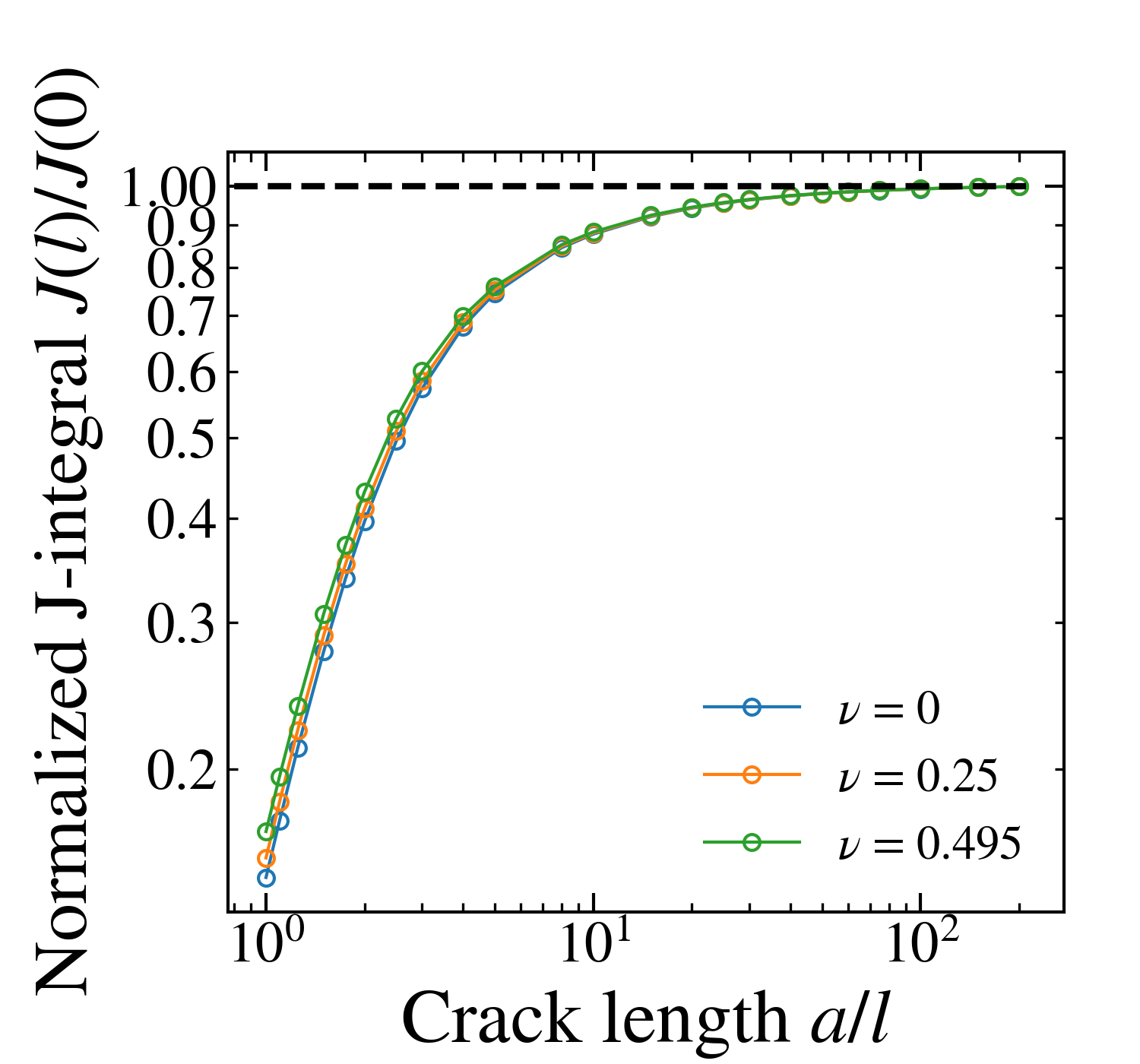}
\caption{Validation of calculated $J$-integral values: (a) comparison of obtained numerical solution (dots) with the semi-analytical solution of Gourgiotis and Georgiadis~\cite{gourgiotis2009plane} (lines), (b) convergence to the classical LEFM limit of numerical solution.}
\label{fig:j_valid}
\end{figure}

To further validate the numerical implementation, the J-integral \eqref{eq:ji} computed using the enriched elements was compared with the semi-analytical results of Gourgiotis and Georgiadis~\cite{gourgiotis2009plane}. The comparison was performed over a range of the ratios between the length scale parameters of material and the crack length \(l/a\). The results are presented in Fig.~\ref{fig:j_valid}a. In these calculations, we used CCT loading conditions with enriched elements size $R/l = 0.01$ and number $N=26$.  An agreement between the numerical and reference solutions is observed in Fig. 4a, confirming the accuracy of the implemented numerical method.

Finally, convergence of the numerical SGE solution to the classical LEFM solution in terms of the J-integral was examined. This convergence is already visible in the Fig. 4a comparing with the results of Gourgiotis and Georgiadis, where the region of solutions close to classical ones is determined by small ratios $l/a$, and the J-integral computed within SGE ($J(l)$), normalised by the classical value ($J(0)$), tends to 1. To show this convergence of the solution to the classical case more clearly, in Fig. 4b we present the obtained dependence of $J(l)/J(0)$ on the normalised crack length $a/l$.

For long cracks ($a/l>>1$), the SGE solution is expected to asymptotically approach the classical LEFM predictions. The results, shown in Fig.~\ref{fig:j_valid}b, confirm this behaviour of the numerical solution and demonstrate the consistency of the proposed approach with classical fracture mechanics in the limiting case of long cracks from the viewpoint of energy-based fracture criterion.\\

Thus, the presented convergence studies confirm that the proposed enriched $C^1$-FEM formulation, with an appropriate integration order and mesh refinement around the crack tip, provides accurate and reliable results for SGE crack problems. Based on these results, all subsequent computations were performed using enriched elements with 42 integration points and a mesh with $N=26$ enriched elements around the crack tip and $R/l = 0.01$. It should be noted that this configuration was chosen to achieve superior accuracy. For practical applications, acceptable accuracy can be obtained using a coarser mesh with $R/l = 0.1$ and 10--15 elements around the crack tip.

%%%%%%%%%%%%%%%%%%%%%%%%%%%%

\section{Methodology for the length scale parameter identification}
\label{sec:methodology}

The simplest estimates typically assume that the length scale parameter of SGE is of the order of the characteristic size of the material's microstructure (e.g., the grain size) \cite{mindlin1964microstructure}. Refined analytical estimates were obtained in Ref. \cite{gitman2005representative}, where a relation was proposed linking this parameter to the size of a representative volume element of the heterogeneous material structure. More general homogenization methods for different type of inhomogeneities were also developed within SGE (see, e.g. \cite{bacca2013mindlin,ganghoffer2021variational}). 
%Closed-form estimates for the length scale parameters were proposed within the approximations of the dilute inclusion model~\cite{} and the self-consistent approach \cite{}. Numerical estimates within the energy-based homogenization methods were widely implemented \cite{}.

The possibility of relating the length scale parameter of simplified SGE to conventional fracture-toughness characteristics was demonstrated in \cite{askes2015understanding}. Using the implicit gradient method, that study estimated the proportionality between the SGE length scale parameter ($l$) and the LEFM characteristic length ($L_{LEFM}=\frac{K_{Ic}^{2}}{\pi\sigma_{\mathrm{ult}}^{2}}$), yielding the closed-form relation ($l=L_{LEFM}/8$). More recently \cite{vasiliev2021new}, the numerical simulations based on a mixed finite-element formulation, together with an analysis of experimental data, showed that, for quasi-brittle materials, the SGE length scale parameter $l$ may be of the same order of magnitude as $L$, although it can be either smaller or larger depending on the material.

The approaches proposed in \cite{askes2013intrinsic,askes2015understanding,vasiliev2019estimation,vasiliev2021failure,vasiliev2021new} are based on the assumption that, within the strain gradient and non-local elasticity theories, the fracture load of a cracked body can be determined by analysing the stress field near the crack tip and applying a conventional strength criterion. A key distinguishing feature of gradient theories is that the crack-tip stress field remains regular, i.e. nonsingular. In the present study, we adopt the same approach to evaluate fracture loads and derive regression relations that can subsequently be used to process experimental data and identify the SGE length scale parameters of various materials. In essence, this is a stress-based approach founded on classical strength analysis.
Specifically, the maximum principal stress at the crack tip (corresponding to the maximum hoop stress under Mode I loading) is compared with the ultimate strength of the material. Note that in conventional failure analysis, the maximum principal stress criterion provides a reasonably accurate description of brittle failure under highly triaxial stress states, such as those arising under pure Mode I loading. Here, we extend this criterion to cracked bodies using the regular crack-tip solutions provided by SGE .
More sophisticated failure criteria may be required for complex stress states and mixed-mode loading \cite{vasiliev2021new, Solyaev2026}. Also, more general constitutive relations of SGE then those given by Eq. \eqref{eq:stress} may be required. The present study is therefore restricted to pure Mode I within the simplified SGE, while the development of more general identification procedures for mixed-mode conditions is left for future investigation.

%In this paper, we employ the last approach and develop two approaches for identification of the length scale parameter \(l\) based on numerical SGE solutions and experimental data for brittle and quasi-brittle materials. The first method, described below in Section~\ref{sec:size_effect}, exploits the equivalence between SGE and classical LEFM predictions for long cracks to establish a direct relationship between the length scale parameters of the two theories ($l$ and $L_{LEFM}$). The second method, presented in Section~\ref{sec:regression}, provides regression approximations of the nominal strength for three standard fracture mechanics specimens, enabling parameter identification without the need for additional numerical simulations.

In this study, we develop two approaches to SGE length scale parameter identification. The first is intended for ideally brittle materials with crack lengths substantially greater than the length scale parameter (Section~\ref{sec:size_effect}). Within this approach, we demonstrate that Irwin's K-based LEFM criterion and the stress-based SGE analysis yield equivalent predictions if we use the appropriate proportionality relation between their respective length scale parameters. Accordingly, we refine the previously proposed proportionality coefficient relating $L_{LEFM}$ and $l$ in Ref. \cite{askes2015understanding}. To this end, we perform high-precision numerical simulations using a $C^1$-continuous FEM, with all boundary conditions explicitly taken into account. The second method, presented in Section~\ref{sec:regression}, provides regression approximations of the nominal strength for three standard fracture mechanics specimens, enabling parameter identification without the need for additional numerical simulations.

The numerical analysis was performed as follows. For each specimen type (Fig. \ref{fig:model}b), a series of finite element simulations were performed to obtain the dependence of the stress concentration factor $K_t$ (for regularised stresses $\tau_{ij}$ \eqref{eq:stress}), the nominal strength $\sigma_n$, the amplitude factors $K_1$, $K_2$, and the J-integral values \eqref{eq:ji} on the normalised crack length \(a/l\) for different ratios \(a/b\) ($b$ is the specimen width, see Fig. \ref{fig:model}b). In all models, the crack size was fixed, while the length scale parameter \(l\) and the specimen width \(b\) were varied to achieve different \(a/l\) and \(a/b\) values. The ratios of the specimen length to its width were kept constant throughout the study: for the CCT and SENT specimens, the length to width ratio of the domain was \(s/b = 3\), while for the SENB specimen it was \(s/b = 2\). These values were chosen in accordance with the standard geometric requirements for the considered fracture mechanics specimens.

The stress concentration factor at the crack tip is evaluated as $K_t=\tau_{22}/\sigma_{ref}$ for each considered specimen geometry and crack length.
The reference stress $\sigma_{ref}$ in CCT and SENT tests equals the applied far-field stress $\sigma_{ref}=\sigma_0$. For the SENB test the reference stress is calculated taking into account the non-uniform distribution of stress under bending: $\sigma_{ref}=3\sigma_0sd/b^2$. It defines the maximum stress under bending in the specimen without crack. Similar normalization is also used in classical LEFM analysis for bending tests \cite{tada2000stress}.

In the considered experiments with the Mode I loading, the maximum principal stress at the crack tip is governed entirely by the $\tau_{22}$ stress component (see, e.g. \cite{papanicolopulos2010crack,solyaev2024higher}). By applying the maximum principal stress criterion, we implicitly assume that fracture occurs once this stress reaches the tensile strength of the material, i.e., $\tau_{22}=\sigma_{ult}$. Accordingly, if we define the normalized nominal strength $\sigma_n$ as the ratio of the actual critical value of applied stress ($\sigma_0$) of the cracked specimen to the intrinsic tensile strength of the material ($\sigma_{ult}$), it follows that this nominal strength is equal to the inverse of the stress concentration factor: $\sigma_n = 1/K_t$ ~\cite{askes2015understanding, vasiliev2021failure}.
%
%\begin{equation}\label{eq:nominal_strength_GTE}
%\sigma_n = \frac{1}{K_t}
%\end{equation}

\begin{figure}[t!]
\centering
(a)\includegraphics[width=0.45\textwidth]{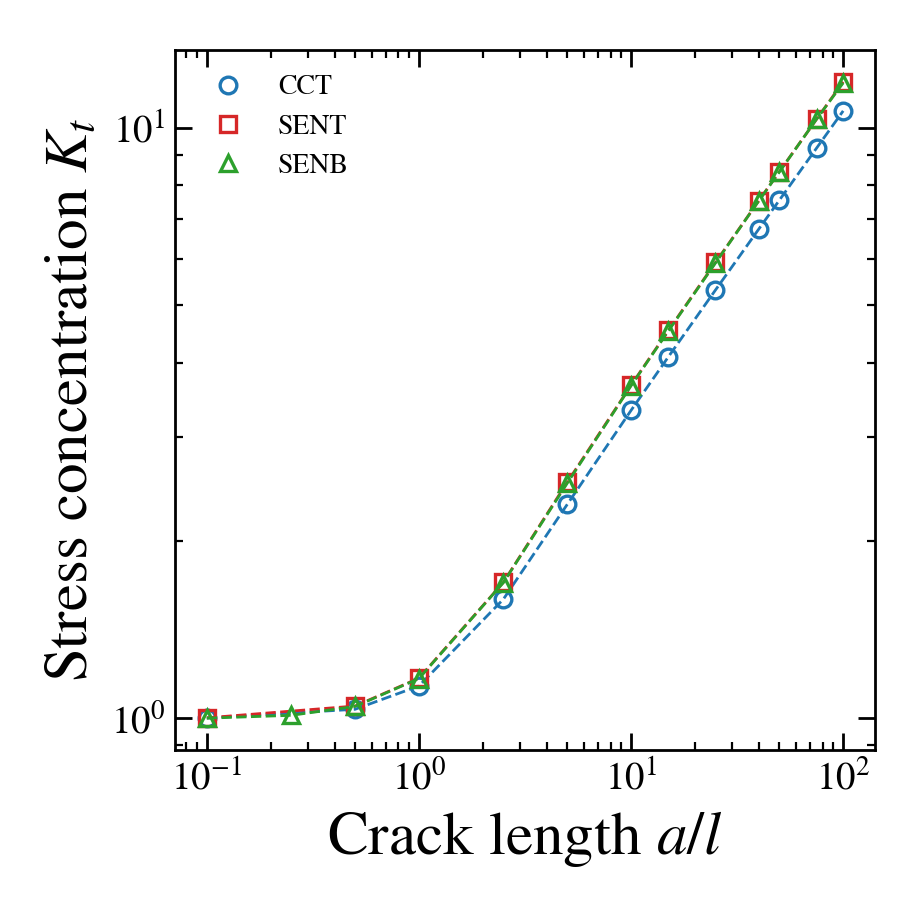}
(b)\includegraphics[width=0.45\textwidth]{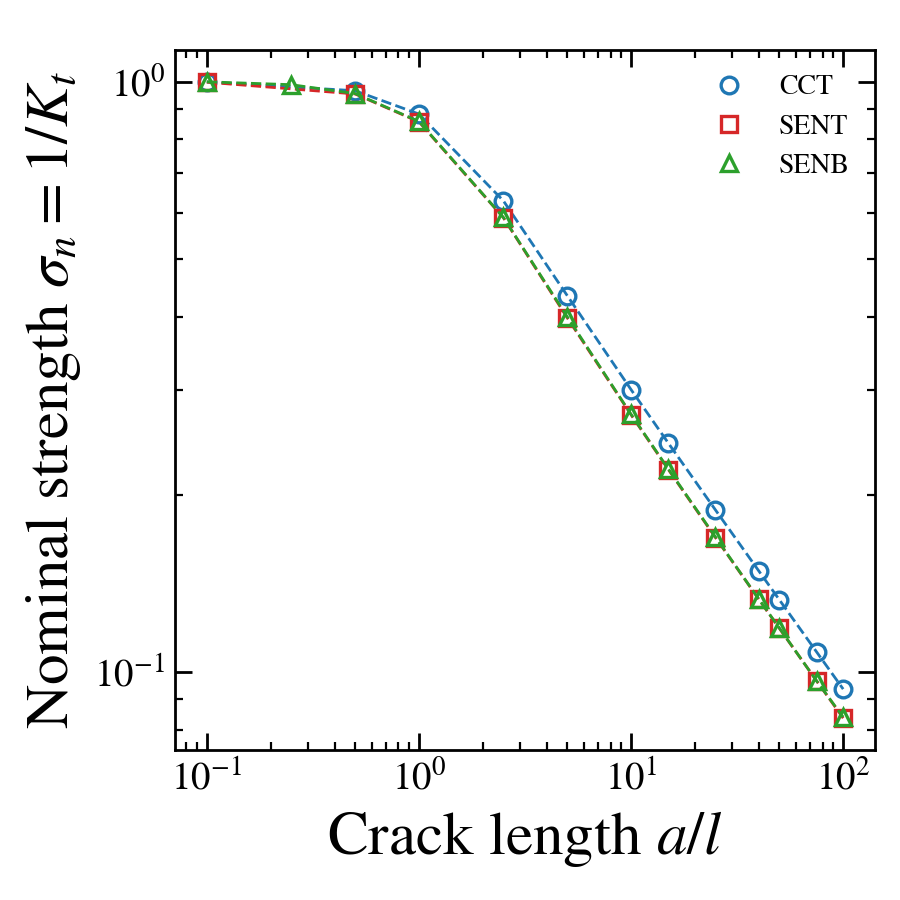}
\caption{Typical dependence of stress concentration factor $K_t$ (a) and nominal strength $\sigma_n$ (b) on the normalised crack length \(a/l\) for different specimens in fracture mechanics tests evaluated within SGE for $a/b=0.1$.}
\label{fig:numerical_result}
\end{figure}

Typical examples of the calculated dependences of $K_t$ and $\sigma_n$ on crack length are presented in Fig. \ref{fig:numerical_result}. As can be seen, these dependences reveal two distinct regions -- one corresponding to long cracks and the other to relatively short cracks. For long cracks ($a/l>>1$), the linear relationship between $\sigma_n$ and crack length in logarithmic coordinates allows these predictions to be directly compared with LEFM. This is the basis for the first identification procedure we have implemented for determining the length scale parameter of brittle materials (Section 5.1). For short crack lengths, the dependence of $\sigma_n$ on $a/l$ becomes nonlinear and represents a variant of the non-classical size effect for cracks that can be described within SGE to capture the transition from the long-crack to the short-crack regimes \cite{askes2015understanding, vasiliev2021failure}. This transition crack-size range is addressed in the second identification procedure, which is based on the use of the regression relations we have constructed (Section 5.2).  

In the following sections, we present the dependence of the calculated nominal strength $\sigma_n$ on the crack size and specimen dimensions. The comparison with LEFM predictions and the regression approximation is performed with the use of these calculated values of $\sigma_n$. The calculated amplitude factors $K_n$ and J-integrals for all considered tests are presented in Appendix B.

\subsection{Size effect in brittle materials (long cracks)}
\label{sec:size_effect}
The first method for determining the length scale parameter \(l\) is based on the fact that the dependence of the nominal strength within SGE, in the long-crack regime, asymptotically follows the same power law as the classical size effect law of LEFM, i.e. $\sigma_n\sim a^{-1/2}$ ~\cite{vasiliev2021failure}.
Within LEFM, the critical fracture stress for an infinite plate with a central crack is given by:
\begin{equation}\label{eq:LEFM_critical_stress}
	\sigma_c = \frac{K_{Ic}}{\sqrt{\pi a}}.
\end{equation}

The nominal strength of the classical solution can then be expressed as
\begin{equation}\label{eq:relative_LEFM}
	\sigma_n^{\text{LEFM}} = \frac{\sigma_c}{\sigma_{ult}} = \frac{K_{Ic}}{\sigma_{ult}\sqrt{\pi a}} = \Big(\frac{a}{L_{LEFM}}\Big)^{-1/2},\qquad
	L_{LEFM} = \frac{1}{\pi}\left(\frac{K_{Ic}}{\sigma_{ult}}\right)^{\!2}
\end{equation}

\begin{figure}[b!]
	\centering
	(a)\includegraphics[width=0.45\textwidth]{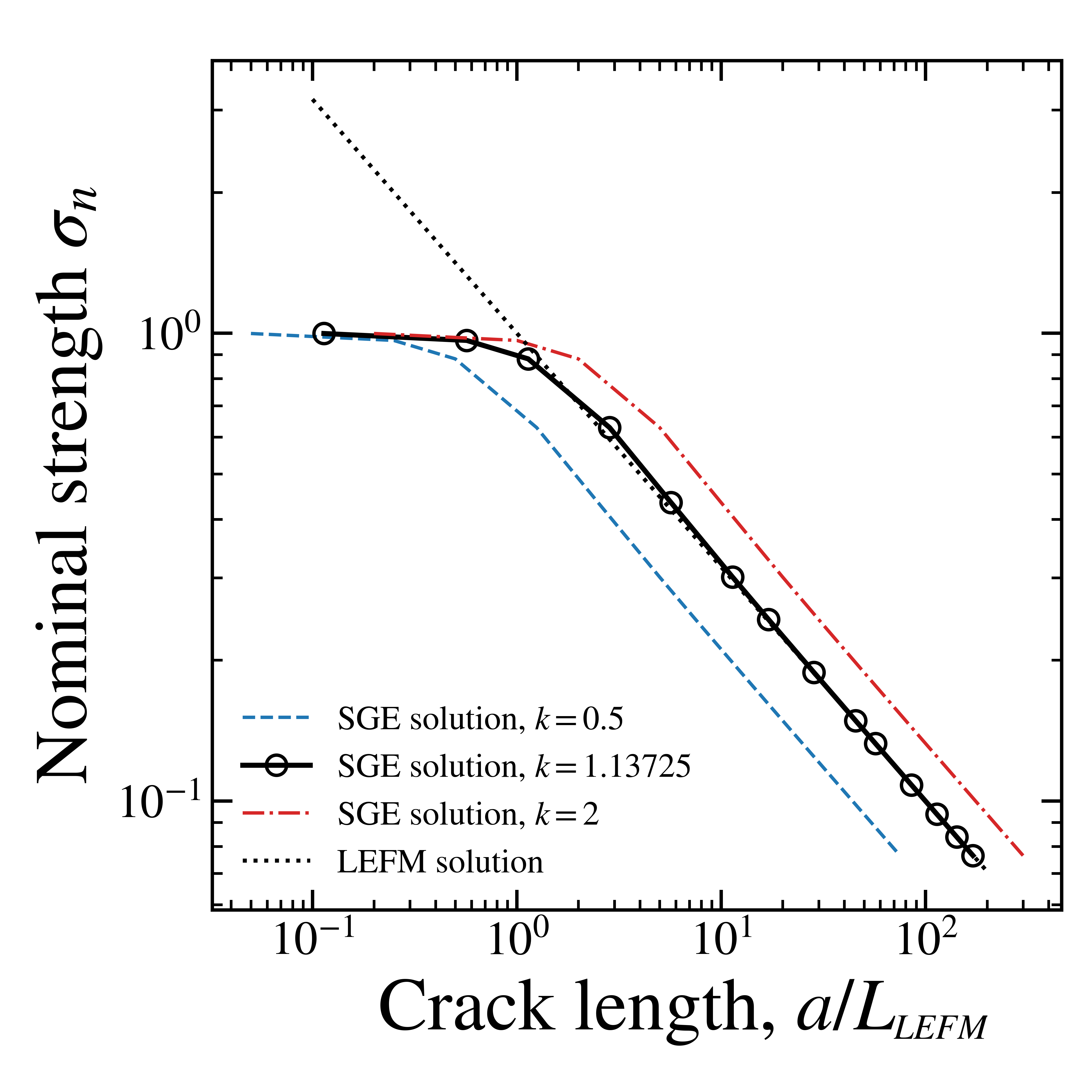}
	(b)\includegraphics[width=0.45\textwidth]{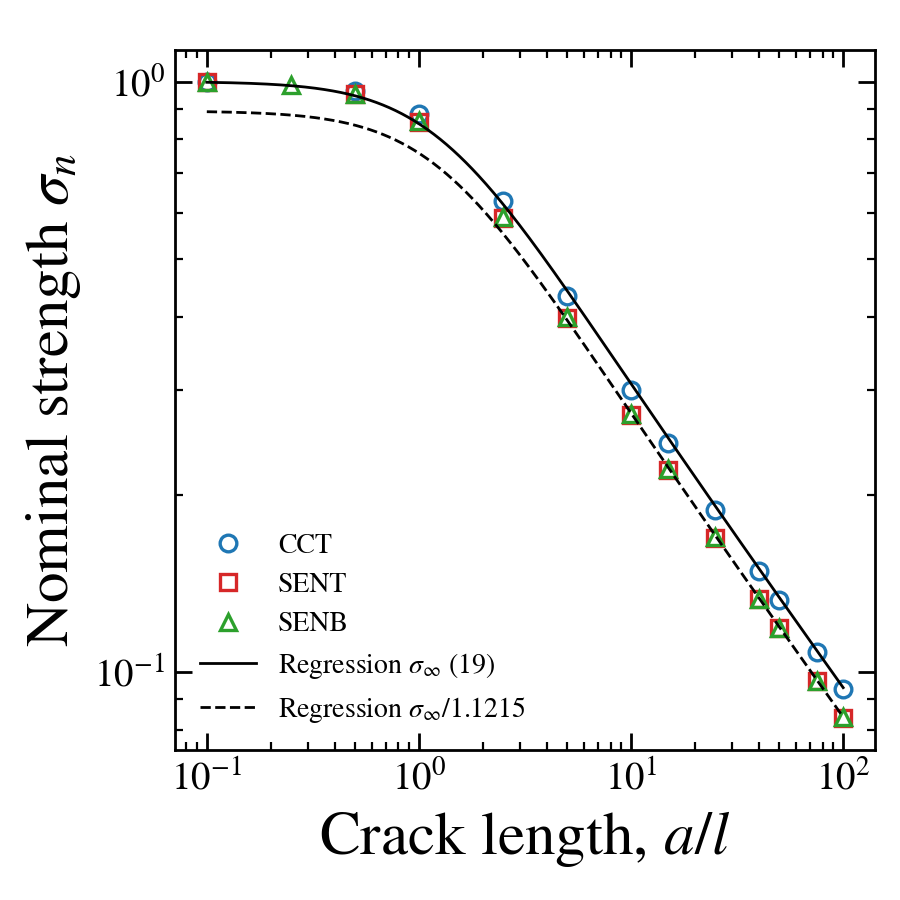}
	\caption{Nominal strength for a central crack in an infinite plate: (a) illustration of the SGE length scale parameter identification procedure, in which the different values of proportionality coefficient is used in Eq. \eqref{eq:ell_vs_l}, (b) comparison between regression relation \eqref{eq:sigmainf} and the numerical SGE solution}
	\label{fig:nomstr}
\end{figure}

Within SGE, we then consider a CCT model with a very large ratio between the plate width $b$ and the crack length $a$, approximating an infinite plate. Specifically, we use the ratio $b/a=1000$ and determine the dependence of nominal strength on the crack length for different values of the length scale parameter $l$. The comparison of classical LEFM analytical solution \eqref{eq:relative_LEFM} and SGE predictions, obtained based on the numerical simulations, is presented in Fig. \ref{fig:nomstr}a. It can be seen that for the large values of the ratio \(a/l\) the nominal strength within SGE asymptotically obeys the same power law \(\sim a^{-1/2}\) as in classical fracture mechanics. 
We can find the value of $l$ that provides a one-to-one correspondence between LEFM and SGE predictions for nominal strength, i.e. the mapping between dashed line (LEFM solution) and solid line (SGE solution) in Fig.~\ref{fig:nomstr}a. This correspondence is obtained when we use the following value of the length scale parameter:
\begin{equation}\label{eq:ell_vs_l}
	l = kL_{LEFM} = \frac{k}{\pi} \left( \frac{K_{Ic}}{\sigma_{\text{ult}}} \right)^{\!2}
	= 0.361999 \left( \frac{K_{Ic}}{\sigma_{\text{ult}}} \right)^{\!2}
\end{equation}
in which $k = 1.137254$ is the proportionality coefficient that provides the best fit between SGE predictions and LEFM analytical solution with a mean absolute percentage error (MAPE) $0.24\%$ for $a/L>10$ and $0.14\%$ for $a/L>100$. The value of this error measure is evaluated as 
$$\mathrm{MAPE}=\frac{100\%}{N_p}\sum_{i=1}^{N_p}\left|\frac{(\sigma_{n}^{\mathrm{SGE}})_i-(\sigma_{n}^{\mathrm{LEFM}})_i}{(\sigma_{n}^{\mathrm{LEFM}})_i}\right|$$ 
where $N_p=3...8$ is the number of points at which the SGE and LEFM solutions were compared.

Thus, within Mode I, the predictions for long cracks obtained from SGE using the maximum stress criterion coincide with those of LEFM. This was first pointed out in Refs. \cite{gourgiotis2009plane, askes2015understanding, vasiliev2019estimation} and  validated also based on generalized Eshelby's equivalent inclusion method for a penny-shaped crack in SGE in Ref. \cite{solyaev2025energy}. Given the established relationship between the length scale parameter $l$ and $K_{Ic}$, $\sigma_{\text{ult}}$ \eqref{eq:ell_vs_l}, we can assert that this length scale parameter of simplified SGE can be estimated for any material for which these classical LEFM characteristics are known. The resulting relationship differs in the proportionality coefficient from that previously reported in Ref. \cite{askes2015understanding}. This refinement is attributed to the detailed numerical simulations performed using the enriched $C^1$-continuous FEM with rigorous satisfaction of all boundary conditions. 

\begin{table}[b!]
	\centering
	\caption{Mechanical properties and calculated length scale parameters of brittle materials}
	\label{tab:lefm_lengths}
	
	\scriptsize
	\renewcommand{\arraystretch}{1.25}
	\setlength{\tabcolsep}{3.5pt}
	
	\resizebox{\textwidth}{!}{%
		\begin{tabular}{ccccccc}
			\hline
			\makecell[c]{Material} &
			\makecell[c]{$K_{\mathrm{Ic}}$\\$\mathrm{MPa}\sqrt{\mathrm{m}}$} &
			\makecell[c]{$\sigma_{\mathrm{ult}}$\\$\mathrm{MPa}$} &
			\makecell[c]{$L_{\mathrm{LEFM}}$\\$\mu\mathrm{m}$} &
			\makecell[c]{$l$\\$\mu\mathrm{m}$} &
			\makecell[c]{Grain or RVE\\ size, $\mu\mathrm{m}$} &
			\makecell[c]{Reference} \\
			\hline
			
			Annealed soda--lime--silica float glass
			&
			0.75 & 45.0 & 88.4 & 100.5 &
			-- & \citep{yankelevsky2014,karlsson2023}
			\\
			\hline
			
			AISI 4340 steel, quenched
			&
			37.5 & 2217 & 91.2 & 103.7 &
			$24$--$32$ & \citep{ritchie1976}
			\\
			\hline
			
			AISI 4340 steel, austenitized and quenched
			&
			70.2 & 2193 & 325.9 & 370.6 &
			$254$--$360$ & \citep{ritchie1976}
			\\
			\hline
			
			Ceramics $\alpha$-$\mathrm{Al}_{2}\mathrm{O}_{3}$
			&
			3.5 & 267 & 54.7 & 62.2 &
			$\sim 5$ & \citep{munro1997alumina}
			\\
			\hline
			
			Ceramics $\alpha$-$\mathrm{SiC}$
			&
			3.1 & 250 & 48.9 & 55.6 &
			$6\pm2$ & \citep{munro1997sic}
			\\
			\hline
			
			PMMA
			&
			1.62 & 56.8 & 257.9 & 293.3 &
			-- & \citep{braun2021}
			\\
			\hline
			
%			Unfilled epoxy resin
%			&
%			0.597 & 53.12 & 40.2 & 45.7 &
%			-- & \citep{li2022}
%			\\
%			\hline
			
			Unfilled epoxy resin
			&
			0.66 & 82.6 & 20.3 & 23.1 &
			-- & \citep{liu2008tailoring}
			\\
			\hline
			
			Epoxy with $6$ vol.\% elastomeric nanoparticles
			&
			1.26 & 69.1 & 105.8 & 120.3 &
			$\sim$0.2 & \citep{liu2008tailoring}
			\\
			\hline
			
			Epoxy with $6$ vol.\% nanosized SiO$_2$
			&
			0.98 & 89.2 & 38.4 & 43.7 &
			$\sim$ 0.3 & \citep{liu2008tailoring}
			\\
			\hline
			
			Freshwater ice
			& 0.083
			& 1.77
			& 700.0
			& 796.1
			& 1000--3000
			& \cite{timco1982fresh,frederking1983flexural} 
			\\
			\hline
			
			Sea ice
			& 0.120
			& 0.80
			& 7162
			& 8145
			& 1000--7000
			& \cite{timco1983sea}
			 \\
			\hline
			
			Nuclear-grade isotropic graphite IG-110
			&
			0.89 & 25 & 403.4 & 458.8 &
			$\sim 20$ & \citep{yamada2014,fujita2013}
			\\
			\hline
			
			Nuclear-grade isotropic graphite IG-430
			&
			1.031 & 37 & 247.2 & 281.1 &
			$\sim 10$--$20$ & \citep{yamada2014,fujita2013,simos2020}
			\\
			\hline
		\end{tabular}%
	}	
\end{table}

The examples of estimated values of $l$ for different brittle materials are presented in Table \ref{tab:lefm_lengths}. 
In the calculations, we employed the known properties of various brittle materials and relation \eqref{eq:ell_vs_l}. For certain materials, the ultimate strength values used were those obtained from three-point bending tests (glass) and four-point bending tests (ice). In Table~\ref{tab:lefm_lengths} we also report the grain or representative volume element (RVE) size of the materials considered. For graphites, this refers to the domain size. For filled epoxy, this refers to the RVE size estimated within the sphere-in-cube model: $g_{RVE} = d_m (\pi/(6f))^{1/3}$ (where $d_m$ is the mean particle size and $f$ is the volume fraction of particles). The values of $f$ are given in Table~\ref{tab:lefm_lengths}. The mean size of particles was $d_m=100$ nm (elastomeric particles) and $d_m=150$ nm (clusters of SiO$_2$ particles).

As can be seen, the estimated length scale parameters range from tens to hundreds of microns. However, there is no evidence that these values are close to the material grain size or RVE size, especially for nanocomposites. This could be a peculiarity of the simplified SGE under consideration, or it may indicate that the estimation of $l$ based on grain/RVE size is not reliable in general case of arbitrary microstructure.

It is important that the established relation \eqref{eq:ell_vs_l} remains valid for an edge crack in a half-space as well. In particular, we considered the SENT and SENB models (Fig. 1b) with a width-to-crack length ratio of $b/a = 100$ for different length scale parameters and found that the agreement of the predictions for the nominal strength for relatively long cracks based on SGE and LEFM persists if relation \eqref{eq:ell_vs_l} is used and if we take  into account that for an edge crack in the LEFM solution \eqref{eq:LEFM_critical_stress}, the value of $K_{Ic}$ should be replaced by $K_{Ic}/1.1215$ (according to classical LEFM \cite{tada2000stress}). The corresponding behaviour of the SGE solution for central versus edge cracks is clearly seen in Fig. \ref{fig:numerical_result}, where, in the long-crack region, the CCT curve and the SENT/SENB curves run parallel (their $y$-axis coordinates differ exactly by a factor of 1.1215).

For the convenience of subsequent analysis, the obtained SGE solution, which gives the predictions for the nominal strength of an infinite plate with a central crack, is approximated by the following regression.
\begin{equation}\label{eq:sigmainf}
	\sigma_{\infty}\!\left(a/l\right)
	=
	\left(
	1+k\,\frac{(a/l)^2}{a/l+c_0}
	\right)^{-1/2}
\end{equation}
where $\sigma_{\infty}(a/l)$ denotes the nominal strength of an infinite plate. The coefficient $k = 1.137254$ was defined above based on LEFM solution approximation \eqref{eq:ell_vs_l} and $c_0 = 2.088$  is the fitted value of an additional coefficient that provides the best fit of this regression to SGE predictions over a wide range of crack lengths $10^{-1}<a/l<10^2$ with $MAPE=1.6\%$ (see Fig.~\ref{fig:nomstr}b, solid line) for CCT model.

The introduction of regression \eqref{eq:sigmainf} is important for the subsequent construction of more general regression relations that account for the finite dimensions of the plate and small crack length, which are considered in the next section. These more complex regressions should reduce to \eqref{eq:sigmainf} in the case of large relative plate width. Moreover, regression \eqref{eq:sigmainf} itself has the following important features. It is chosen to reproduce two asymptotic limits: for long cracks (\(a/l \to \infty\)) it follows the classical power law \(\sigma_{\infty} = (ka/l)^{-1/2}=(a/L_{LEFM})^{-1/2}\), recovering the LEFM asymptote \eqref{eq:relative_LEFM}. For short cracks (\(a/l \to 0\)) it tends to unity, so that the nominal strength approaches the ultimate strength \(\sigma_{ult}\) of the material -- the so-called "strength plateau". The chosen regression form is similar to those used in Bazant's size-effect theory \cite{bazant1990determination, bazant1998fracture}.

It should be noted that the presented regression relation \eqref{eq:sigmainf} also covers the SENT and SENB cases for long cracks, with the classical correction factor 1.1215 taken into account (Fig.~\ref{fig:nomstr}b, dashed line). However, for short edge cracks this correction is no longer valid, and to describe such solutions, in the next section we construct more general regression relations that are applicable over a wide range of crack lengths.

%\begin{figure}[b!]
%	\centering
%	\includegraphics[width=0.5\textwidth]{figures/sigma_inf_reg.png}
%	\caption{Regression relation for a central crack in an infinite plate.}
%	\label{fig:inf_reg}
%\end{figure}

\subsection{Size effect in quasi-brittle materials}
\label{sec:regression}
%The asymptotic matching described above provides a direct relationship between the SGE length scale parameter \(l\) and the classical LEFM characteristic length \(L\) for sufficiently long cracks. However, this approach is limited to the idealised case of a crack in an infinite plate under remote tension for brittle material (or a very long crack in quasi-brittle material, \(a/l>>1\)). 
In practical applications, fracture tests are conducted on specimens of finite size and various geometries, for which the nominal strength depends not only on the ratio \(a/l\), but also on the specimen type and its width \(b\). 
%This dependence is usually taking into accoun through the introduction of the geometric correction factor \(F(a/b)\) in the definition fo nominal strength . 
Moreover, for short cracks and small \(a/l\) ratios, the asymptotic equivalence to LEFM no longer holds, and the full SGE solution must be employed.

To address these issues, a second approach is developed in this section. It relies on regression relations that approximate the nominal strength obtained from the numerical SGE solutions for three standard specimens (Fig. 1b) with finite dimensions, covering the full range of crack lengths from short to long. These regression relations are intended to facilitate the processing of experimental data and the identification of the SGE length scale parameter without the need for additional numerical simulations.

To approximate the numerical SGE solutions for nominal strength, we propose the following regression relation::
\begin{equation}
	\sigma_n = \sigma_{\infty}\!\left(\frac{a}{l}\right) \, F\!\left(\frac{a}{b}\right)^{-1}  G\!\left(\frac{a}{b}, \frac{a}{l}\right)
	\label{eq:regression_general}
\end{equation}
where \(\sigma_{\infty}(a/l)\) is the regression for nominal strength in infinite plate \eqref{eq:sigmainf}; \(F(a/b)\) is the classical geometric correction factor used in LEFM for the processing of experimental data for the specimens of finite size; and \(G(a/b, a/l)\) is an additional correction factor for SGE solution that refines the effect of specimen dimensions at small crack lengths, i.e., it accounts for the mutual influence of the non-classical size effect and the specimen size on the nominal strength.

Note that the inverse correction factor ($F^{-1}(a/b)$) appears in   \eqref{eq:regression_general}, since in classical LEFM it is used to estimate stress intensity factors that are inverse to nominal strength. The functions $F(a/b)$ we use in the analysis are presented in Table~\ref{tab:F_factors}. These are commonly accepted correction factors that prove to be valid over a wide range of specimen sizes \cite{tada2000stress}.

\begin{table}[t!]
	\centering
	\scriptsize
	\caption{Geometric correction factors \(F(a/b)\) according to classical LEFM \cite{tada2000stress}}
	\label{tab:F_factors}
	\begin{tabular}{llc}
		\hline\\
		Specimen type & Expression \(F(a/b)\) & Range of $a/b$\\[5pt]
		\hline\\
		CCT  & $\displaystyle \sqrt{\sec\!\left(\frac{\pi a}{2b}\right)}$ & 0...0.8 \\[10pt]
		SENT   & $\displaystyle \sqrt{\frac{2b}{\pi a} \tan\!\left(\frac{\pi a}{2b}\right)} \,\, \frac{0.752 + 2.02\frac{a}{b} + 0.37\bigl[1 - \sin(\frac{\pi a}{2b})\bigr]^3}{\cos(\frac{\pi a}{2b})}$ & any $a/b$ \\[15pt]
		SENB   & $ \frac{1}{\sqrt{\pi}}\,
			\frac{1.99-\frac{a}{b}\left(1-\frac{a}{b}\right)\left[2.15-3.93\frac{a}{b}+2.7\left(\frac{a}{b}\right)^2\right]}
			{\left(1+2\frac{a}{b}\right)\left(1-\frac{a}{b}\right)^{3/2}}$ & any $a/b$ \\[5pt]
		\hline
	\end{tabular}
\end{table}
 
The additional correction factor \(G(a/b, a/l)\) in \eqref{eq:regression_general} must satisfy the following properties. 
As we have shown in the previous section, the SGE solutions fully agree with the classical LEFM predictions for long cracks. 
Therefore, the regression in \eqref{eq:regression_general} should simply reflect this remarkable feature of SGE. 
Specifically, for long cracks (\(a/l \to \infty\)), we must have \(G \to 1\), because the predictions should tend to those of classical LEFM. 
In the proposed regression \eqref{eq:regression_general}, this condition is ensured by the properties of the already introduced relation for \(\sigma_{\infty}\) \eqref{eq:sigmainf} and by using the classical correction factor \(F(a/b)\) to account for the finite specimen size.
For short cracks, on the other hand, \(G(a/b, a/l)\) must ensure that the nominal strength approaches unity (\(\sigma_{\infty} \to 1\)), which corresponds to the case of very short cracks (\(a/l \to 0\), see Figs. 5b, 6). In this case, the strength of a material with a crack of very small size is determined by its ultimate strength. Consequently, the following formal requirements are imposed on the choice of \(G(a/b, a/l)\) in \eqref{eq:regression_general} for all considered specimen types:
\begin{equation}
\begin{aligned}
	a/l \to \infty:& \quad
	\quad G \to 1 \implies \sigma_n\to\sigma_n^{\text{LEFM}} \\
	a/l \to 0:& \quad 
	\quad G \to F \implies \sigma_n\to1 \\
\end{aligned}
	\label{eq:g}
\end{equation}
where we take into account that $\sigma_\infty\to 1$, as $a/l \to 0$ (see \eqref{eq:sigmainf}). 

This dual limiting behaviour of the factor \(G(a/b,a/l)\) in
Eq.~\eqref{eq:g} ensures a smooth transition between the
strength-dominated regime for short cracks and the toughness-dominated
regime for long cracks.

For the CCT and SENT specimens, we suggest the following form for the function \(G\):
\begin{equation}
	G\left(\frac{a}{b},\frac{a}{l}\right)
	=
	\frac{
		X^{c_1}+c_2
	}{
		X^{c_1}+c_2/F
	},
	\qquad
	X
	=
	\frac{a}{l}\left(1-\frac{a}{b}\right)
	\label{eq:G_cct_sent}
\end{equation}
where \(F(a/b)\) is the classical geometry correction factor (Table~\ref{tab:F_factors}). The coefficients \(c_1\) and \(c_2\) jointly control the shape and
characteristic scale of the transition. The same form of the regression function is used for the CCT and SENT specimens, while the coefficients \(c_1\) and \(c_2\) are identified independently for each specimen geometry.

For the SENB specimen, an additional multiplicative term is introduced
to reproduce the more complex behaviour observed in the transition
region. In this case, the regression function is written as
\begin{equation}
	G\left(\frac{a}{b},\frac{a}{l}\right)
	=
	\frac{
		X^{c_1}+c_2
	}{
		X^{c_1}+c_2/F
	}
	\left(
	1+c_3\,\frac{a}{b}\,\frac{a}{l}\,
	\text e^{-c_4a/l}
	\right),
	\quad
	X
	=
	\frac{a}{l}\left(1-\frac{a}{b}\right)
	\label{eq:G_senb}
\end{equation}
where \(F(a/b)\) is the classical geometry correction factor (Table~\ref{tab:F_factors}), and \(c_i\) ($i=1...4$) are the regression coefficients that are determined by fitting the regression to the numerical SGE solutions for SENB tests.%where the additional term is defined as
%\begin{equation}
%	H\left(\frac{a}{b},\frac{a}{l}\right)
%	=
%	c_3\frac{a}{b}\frac{a}{l}
%	\exp\left(
%	1-\frac{a/l}{c_4}
%	\right).
%	\label{eq:H_senb}
%\end{equation}
%For fixed \(a/b\), the coefficient \(c_4\) determines the position of
%the maximum of the additional correction along the \(a/l\) axis. The
%maximum is attained at \(a/l=c_4\), and its value is
%\(H_{\max}=c_3c_4(a/b)\). Therefore, \(c_3\) controls the magnitude of
%the correction, while \(c_4\) determines its characteristic scale and
%the position of its maximum.

\begin{figure}[t!]
	\centering
	(a)\includegraphics[width=0.4\textwidth]{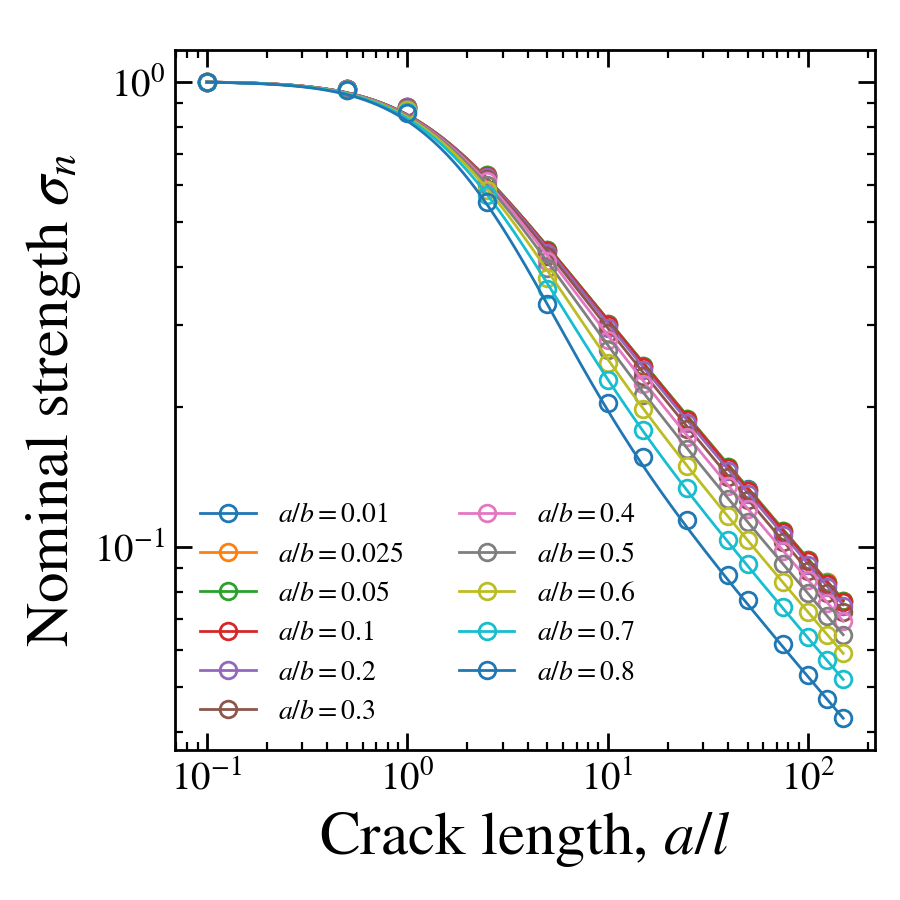}
	(b)\includegraphics[width=0.4\textwidth]{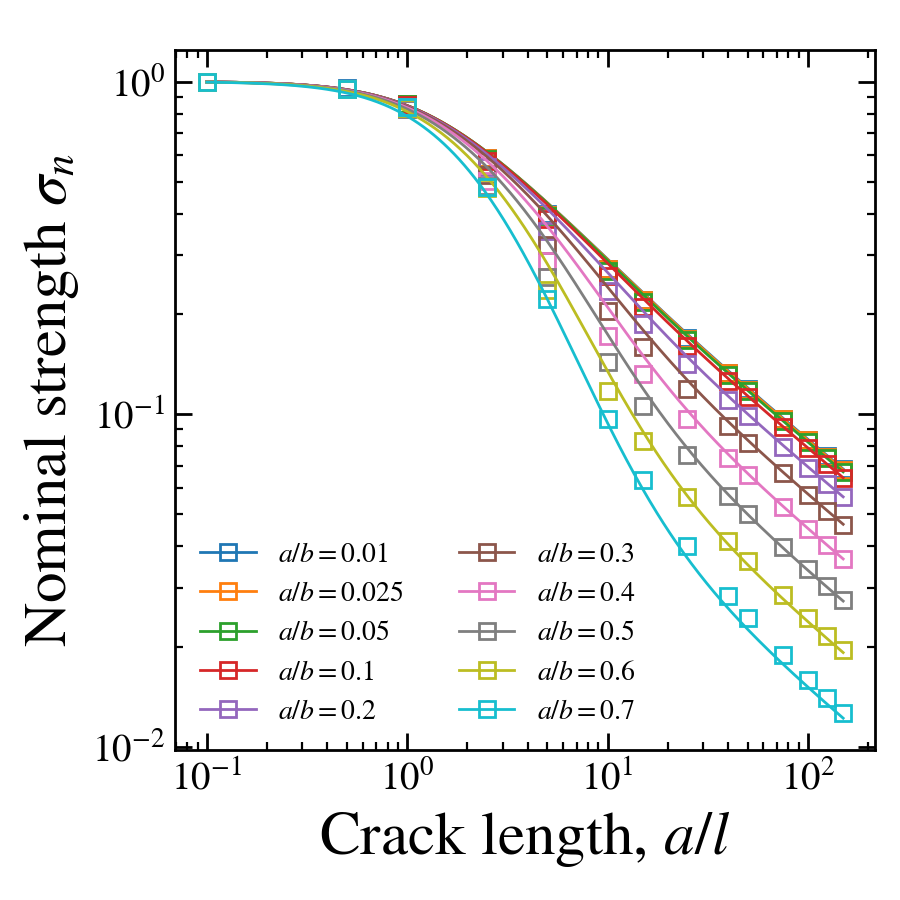}\\
	\vspace{10pt}
	(c)\includegraphics[width=0.4\textwidth]{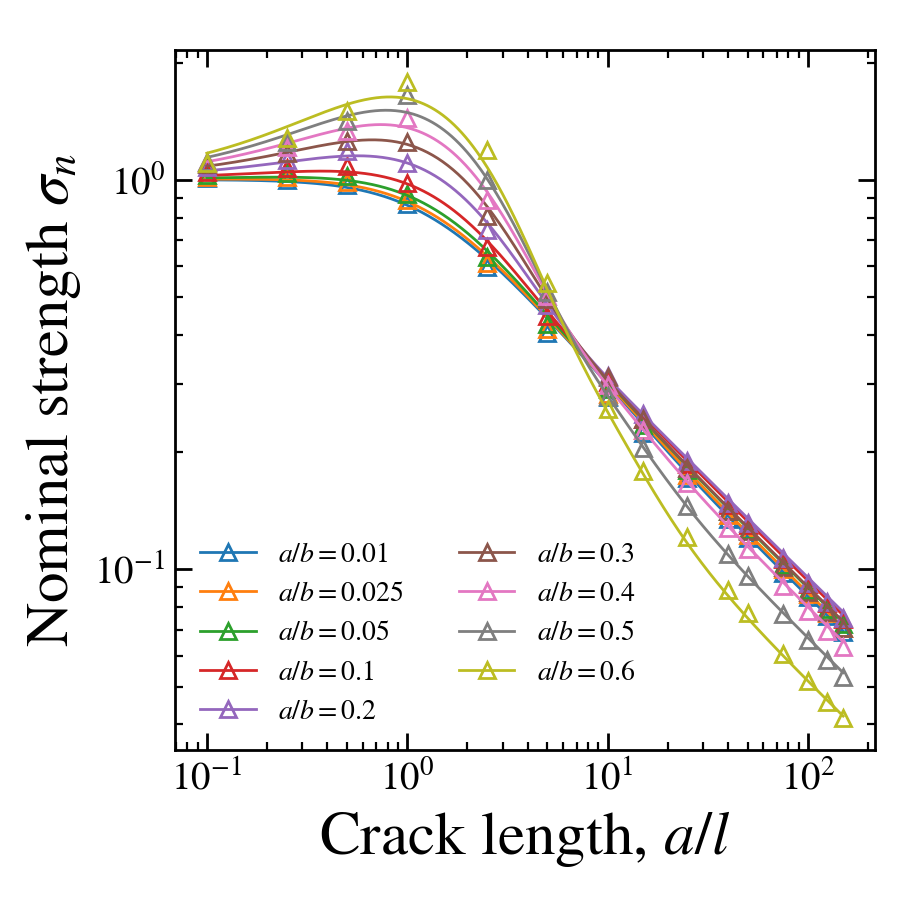}
	\caption{Mapping between the suggested regression relations (lines) and numerical solutions of SGE (dots) for the nominal strength of the specimens of the three configurations: (a) CCT, (b) SENT, (c) SENB}
	\label{fig:regression_fits}
\end{figure}

Note that the chosen forms of correction factor $G\!\left(\frac{a}{b}, \frac{a}{l}\right)$ \eqref{eq:G_cct_sent}, \eqref{eq:G_senb} satisfies requirements \eqref{eq:g}. 
Thus, in total, we have two independent coefficients ($c_1,c_2$) to describe SGE solutions for CCT and SENT tests based on the regression \eqref{eq:regression_general} with definitions \eqref{eq:sigmainf} and \eqref{eq:G_cct_sent}. For the SENB solution, four independent coefficients ($c_1,c_2,c_3,c_4$) are used in the regression \eqref{eq:regression_general} with definitions \eqref{eq:sigmainf} and \eqref{eq:G_senb}. 
%The calibration of these regressions was performed based on the least square method.
The regression coefficients were identified by minimising the maximum
absolute relative deviation between the proposed regressions and the
corresponding numerical SGE solutions.
%\begin{equation}
%	\mathbf{c}^{*}
%	=
%	\min_{\mathbf{c}}
%	\max_i
%	\left|
%	\frac{
%		\sigma_{n,i}^{\mathrm{reg}}
%		-
%		\sigma_{n,i}^{\mathrm{SGE}}
%	}{
%		\sigma_{n,i}^{\mathrm{SGE}}
%	}
%	\right|
%	\label{eq:regression_minimax}
%\end{equation}
This minimax criterion controls the largest local regression error over the entire considered ranges of \(a/b\) and \(a/l\). The identified regression coefficients are listed in Table~\ref{tab:regression_coeffs}. The obtained mapping between these regressions and numerical solutions of SGE is presented in Fig. \ref{fig:regression_fits}.

\begin{table}[b!]
	\centering
	\footnotesize
	\caption{Calibrated regression coefficients}
	\label{tab:regression_coeffs}
	\begin{tabular}{ccccc}
		\hline
		{Coefficient} & {Infinite plate} & {CCT} & {SENT} & {SENB} \\
		\hline
		$k$   & 1.1373 & 1.1373 & 1.1373 & 1.1373 \\
		$c_0$ & 2.0880 & 2.0880 & 2.0880 & 2.0880 \\
		$c_1$ & ---    & 2.0432 & 1.6434 & 2.6439 \\
		$c_2$ & ---    & 1.7473 & 9.3625 & 57.1947 \\
		$c_3$ & ---    & ---    & ---    & 0.9805 \\
		$c_4$ & ---    & ---    & ---    & 1.5431 \\
		\hline
	\end{tabular}
\end{table}
\begin{figure}[b!]
	\centering
	(a)\includegraphics[width=0.4\textwidth]{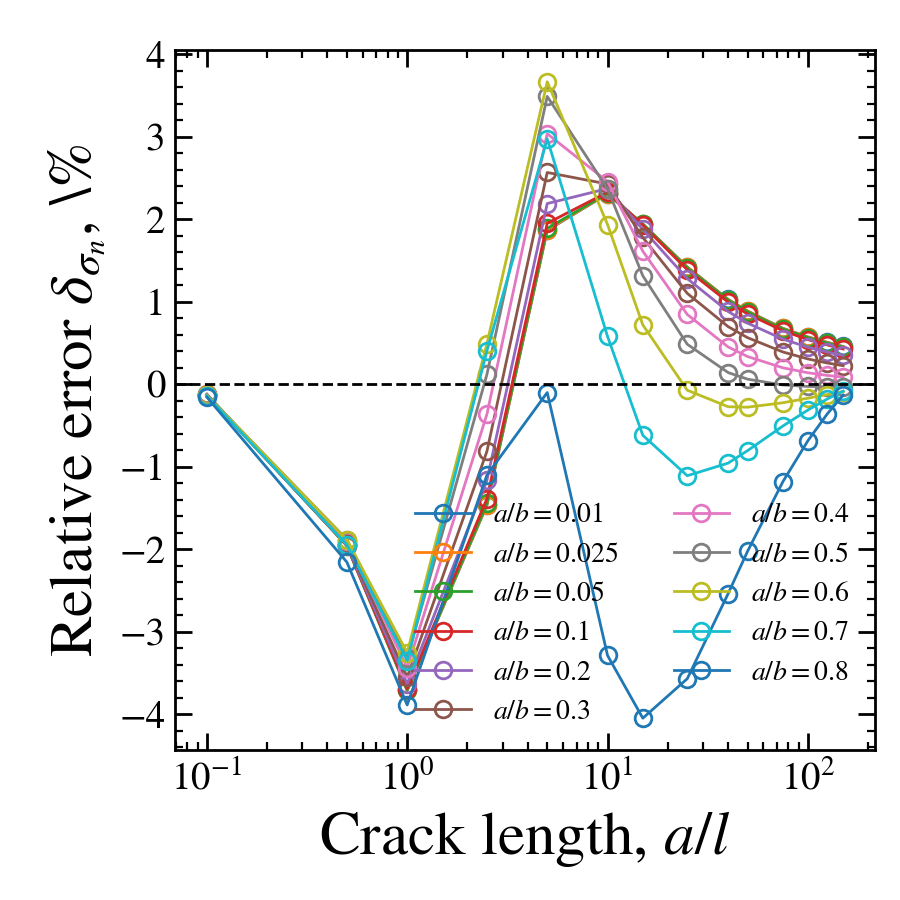}
	(b)\includegraphics[width=0.4\textwidth]{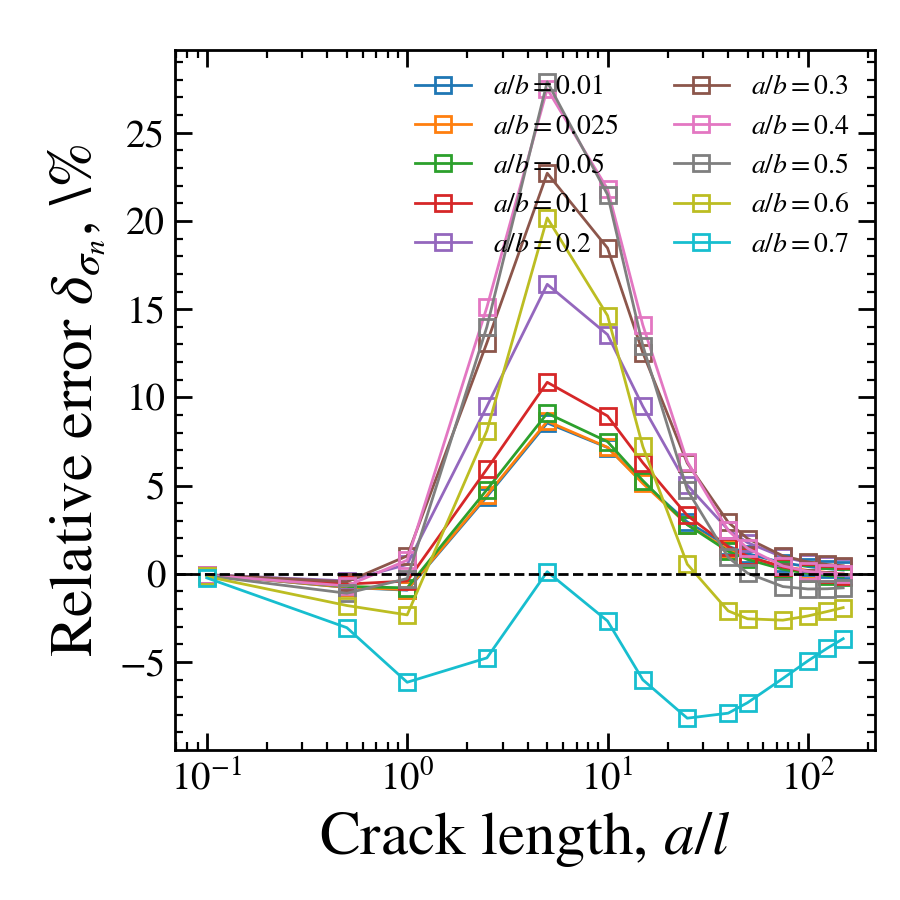}\\
	\vspace{10pt}
	(c)\includegraphics[width=0.4\textwidth]{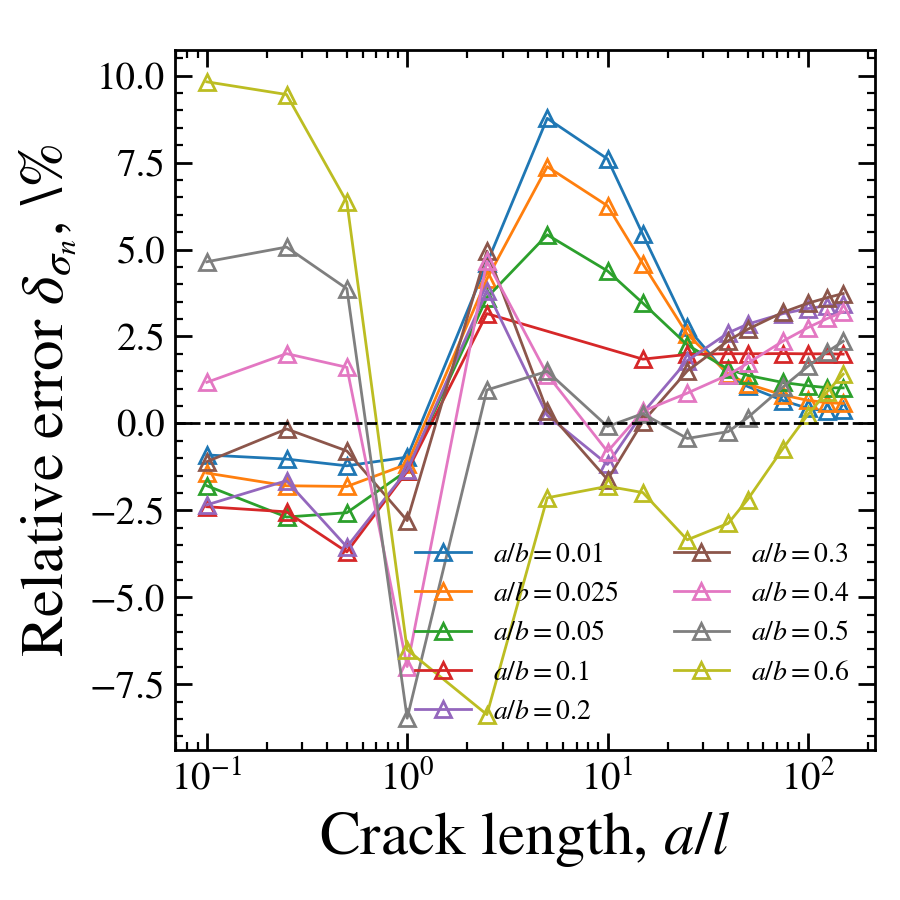}
	\caption{Relative error of the suggested regression relations and numerical solutions of SGE for the nominal strength of the specimens of the three configurations: (a) CCT, (b) SENT, (c) SENB. Plot legends in this figure correspond to Fig. 7.}
	\label{fig:regression_fits_err}
\end{figure}

%The proposed forms of
%\(G\left(a/b,a/l\right)\),
%Eqs.~\eqref{eq:G_cct_sent} and \eqref{eq:G_senb}, satisfy the limiting
%%requirements given in Eq.~\eqref{eq:g}. For \(a/l\rightarrow0\),
%%the geometry correction tends to \(F(a/b)\), eliminating the influence
%%of the crack geometry on the nominal strength. For
%%\(a/l\rightarrow\infty\), the correction tends to unity, recovering
%%the classical geometry dependence associated with the long-crack
%%regime. The additional term \(H\) tends to zero in both limiting cases.
%The CCT and SENT regressions contain two geometry-dependent
%coefficients, \(c_1\) and \(c_2\), whereas the SENB regression contains
%four coefficients, \(c_1\)--\(c_4\). The identified regression
%coefficients are listed in Table~\ref{tab:regression_coeffs}. The
%agreement between the regression relations and the numerical SGE
%solutions is presented in Fig.~\ref{fig:regression_fits}.

The accuracy of the proposed regression relations is additionally
assessed using the mean absolute percentage error with respect to the
numerical SGE solutions. Over the entire considered range of \(a/l\)
and \(a/b\), \(MAPE=1.37\%\) for CCT, \(MAPE=2.10\%\) for SENT, and
\(MAPE=2.59\%\) for SENB. The corresponding maximum absolute relative
errors are \(4.07\%\), \(8.37\%\), and \(8.56\%\), respectively. The relative errors of the regressions with respect to the SGE numerical solution for all considered parameter values are presented in Fig. 8. These results indicate that the proposed rather simple regression relations reproduce the numerical SGE solutions with sufficient accuracy for subsequent processing of experimental data and identification of the SGE length scale parameter.

%%%%%%%%%%%%%%%%%%%%%%%%%%%%

\section{Identification of the length scale parameter for quasi-brittle materials}
\label{sec:identification_results}

The proposed regression relations were applied to identify the SGE length scale parameter \(l\) for several quasi-brittle materials using experimental data from both published studies and tests conducted by the authors.

\subsection{SENT test of random chopped fiber composites}

The first data set was taken from Ko et al.~\cite{ko2019platelet}, who studied the size effect in discontinuous fiber-reinforced composite specimens with chopped carbon fibers and epoxy matrix. In these composites, isotropic properties are realised in the plane of the specimens at the macroscopic level due to the random orientation of chopped fibers. In the out-of-plane direction, the properties may differ from the in-plane characteristics. However, when testing standard specimens, this effect can be considered negligible, and models for isotropic materials can be used to describe the in-plane failure characteristics.

Three chopped-fiber tape sizes were used in the random layup: 75$\times$12~mm, 50$\times$8~mm, and 25$\times$4~mm ~\cite{ko2019platelet}. Five specimen widths were tested for each tape size: $b = \{6.3, 20, 40, 80, 120\}$ mm. The crack length-to-width ratio was fixed: \(a/b = 0.2\).  
A clear size effect was observed for each tape size: as the specimen dimensions decreased, the nominal strength increased. This trend was consistent across all three tape sizes and follows the classical size effect behaviour typical of quasi-brittle materials.
In Ref.~\cite{ko2019platelet}, the size effect was described using Bazant's size effect theory \cite{bazant1990determination, bazant1998fracture}. In the present study, the SGE approach was employed, using the proposed regression relations to identify the length scale parameter. 

To process these experimental data, we used regression for SENT specimens \eqref{eq:sigmainf}, \eqref{eq:regression_general}, \eqref{eq:G_cct_sent}. The ratio $a/b=0.2$ was fixed in accordance with the experimental data. The critical fracture stresses, which are known from the experiment for different specimen sizes and crack lengths, were theoretically calculated through the product of $\sigma_n$ (calculated based on regression) and $\sigma_{ult}$ (the material strength):

\begin{equation}
	\sigma_{\mathrm{c}}^{\mathrm{SGE}}
	=
	\sigma_n\left(\frac{a}{l},\frac{a}{b}\right)
	\sigma_{\mathrm{ult}},
	\label{eq:reg_frac_s}
\end{equation}

 The values of $\sigma_{ult}$ for the specimens without cracks were not reported in ~\cite{ko2019platelet}. For the smallest crack length, the critical stress was found to reach the
 "strength plateau". Therefore, we used these experimental values in the calculations of critical stresses (for the smallest specimens) as $\sigma_{ult}$. These values are listed in Table \ref{tab:sent_identification}. Normalised root-mean-square error (NRMSE) was used as an additional accuracy metric to account for scatter in the experimental data according to the following definition:

$$\mathrm{NRMSE}=\frac{\sqrt{\frac{1}{N_p}\sum_{i=1}^{N_p}\left(\sigma_{c,i}^{\mathrm{SGE}}-\sigma_{c,i}^{\mathrm{exp}}\right)^2}}{\overline{\sigma}_c^{\mathrm{exp}}}\,100\%.$$
where $N_p$ is the number of experimental data points; $\sigma_{c,i}^{\mathrm{SGE}}$ and $\sigma_{c,i}^{\mathrm{exp}}$ are the critical fracture stresses calculated using the SGE and measured experimentally in the $i$-th test for a given crack length, respectively; and $\overline{\sigma}_c^{\mathrm{exp}}$ is the mean fracture stress obtained from repeated experiments for that crack length. The use of NRMSE allows estimating the accuracy of the regression description of the dependence of the average nominal strength on crack length.

\begin{figure}[t!]
\centering
(a)\includegraphics[width=0.4\textwidth]{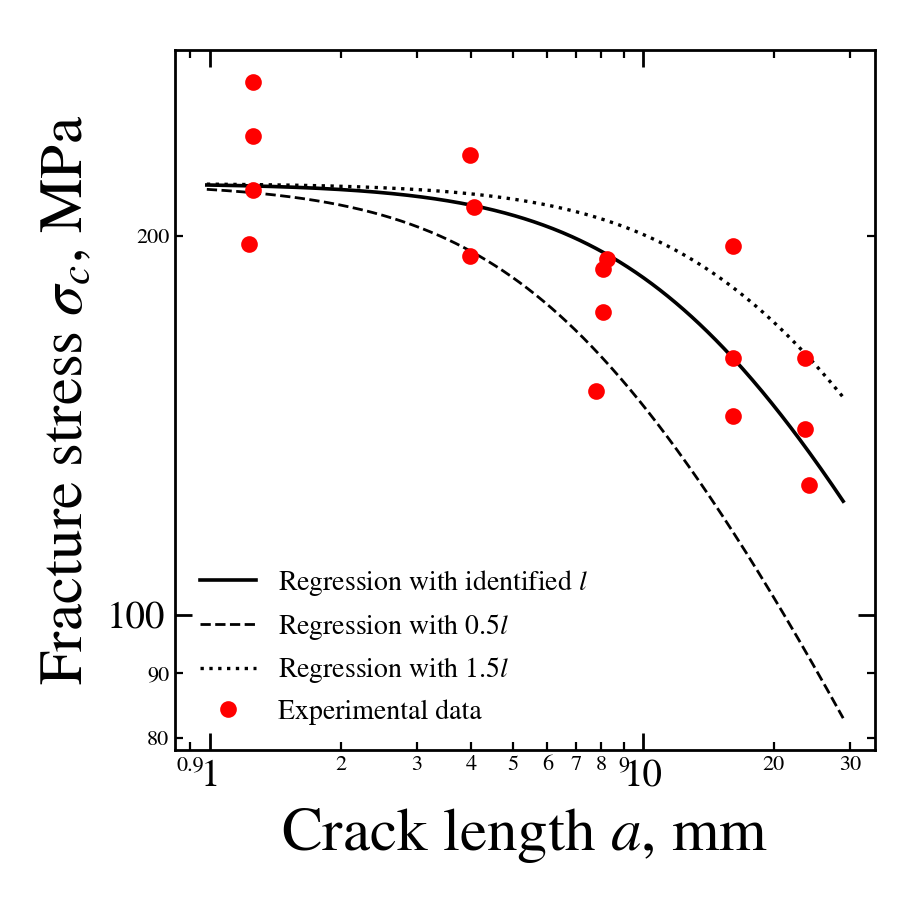}
(b)\includegraphics[width=0.4\textwidth]{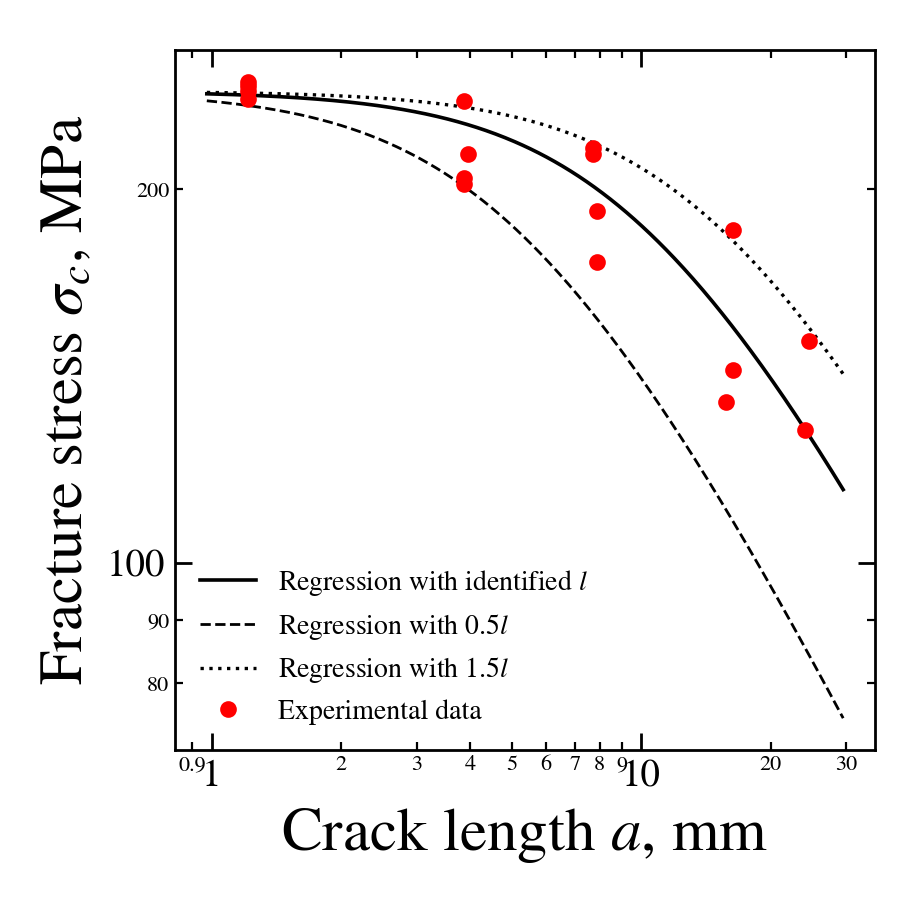}
(c)\includegraphics[width=0.4\textwidth]{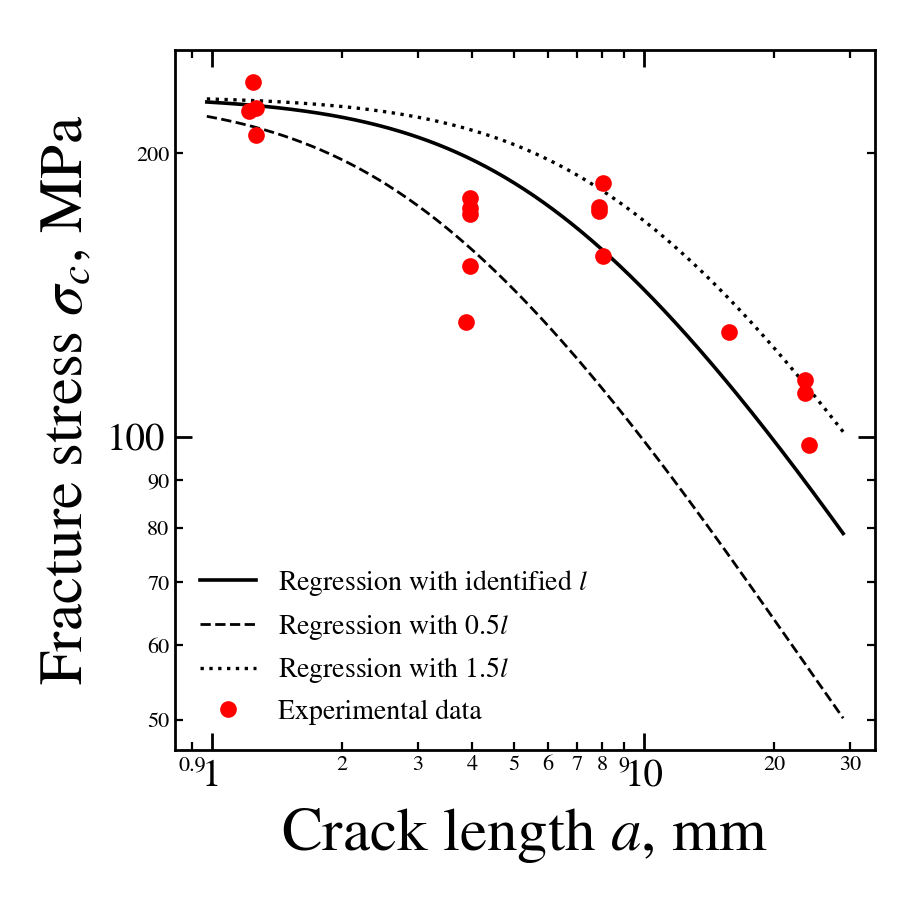}
\caption{Dependence of critical fracture stress on the crack length in chopped fiber composites: (a) 75 x 12 mm, (b) 50 x 8 mm, (c) 25 x 4 mm. Experimental data (dots) from Ref. \cite{ko2019platelet}. SGE predictions from established regression relations are shown by lines. Solid lines represent the best fit.}
\label{fig:sent_id}
\end{figure}

The single unknown parameter that enters the definition of $\sigma_n$ \eqref{eq:regression_general} is the length scale parameter of SGE $l$. It was determined from the condition of the best agreement between the theoretical predictions of regression \eqref{eq:regression_general} and the experimental data for critical stress (Fig. \ref{fig:sent_id}). The fitting was performed by minimizing NRMSE. For the composites under consideration, the identified length scale parameters are presented in Table~\ref{tab:sent_identification}. It was found that the identified length scale parameter $l$ is close to the minimum dimension of the tape (i.e. the width of chopped tapes).

\begin{table}[h!]
	\centering
	\caption{Identified SGE length scale parameters for the chopped fiber composites~\cite{ko2019platelet}.}
	\label{tab:sent_identification}
	
	\resizebox{\linewidth}{!}{%
		\begin{tabular}{ccccc}
			\hline
			\makecell{Tape size,\\mm} &
			\makecell{Ultimate strength,\\$\sigma_{ult}$, MPa} &
			\makecell{SGE length scale\\parameter, $l$, mm} &
			\makecell{MAPE,\\\%} &
			\makecell{NRMSE,\\\%} \\
			\hline
			$75\times12$ & 220 & 10.7 & 9.6 & 10.6 \\
			$50\times8$  & 240 & 8.2 & 8.0 & 8.9 \\
			$25\times4$  & 230 & 5.0 & 15.3 & 17.0 \\
			\hline
		\end{tabular}%
	}
\end{table}

\newpage
\subsection{SENB test of the porous SiO$_2$ ceramics}

\begin{figure}[h!]
\centering
(a)\includegraphics[width=0.45\textwidth]{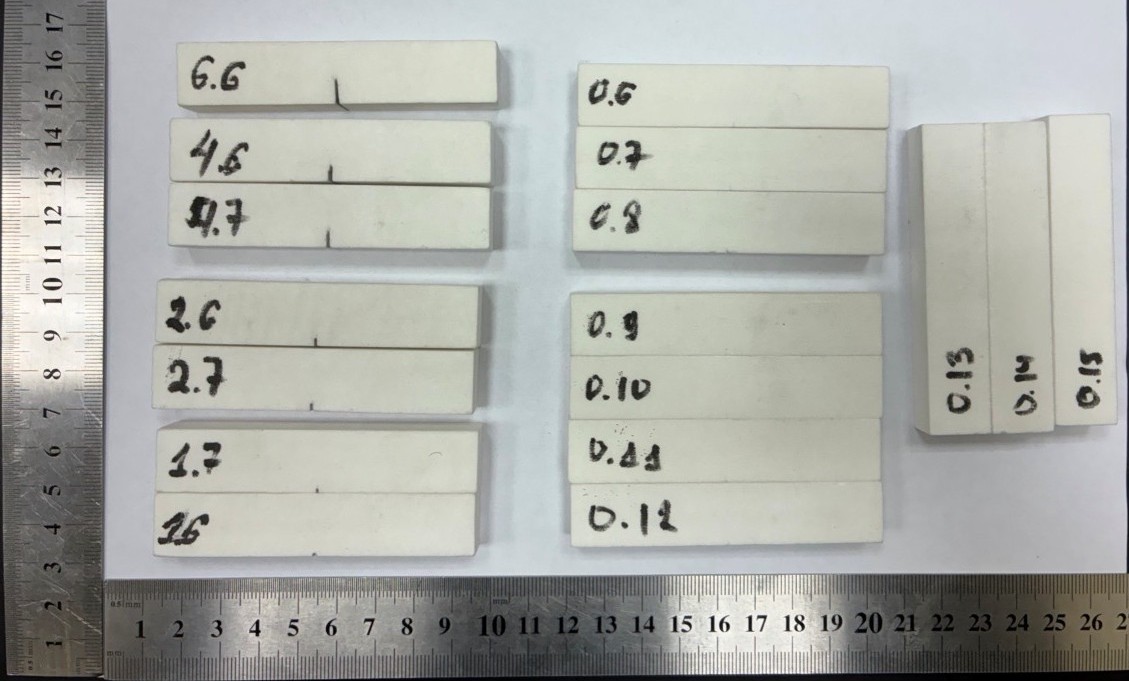}
(b)\includegraphics[width=0.45\textwidth]{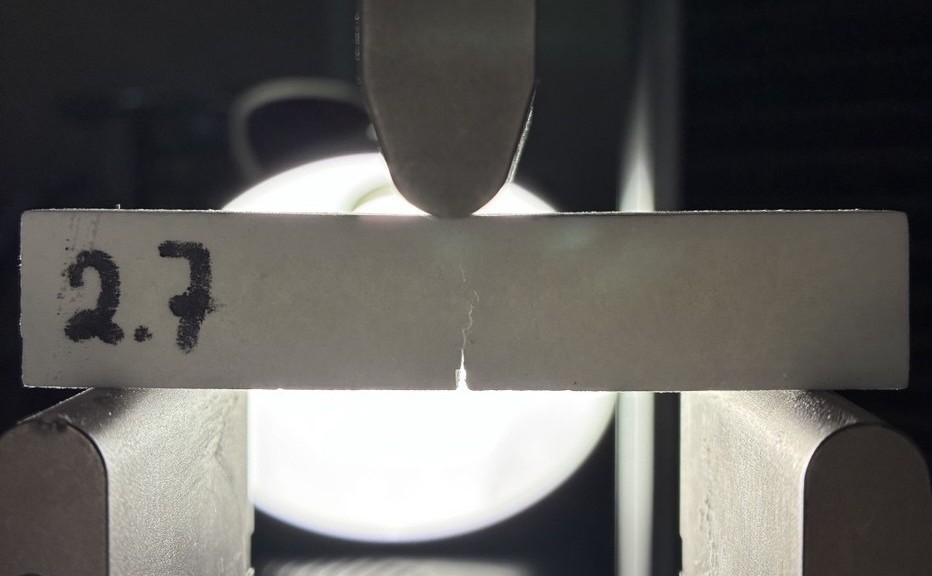}
(c)\includegraphics[width=0.45\textwidth]{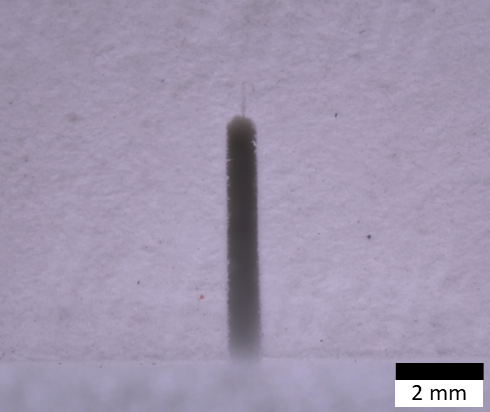}
(d)\includegraphics[width=0.45\textwidth]{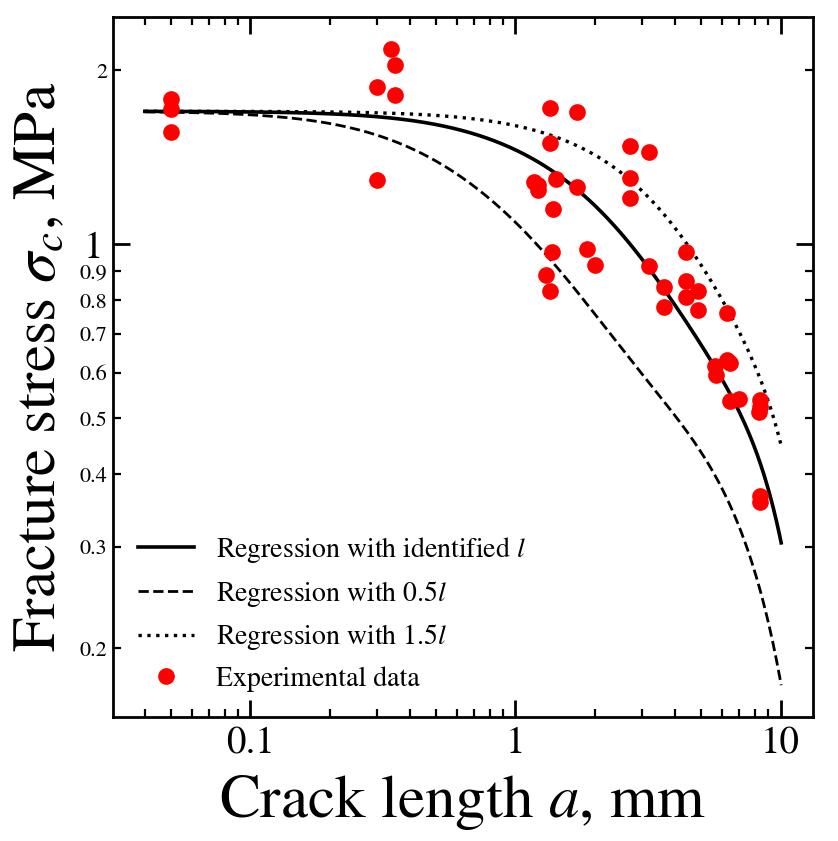}
\caption{Experiments on porous SiO$_2$ ceramics: (a) material specimens, (b) fracture during a three-point bending test, (c) optical microscopy of the notch with precrack in the specimen, (d) dependence of the critical fracture stress on crack length. Dots denote experimental data and lines denote the SGE regression predictions. The best fit is shown by the solid line ($l=0.76$ mm).}
\label{fig:author_id_graph}
\end{figure}

These bending tests were conducted by the authors using single-edge notched beam specimens made of highly porous quartz (SiO$_2$) ceramics. This ceramic is used in thermal-insulation applications and its measured porosity is 92\%. The specimens were machined from a single plate of sintered ceramics using circular diamond saw blades. Preliminary measurements confirmed that this material was homogeneous and isotropic. This was established by measuring the density and compressive strength of the samples machined from different parts of the plate and in different directions. The dimensions of specimens for SENB tests were as follows: height \(b = 16\)~mm, thickness \(h = 12\)~mm, and total length \(L = 96\)~mm (Fig. \ref{fig:author_id_graph}a). The span-to-height ratio was 4 and the corresponding span in the three-point-bending was \(2s = 64\)~mm (Fig. \ref{fig:author_id_graph}b). 

Cracks in the specimens were introduced manually using a sharp blade with preliminary notches machined by the saw blade (Fig. \ref{fig:author_id_graph}c). The crack length \(a\) was varied from 0 to 9~mm while the specimen dimensions were kept constant. The size of crack was controlled by optical microscopy on both sides of each specimen (Fig. \ref{fig:author_id_graph}c). In total, 30 specimens with different crack lengths were tested. The tests were performed using an Instron 4959 universal testing machine. The radii of the loading nose and supports in the equipment for the three-point-bending test were 5 mm. The crosshead speed was 1 mm/min in all tests. The fracture load (peak load recorded during the test) was determined for all specimens. 

The experimental critical fracture stress was calculated as usual in the fracture mechanics analysis of three-point-bending tests~\cite{tada2000stress}:
\begin{equation}
\sigma_{c}^{EXP} = \frac{3 P_{frac} s}{h b^2}
\label{eq:frac_s}
\end{equation}
where \(P_{frac}\) is the fracture load recorded during the test.

SGE predictions for the fracture stress were calculated using Eq.~\eqref{eq:reg_frac_s}, in which the nominal strength $\sigma_n$ was calculated based on the developed regression relations for the three-point-bending test \eqref{eq:sigmainf}, \eqref{eq:regression_general}, \eqref{eq:G_senb}. The ultimate strength $\sigma_{ult} = 1.7$ MPa used in \eqref{eq:reg_frac_s} was measured experimentally for the specimens without cracks.

The comparison between the experimental and calculated critical stress is presented in Fig.~\ref{fig:author_id_graph}d. The value of the length scale parameter of SGE was used to fit SGE predictions to experimental data.
The fitting was performed using the least-squares method. The identified value of $l$ is presented in Table~\ref{tab:author_comparison} together with the regression accuracy metrics and the identification results for the other porous ceramics discussed in the next subsection. A fairly large value of $l=0.76$ mm has been obtained. Note that the ceramic under consideration exhibits open porosity with a complex internal pore geometry, where the pores take the form of interconnected and extended channels. This can lead to the occurrence of significant nonlocal effects and affect the value of the length scale parameter that determines the characteristic crack size for which substantial non-classical size effects are realized.

\begin{table}[h!]
	\centering
	\caption{Identified SGE length scale parameters for porous ceramics}
	\label{tab:author_comparison}
	
	\resizebox{\linewidth}{!}{%
		\begin{tabular}{lcccccc}
			\hline
			\makecell{Material} &
			\makecell{Porosity, \%} &
			\makecell{Ultimate \\strength,\\$\sigma_{ult}$, MPa} &
			\makecell{SGE length \\scale para-\\meter, mm} &
			\makecell{MAPE,\\\%} &
			\makecell{NRMSE,\\\%} &
			\makecell{Experimental \\data} \\
			\hline
			SiO$_2$ & 92& 1.7 & 0.76 & 17.1 & 20.8 
			& present study\\
			SiO$_2$ & 12&8 & 1.3 & 12 & 17.4 &\cite{bazant1990size, mckinney1981specimen} \\
			SiC CN-163 &15& 24 & 1.5 & 9.4 & 11.7 &\cite{bazant1990size, mckinney1981specimen}\\
			\hline
		\end{tabular}%
	}
\end{table}

\subsection{SENB test of the porous SiO$_2$ and SiC ceramics~\cite{bazant1990size}}

The experimental data for slip-cast fused silica SiO$_2$ and silicon carbide refractory SiC CN-163 ceramics were analyzed by Bazant and Kazemi~\cite{bazant1990size} within the strength size effect theory. The original experiments were conducted by McKinney and Rice using single-edge-notched beams subjected to three-point bending~\cite{mckinney1981specimen}.  

The slip-cast fused silica consisted of approximately 99\% SiO$_2$ and had a porosity of 12\%, a grain size of 10--20~$\mu$m, and a Young's modulus of 57.9~GPa. The SiC CN-163 material consisted of approximately 85\% SiC and 13\% Si$_2$ON$_2$. It had a porosity of 15\%, a maximum grain size of approximately 2~mm, and a Young's modulus of 140~GPa \cite{bazant1990size}. As noted in \cite{bazant1990size}, the relatively coarse and porous microstructures of these materials lead to a finite fracture process zone and, consequently, to a pronounced size effect.

All specimens were single-edge-notched beams tested under three-point bending. The beam height $b$ ranged from 4.8 to 31.7~mm for SiO$_2$ and from 6.7 to 37.3~mm for SiC CN-163. The corresponding spans $2s$ ranged from 19.3 to 127.0~mm and from 28.7 to 149.4~mm, respectively. Although the specimens were not exactly geometrically similar, the span-to-depth ratio was close to $2s/b=4$. The relative crack length was varied between $a/b=0.19$ and 0.38 for SiO$_2$, and between $a/b=0.13$ and 0.22 for SiC CN-163.

A clear size effect was observed for both materials. Bazant and Kazemi described this behaviour using the size effect law~\cite{bazant1990size}. In the present study, the same experimental data were processed using the SGE approach and the regression relations developed for SENB specimens.
The experimental fracture stress was calculated using \eqref{eq:frac_s} and the reported fracture-load data $P_{frac}$ from Ref.~\cite{bazant1990size}. SGE predictions for fracture stress were obtained using SENB regression relations and the same approach as that described in the previous subsection.  Since the relative crack length was different for individual specimens, the actual value of $a/b$ was used for each experimental data point.

The strengths of uncracked specimens were not reported together with the fracture data. However, the nominal strengths of the smallest specimens approached a "strength plateau". Therefore, representative plateau values of $\sigma_{\mathrm{ult}}=8$~MPa for SiO$_2$ and $\sigma_{\mathrm{ult}}=24$~MPa for SiC CN-163 were adopted in the calculations.

\begin{figure}[t!]
\centering
(a)\includegraphics[width=0.45\textwidth]{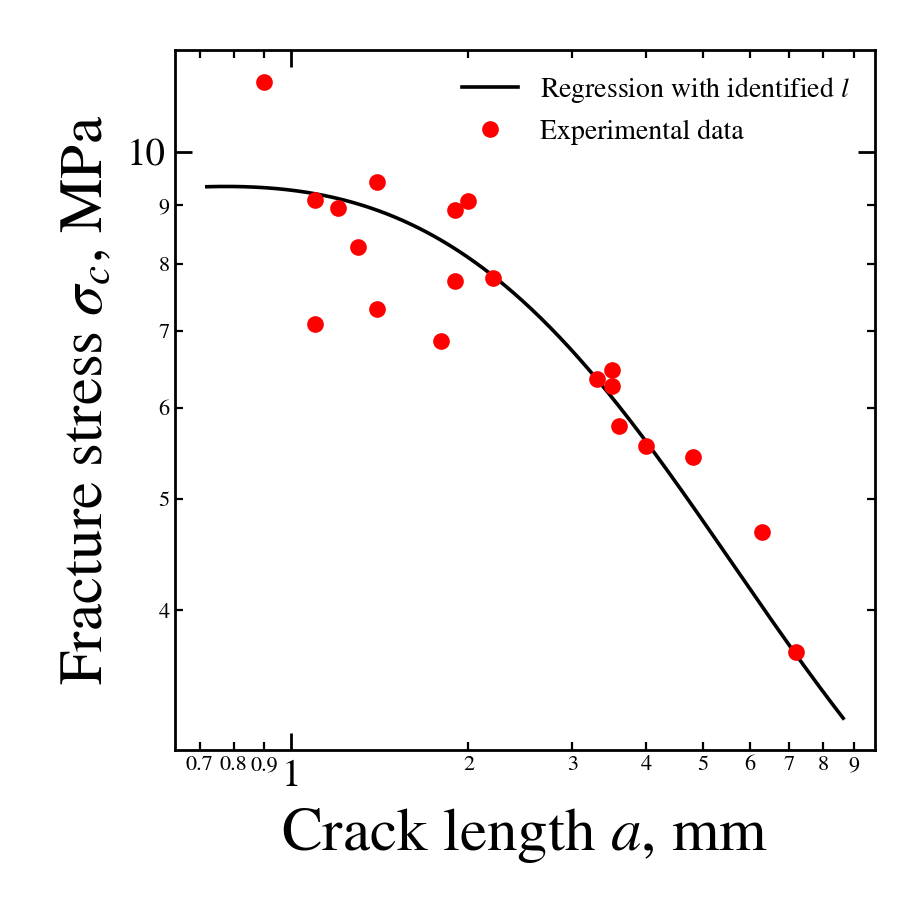}
(b)\includegraphics[width=0.45\textwidth]{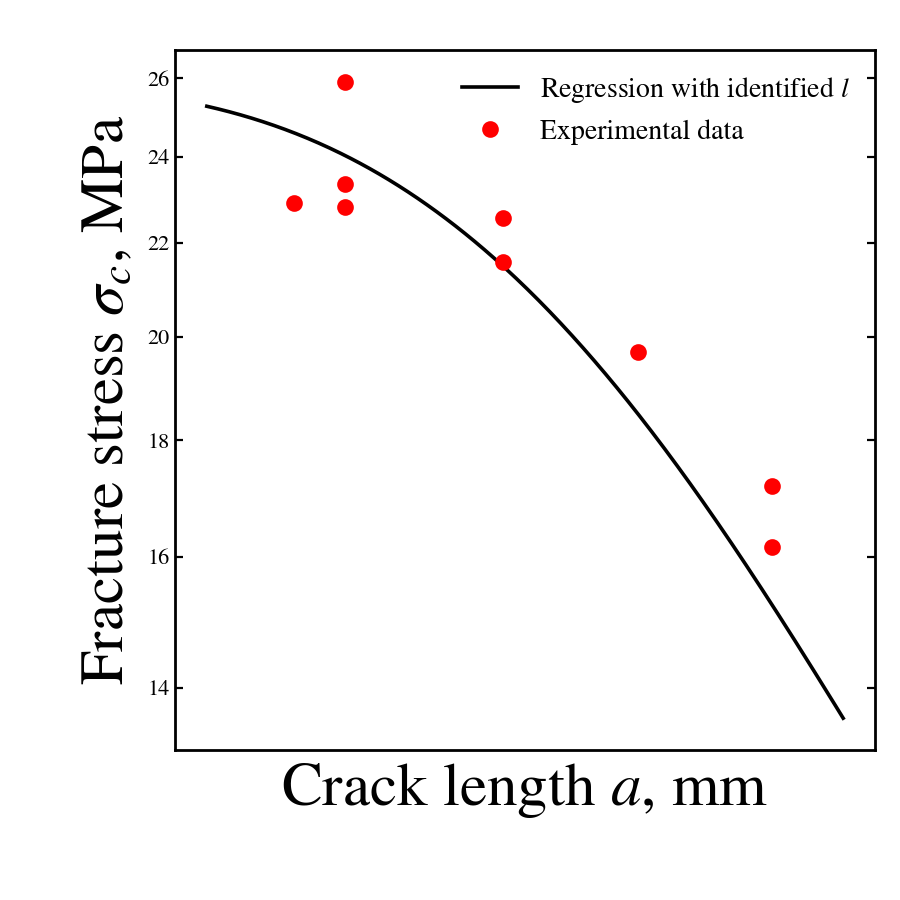}
\caption{Identification of the SGE length scale parameter for SENB specimens from~\cite{bazant1990size, mckinney1981specimen}. Dots -- experimental data, lines -- regression relations of SGE. The best fit is shown by the solid line.}
\label{fig:Bazant_id}
\end{figure}

The comparison between the experimental data and SGE predictions is shown in Fig.~\ref{fig:Bazant_id}. The identified length scale parameters and regression metrics are presented in Table~\ref{tab:author_comparison}.
For the porous, coarse-grained ceramics considered, the identified length scale parameters are relatively large ($>1$ mm). This may be related to the influence of grain size, pore size, and pore geometry. The results obtained for these ceramics with closed porosity can be used for validation of the generalized homogenization methods developed in SGE \cite{bacca2013mindlin, ganghoffer2021variational}. In these methods, the pores can be considered as inhomogeneities that give rise to nonlocal effects in a macroscopically homogeneous effective medium.

%\begin{table}[h!]
%	\centering
%	\caption{Identified SGE length scale parameters for SiO2 and SiC CM-163~\cite{}.}
%	\label{tab:Bazant_comparison}
%	
%	\resizebox{\linewidth}{!}{%
%		\begin{tabular}{lccccc}
%			\hline
%			\makecell{Tape size,\\mm} &
%			\makecell{Ultimate strength,\\MPa} &
%			\makecell{SGE length scale\\parameter, mm} &
%			$R^2$ &
%			\makecell{MAPE,\\\%} &
%			\makecell{NRMSE,\\\%} \\
%			\hline
%			SiO$_2$ & 8 & 1.34 & 0.54 & 11.3 & 16.6 \\
%			SiC CM-163 & 24 & 1.56 & 0.39 & 8.4 & 10.6 \\
%			\hline
%		\end{tabular}%
%	}
%\end{table}

\subsection{SENB tests of the dense ceramics~\cite{usami1986strength}}

The experimental data considered in this subsection were taken from Usami et al.~\cite{usami1986strength}, who investigated the effects of small flaws and material microstructure on the fracture strength of
polycrystalline ceramics. Four structural ceramic materials were considered: Al$_2$O$_3$, SiC, Si$_3$N$_4$, and Sialon.

%Unlike a conventional size-effect test employing geometrically similar specimens, the data analyzed by Usami et al.~\cite{usami1986strength} were obtained for specimens containing flaws of different shapes and origins. The considered defects included surface scratches, large grains, pores, controlled surface flaws produced by Vickers or Knoop indentation, and machined notches.

The experimental results demonstrate a pronounced short-crack effect. For relatively large flaws, the fracture stress follows the conventional LEFM relation governed by the long-crack fracture toughness $K_{\mathrm{Ic}}$. As the equivalent crack length decreases, however, the experimental strength deviates from the LEFM prediction and approaches a material-dependent "strength plateau". Usami et al.~\cite{usami1986strength} related this transition to the interaction between the crack-tip stress field and the characteristic size of the ceramic microstructure.

\begin{figure}[t!]
\centering
\vspace{5mm}
(a)\includegraphics[width=0.45\textwidth]{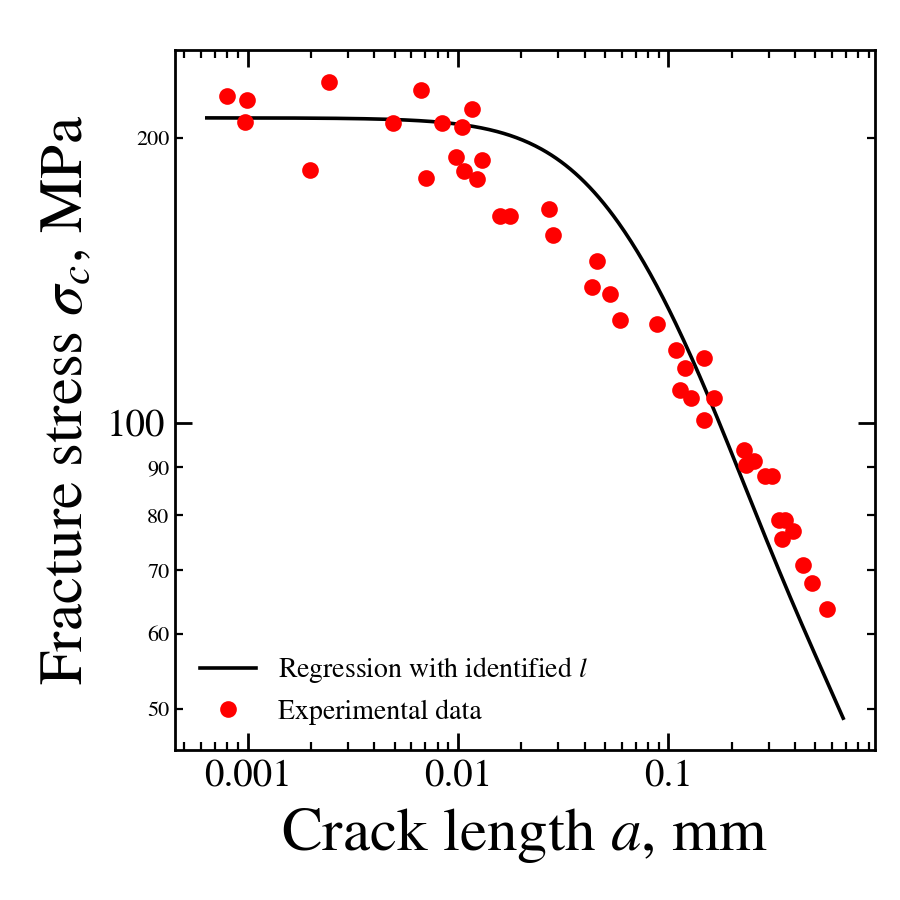}
(b)\includegraphics[width=0.45\textwidth]{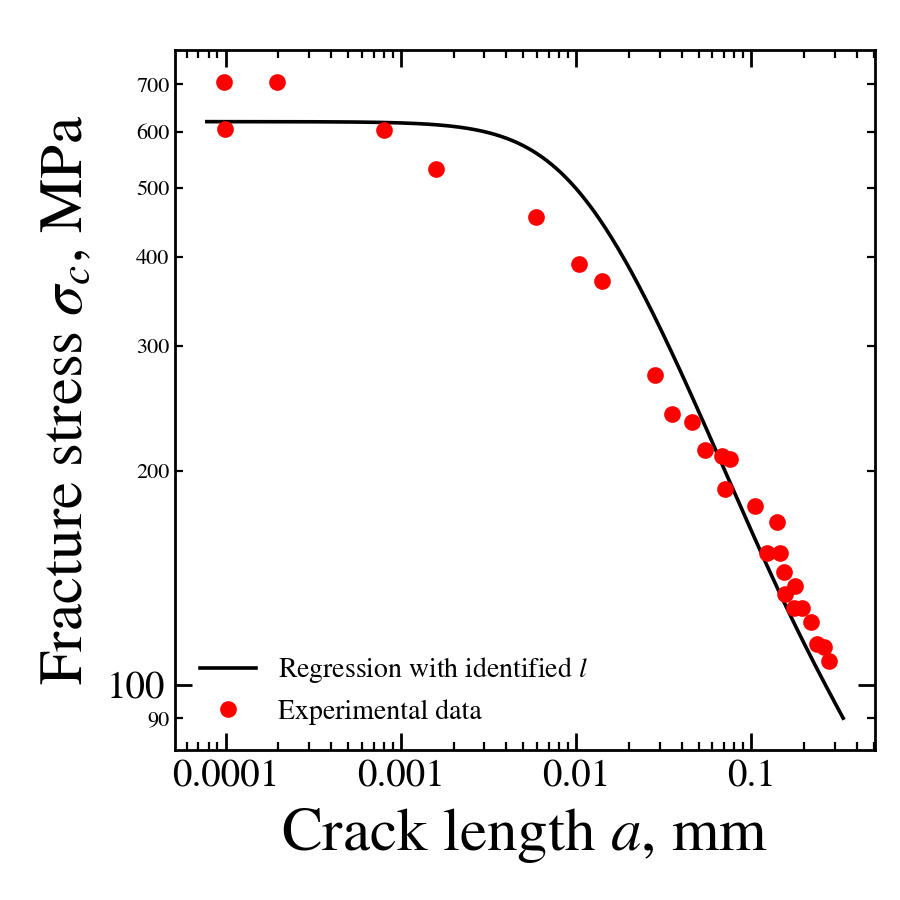}
(c)\includegraphics[width=0.45\textwidth]{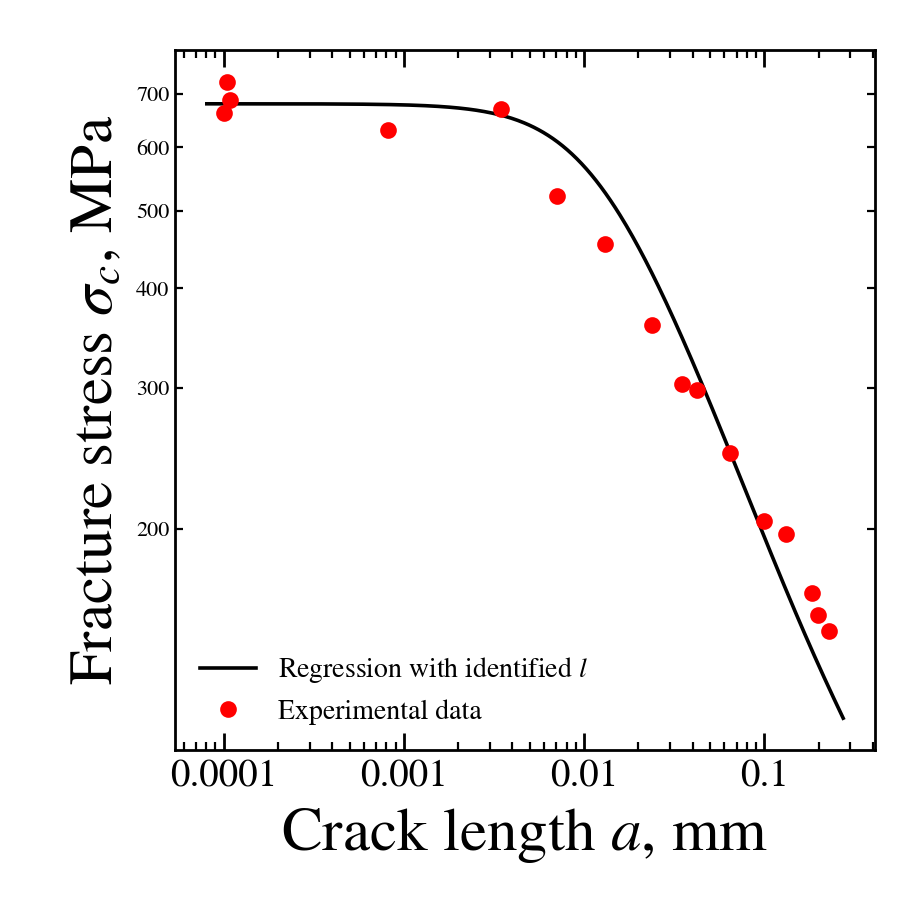}
(d)\includegraphics[width=0.45\textwidth]{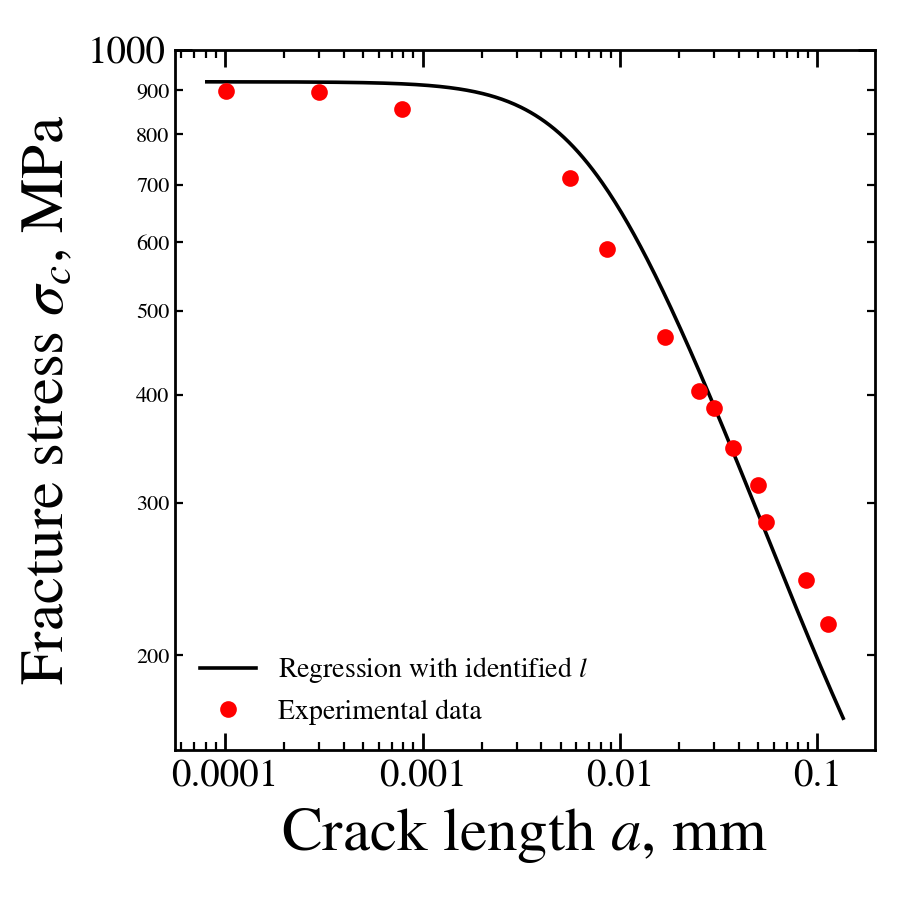}
\caption{Identification of the SGE length scale parameter for SENB specimens ~\cite{usami1986strength}. (a) Al$_2$O$_3$, (b) SiC, (c) Si$_3$N$_4$, (d) Sialon. Dots -- experimental data, lines -- regression relations of SGE. The best fit is shown by the solid line. Blue lines correspond to SGE predictions with $l$ calculated by using Eq. \eqref{eq:ell_vs_l}.}
\label{fig:usami_id}
\end{figure}

The identification of the length scale parameter was carried out based on regression relations for SENB tests, as well as on the basis of the simplified relation \eqref{eq:ell_vs_l} (for the long-crack  fracture toughness). The regression analysis was performed in the same manner as described above for the SENB tests (Section 6.2). The values of $\sigma_{ult}$ used in the regression analysis correspond to the "strength plateau" established in the experiments (see Table~\ref{tab:usami_comparison} and Fig.~\ref{fig:usami_id}). To apply Eq.~\eqref{eq:ell_vs_l}, we used the $K_{Ic}$ values calculated based on the experimental data from \cite{usami1986strength} for the specimens with the largest crack sizes (see Table~\ref{tab:usami_comparison}). A comparison of the regression analysis with the experimental data for critical fracture stresses is presented in Fig. 12. The predictions obtained from the regression relations, in which the length scale parameter is calculated using Eq.~\eqref{eq:ell_vs_l}, are shown in Fig.~\ref{fig:usami_id} as blue lines.

\begin{table}[b!]
	\centering
	\caption{Identified SGE length scale parameters for dense ceramics~\cite{usami1986strength}.}
	\label{tab:usami_comparison}
	
	\resizebox{\linewidth}{!}{%
		\begin{tabular}{lcccccccc}
			\hline
			Material &
			\makecell{Ultimate \\strength, MPa} &
			\makecell{$K_{Ic}$,\\MPa$\sqrt{\mathrm{m}}$} &
			\makecell{$L_{\mathrm{LEFM}}$,\\$\mu$m} &
			\multicolumn{2}{c}{\makecell{SGE length scale\\parameter, $\mu$m}} &
			\makecell{Grain size,\\$\mu$m} &
			\makecell{MAPE,\\\%} &
			\makecell{NRMSE,\\\%} \\
			
			\cline{5-6}\\[-10pt]
			& & & & Regression & Eq.~\eqref{eq:ell_vs_l} & & & \\[3pt]
			\hline
			
			Al$_2$O$_3$ & 210 & 2.7 & 52.6 & 39  & 59 & 20 & 11.5 & 13.0 \\
			SiC          & 620 & 3.3 & 9 & 8.3 & 10 & 3  & 11.0 & 17.1 \\
			Si$_3$N$_4$  & 680 & 4.1 & 11.5 & 9.5 & 13 & 4  & 9.7  & 10.4 \\
			Sialon       & 920 & 4.3 & 7  & 5.8 & 7.8 & 2  & 8.1  & 9.7 \\
			
			\hline
		\end{tabular}%
	}
\end{table}

The length scale parameters identified using the two approaches are given in Table~\ref{tab:usami_comparison}. The values obtained by regression fitting range from $l=5.8~\mu$m for Sialon to $l=39~\mu$m for Al$_2$O$_3$. These values are of the same order ofthe corresponding ceramic grain sizes. More specifically, the ratio $l/d$ varies from approximately 1.95 to 3.00. This result is consistent with the grain-fracture model of Usami et al.~\cite{usami1986strength}, in which the characteristic structural size was taken as $r_0=2d$. Thus, the identified SGE length scale parameter can be interpreted as a microstructure-related characteristic length associated with the interaction between a small flaw and the surrounding grains in the dense ceramics. It appears that a more accurate description of the size effect in the transition region between long and short cracks (Fig. 12) can be achieved using more general SGE constitutive models that include additional length scale parameters (see, e.g. \cite{vasiliev2021new}).\\

Using the simplified relation \eqref{eq:ell_vs_l} to compute the length scale parameter yields good agreement of the SGE predictions for relatively long cracks (see Fig. 12, blue lines). In terms of describing the size effect on strength for short cracks, the value of the length scale parameter obtained in this way is higher than that derived from a more rigorous regression-based analysis (see Table 6). Therefore, when using Eq.~\eqref{eq:ell_vs_l}, it should be borne in mind that for very small crack lengths the predicted fracture loads may be overestimated.

%\subsubsection{Hoover and Ba{\v z}ant (2013)}

%The third data set was taken from Hoover and Ba{\v z}ant~\cite{hoover2013}, who conducted %comprehensive fracture tests on concrete specimens. The identification results are presented in %Fig.~\ref{fig:hoover_id}.

%For this material, the identified SGE length scale parameter is \(l = 18.8\)~mm, while the %classical LEFM characteristic length, computed as \(L = \frac{1}{\pi}(K_{Ic}/\sigma_{ult})^2\), is %\(L = 17.5\)~mm.

%A comparison with the alternative approaches is presented in Table~\ref{tab:hoover_comparison}.

%\begin{table}[h!]
%	\centering
%	\caption{Identified SGE length scale parameters for concrete~\cite{}.}
%	\label{tab:hoover_comparison}
%	
%	\resizebox{\linewidth}{!}{%
%		\begin{tabular}{lccccc}
%			\hline
%			\makecell{Tape size,\\mm} &
%			\makecell{Ultimate strength,\\MPa} &
%			\makecell{SGE length scale\\parameter, mm} &
%			$R^2$ &
%			\makecell{MAPE,\\\%} &
%			\makecell{NRMSE,\\\%} \\
%			\hline
%			500 mm & 5 & 14.9 & 0.75 & 14 & 7.9 \\
%			200 mm & 4 & 14.9 & 0.4 & 16.1 & 4.5 \\
%			\hline
%		\end{tabular}%
%	}
%\end{table}

%\begin{figure}[h!]
%\centering
%\includegraphics[width=0.5\textwidth]{figures/hoover.png}
%\caption{Identification of the SGE length scale parameter for SENB specimens from Hoover and Ba{\v %z}ant~\cite{hoover2013}}
%\label{fig:hoover_id}
%\end{figure}

%\newpage
%%%%%%%%%%%%%%%%%%%%%%%%%%%%
\section{Conclusions}
\label{sec:conclusions}

In this work, we developed an approach that enables the use of standard fracture mechanics test data to identify an additional material constant of the simplified SGE -- the length scale parameter. 
Note that the reliable identification of additional SGE parameters has remained an open problem for a long time. In this work, we have generalized the approaches previously proposed in \cite{askes2015understanding,vasiliev2019estimation}. In particular, the presented results are based on accurate numerical solutions of SGE problems within an enriched $C^1$-continuous finite element method with rigorous satisfaction of all boundary conditions.

The first proposed approach is based on comparing the fracture load predictions for long cracks in SGE using a strength criterion based on the maximum principal stress and in LEFM using a $K_{Ic}$-based criterion. A closed-form expression has been obtained here for calculating the length scale parameter for any brittle material (or quasi-brittle material containing sufficiently long cracks) for which the classical characteristics -- $K_{Ic}$ and $\sigma_{ult}$ -- are known.

In the second approach, we processed the results of a large number of numerical solutions of SGE problems for three standard specimen types for the Mode I fracture tests (CCT, SENT, SENB). The compact regression relations constructed satisfy the main requirements for limit cases (for long and short cracks) and account for the influence of specimen dimensions and the non-classical size effect on strength typical for quasi-brittle materials. The presented identification examples demonstrate the possibility of effectively applying these regressions. We note that in these examples we no longer used numerical modelling but directly processed the experimental data using the proposed regression relations.

The proposed methods can be used to identify the SGE length scale parameter for a wide class of brittle and quasi-brittle materials. Once this parameter for a given material is identified, SGE can be applied to analyze the strength of structural elements with more complex geometry under complex loading conditions. It is assumed that this analysis can be carried out using an approach based on stress analysis and failure criteria, including for bodies containing cracks or other classical singular stress concentrators, for which SGE allows one to obtain regular solutions.

Future work should verify experimentally whether the identified length scale parameters provide reliable predictions for cracks under mixed-mode loading within the adopted failure criterion, as well as for specimens containing other types of stress concentrators, such as sharp notches or small holes. Generally, it may be necessary to employ more general strength criteria or more general constitutive relations within the framework of SGE.

In the present work, we have focused on the application of strength criteria evaluated for regular stress solutions in simplified SGE. Alternatively, an approach can be used in which critical loads are determined using an energy criterion based on the J-integral or more specific criteria formulated in terms of the amplitude factors of asymptotic solutions for cracks in SGE (by analogy with classical K-based criteria). The latter approach has, to date, not been considered at all and may be of significant interest for future research.

The values of the J-integral and amplitude factors that we computed for standard specimens with different crack lengths (presented in Appendix B) can be used to formulate such criteria. In particular, we have obtained an interesting result for the dependence of SGE amplitude factors on the crack length. For long cracks, these amplitude factors grow linearly with crack length, while for short cracks they tend exponentially to zero (see Figs. B.1-B.3). Providing a theoretical justification for this dependence of the amplitude factors on crack length remains an important topic for future research in SGE.

% in examples we did not used numerical simulations, we used regression
 %mixed mode - question, see LJM

%we used only stress, another - K1,K2, J,- Appendix. Increaes ilnearly, decrease exponentially.

\section*{Appendix A. Finite element formulation}
\label{app:fem}

The finite element formulation for the simplified strain gradient elasticity follows the standard procedure. The displacement field within an element is interpolated as
\begin{equation}
\footnotesize
	\mathbf{u} = \begin{Bmatrix} u \\ v \end{Bmatrix} = \mathbf{N}\widehat{\mathbf{u}},
	\label{eq:app_u}
\end{equation}
where $\mathbf{N}$ is the matrix of shape functions and $\widehat{\mathbf{u}}$ is the vector of nodal degrees of freedom.

For the standard three-node triangular element, the vector of nodal degrees of freedom contains the displacements and their first and second derivatives at each node~\cite{papanicolopulos2010crack}:
\begin{equation}
\footnotesize
	\widehat{\mathbf{u}}_{36\times1} = \Bigl\{ u_1,\; u_{1,x},\; u_{1,y},\; u_{1,xx},\; u_{1,xy},\; u_{1,yy},\;
	v_1 \ldots v_{1,yy},\; u_2 \ldots v_{2,yy},\; u_3 \ldots v_{3,yy} \Bigr\}^{T},
	\label{eq:app_dof}
\end{equation}
and the matrix of shape functions has the form
\begin{equation}
\footnotesize
	\mathbf{N}_{2\times 36} = \begin{bmatrix}
		N_1 \ldots N_6 & 0 \ldots 0 & N_7 \ldots N_{12} & 0 \ldots 0 & N_{13} \ldots N_{18} & 0 \ldots 0 \\
		0 \ldots 0 & N_1 \ldots N_6 & 0 \ldots 0 & N_7 \ldots N_{12} & 0 \ldots 0 & N_{13} \ldots N_{18}
	\end{bmatrix}.
	\label{eq:app_N}
\end{equation}
where $N_i$ are the shape functions of the so-called Bell triangle \cite{dasgupta1990higher}.

The strains and the strain gradients are expressed as $\boldsymbol{\varepsilon}_{3\times 1} = \mathbf{B}_{1}\widehat{\mathbf{u}}$ and $\boldsymbol{\kappa}_{6\times 1} = \mathbf{B}_{2}\widehat{\mathbf{u}}$, where $\mathbf{B}_{1}$ and $\mathbf{B}_{2}$ are the derivative matrices containing first and second derivatives of the shape functions, respectively:
\begin{equation}
\footnotesize
	\begin{aligned}
		\mathbf{B}_1 &= \mathbf{L}_1 \mathbf{N}, \qquad
		\mathbf{L}_1 = \begin{pmatrix}
			\frac{\partial}{\partial x} & 0 & \frac{\partial}{\partial y} \\
			0 & \frac{\partial}{\partial y} & \frac{\partial}{\partial x}
		\end{pmatrix}^T \\[6pt]
		\mathbf{B}_2 &= \mathbf{L}_2 \mathbf{N}, \qquad
		\mathbf{L}_2 = \begin{pmatrix}
			\frac{\partial^2}{\partial x^2} & \frac{\partial^2}{\partial x \partial y} & 0 & 0 & \frac{\partial^2}{\partial y \partial x} & \frac{\partial^2}{\partial y^2} \\
			0 & 0 & \frac{\partial^2}{\partial y \partial x} & \frac{\partial^2}{\partial y^2} & \frac{\partial^2}{\partial x^2} & \frac{\partial^2}{\partial x \partial y}
		\end{pmatrix}^T
		\label{eq:app_L}
	\end{aligned}
\end{equation}

The constitutive relations for stress $\boldsymbol{\tau}_{3\times 1} =  \mathbf{C}\mathbf{B}_1\widehat{\mathbf{u}}$ and double stress $\boldsymbol{\mu}_{6\times 1} = \mathbf{A}\mathbf{B}_2\widehat{\mathbf{u}}$
are defined via the constitutive matrices $\mathbf{C}$ and $\mathbf{A}$, which are given by:
\begin{equation}
\footnotesize
	\begin{split}
		\mathbf{C} &= \begin{pmatrix}
			\lambda + 2\mu & \lambda & 0 \\
			\lambda & \lambda + 2\mu & 0 \\
			0 & 0 & \mu
		\end{pmatrix},\\[6pt]
		\mathbf{A} &= l^2 \begin{pmatrix}
			\lambda + 2\mu & 0 & \lambda & 0 & 0 & 0 \\
			0 & \lambda + 2\mu & 0 & \lambda & 0 & 0 \\
			\lambda & 0 & \lambda + 2\mu & 0 & 0 & 0 \\
			0 & \lambda & 0 & \lambda + 2\mu & 0 & 0 \\
			0 & 0 & 0 & 0 & \mu & 0 \\
			0 & 0 & 0 & 0 & 0 & \mu
		\end{pmatrix}.
		\label{eq:app_CA}
	\end{split}
\end{equation}
where $\lambda$ and $\mu$ are the Lam\'e parameters and $l$ is the length scale parameter of the simplified SGE.

For the enriched elements, additional degrees of freedom corresponding to the asymptotic amplitude factors $K_1$, $K_2$ (for the Mode I problems) are introduced~\cite{solyaev2025enriched}:
\begin{equation}
\footnotesize
	\widehat{\mathbf{u}}_{38\times 1} = \Bigl\{ u_1,\; u_{1,x},\; u_{1,y},\; u_{1,xx},\; u_{1,xy},\; u_{1,yy},\;
	\ldots v_{3,yy},
	K_1,\; K_2\Bigr\}^{T},
	\label{eq:app_enr_dof}
\end{equation}
and the approximation of the displacement field takes the form
\begin{equation}
\footnotesize
	\begin{aligned}
		\mathbf{N}_{2\times 38} = \begin{bmatrix}
		N_1 \ldots N_6 & \ldots & 0 \ldots 0 & Q^*_{11} &Q^*_{12}\\
		0 \ldots 0 & \ldots &  N_{13} \ldots N_{18} & Q^*_{21} &Q^*_{22}
	\end{bmatrix}.
	\end{aligned}
\end{equation}
where $Q_{ij}^{*}$ are the enrichment functions developed based on asymptotic SGE solutions for crack problems. These functions  preserve the $C^1$-continuity of the interpolation at the mesh nodes and are given explicitly in Refs.~\cite{solyaev2025enriched, shelkov2026implementation}.

\section*{Appendix B. Calculated J-integrals and amplitude factors $K_1$, $K_2$}
\label{app:jk_app}	

The use of enriched crack-tip elements within the SGE formulation made it possible to determine not only the regular stresses for the CCT, SENT, and SENB specimens, but also the amplitude factors $K_1$, $K_2$ of the asymptotic solution. These factors, together with Eq.~\eqref{eq:ji}, were subsequently used to evaluate the \(J\)-integral.
Figures \(\ref{fig:jk_center_1}\)-\(\ref{fig:jk_center_3}\) present the normalised amplitude factors $\bar{K}_i$ and the normalised \(J\)-integral values $\bar{J}$ for CCT, SENT and SENB. These normalised quantities are defined as follows:
\[
\bar{J}=\frac{EJ}{\sigma_{ref}^{2}\pi a}, \qquad
\bar{K}_1=-\frac{K_1\sqrt{a}}{\sigma_{ref}}, \qquad
\bar{K}_2=-\frac{K_2\sqrt{a}}{\sigma_{ref}}
\]
where $\sigma_{ref}=\sigma_0$ in CCT and SENT tests and $\sigma_{ref}=3\sigma_0sd/b^2$ in SENB.

In the figures for the SENB test, magnified views of the regions corresponding to small crack lengths are additionally provided (Fig. \ref{fig:jk_center_3}b, d, f), where substantially nonlinear dependences of the amplitude factors arise.
The data presented in Figs. \(\ref{fig:jk_center_1}\)-\(\ref{fig:jk_center_3}\) are also given in tables in the Supplementary Material file.

\clearpage
 \renewcommand{\thefigure}{B.\arabic{figure}}
\setcounter{figure}{0}
\begin{figure}[t!]
\centering
(a)\includegraphics[width=0.405\textwidth]{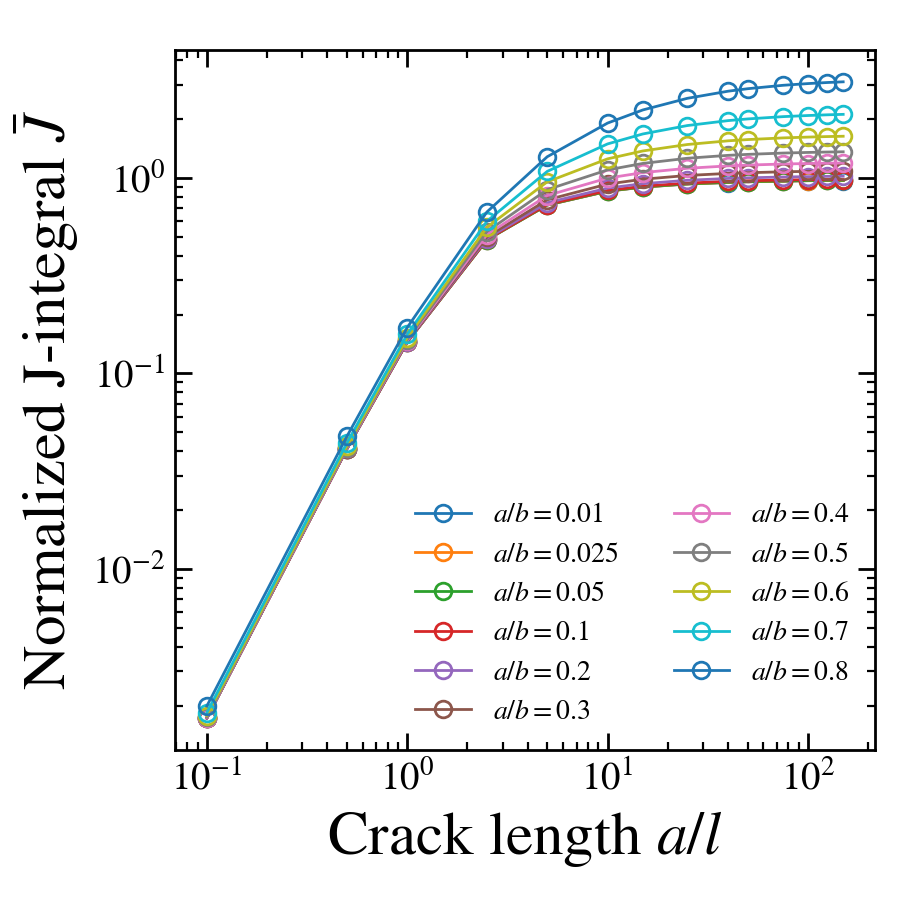}
(b)\includegraphics[width=0.45\textwidth]{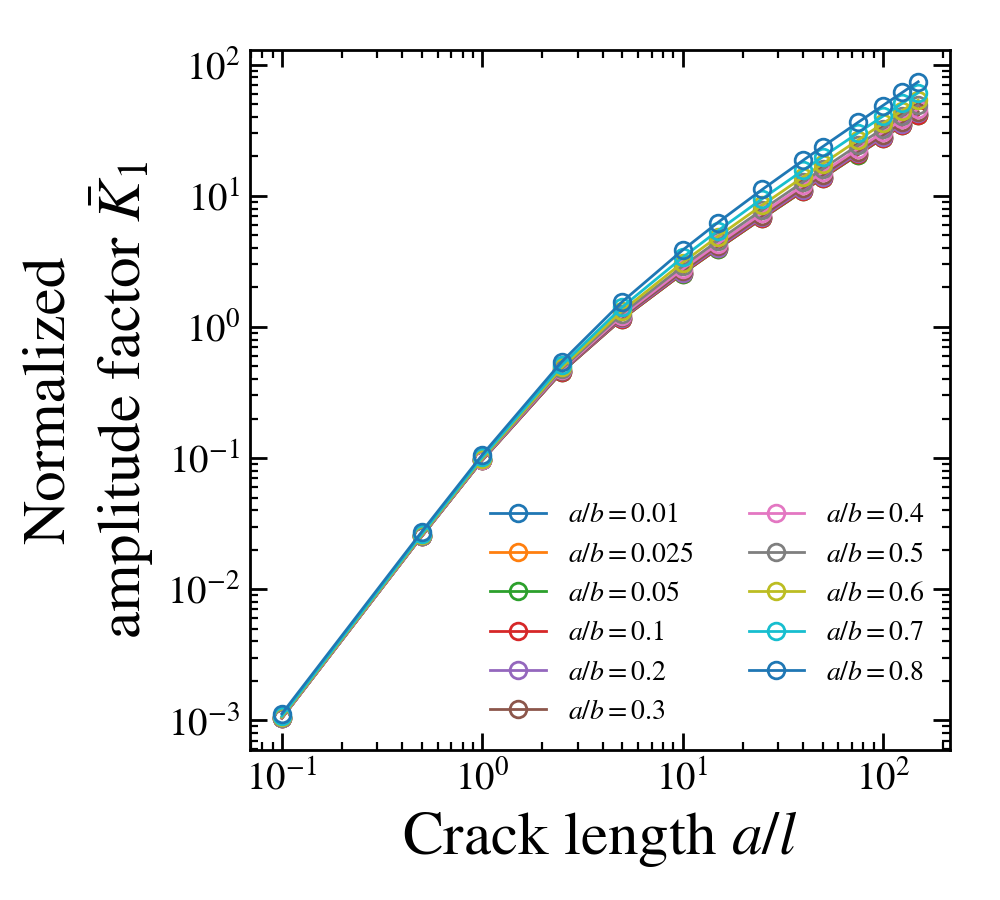}\\
\vspace{10pt}
(c)\includegraphics[width=0.45\textwidth]{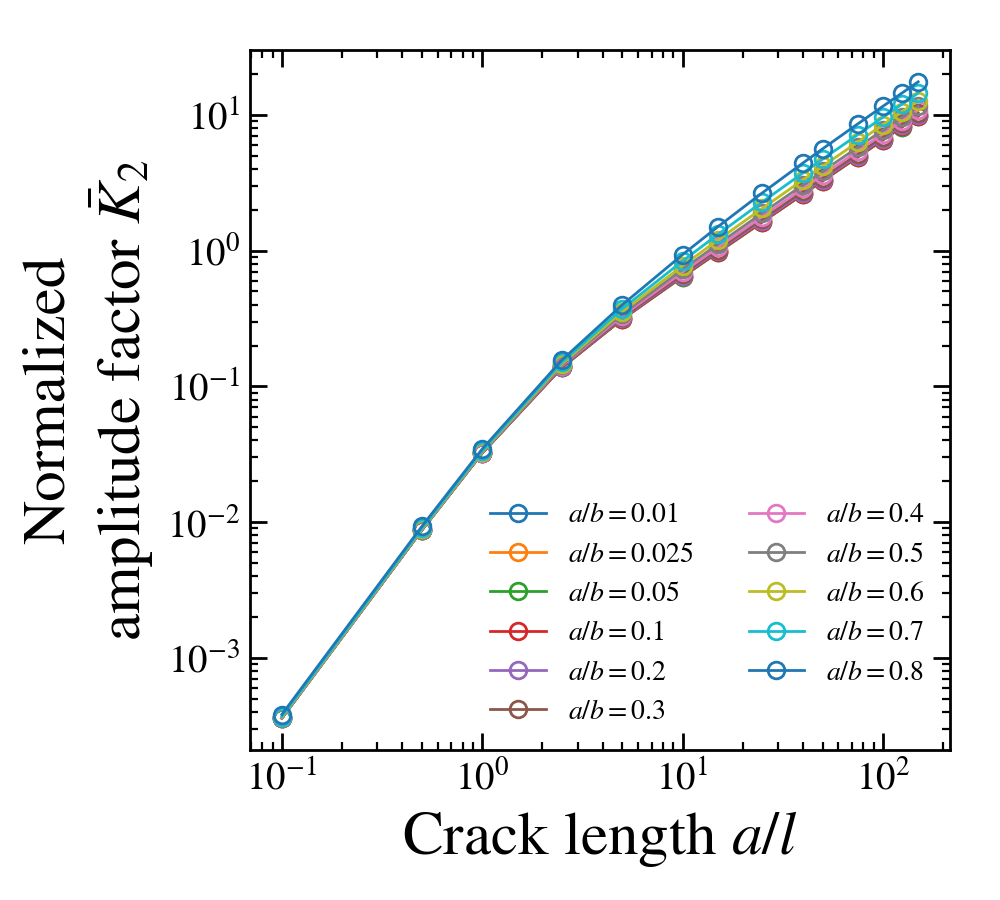}
\caption{Numerical SGE solution for CCT test: (a) J-integral, (b) amplitude factor $K_1$, (c) amplitude factor $K_2$}
\label{fig:jk_center_1}
\end{figure}

\begin{figure}[h!]
\centering
(a)\includegraphics[width=0.405\textwidth]{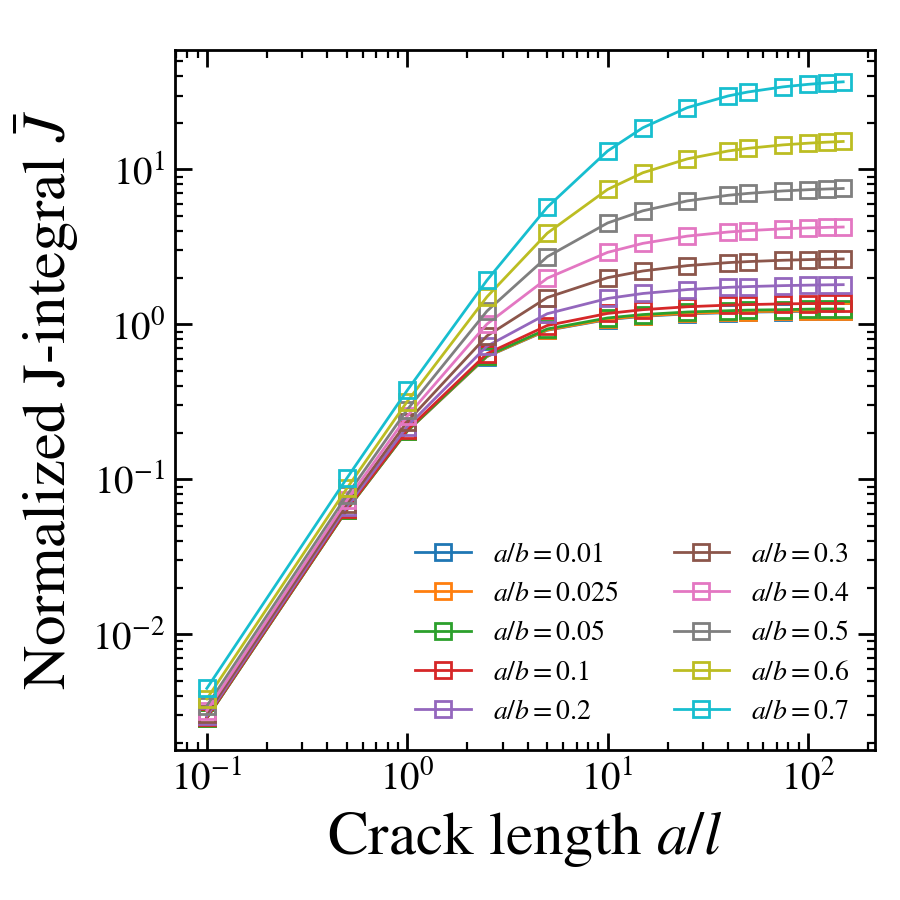}
(b)\includegraphics[width=0.45\textwidth]{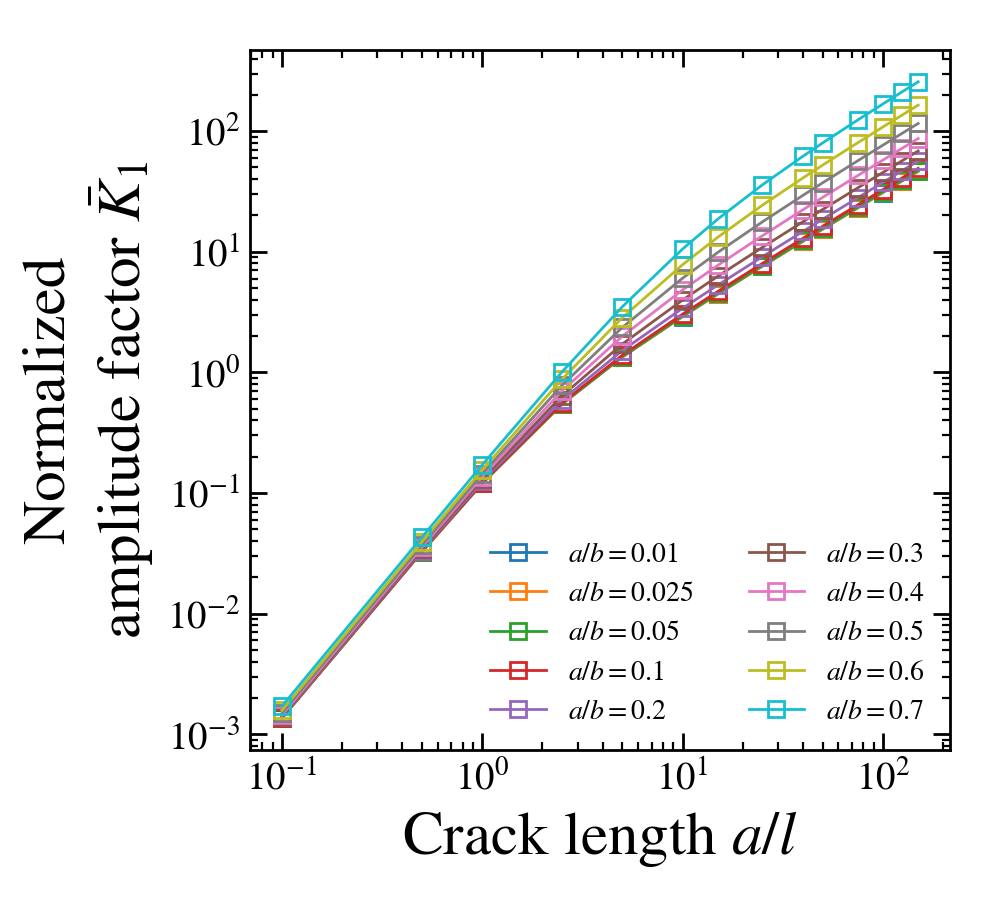}\\
\vspace{10pt}
(c)\includegraphics[width=0.45\textwidth]{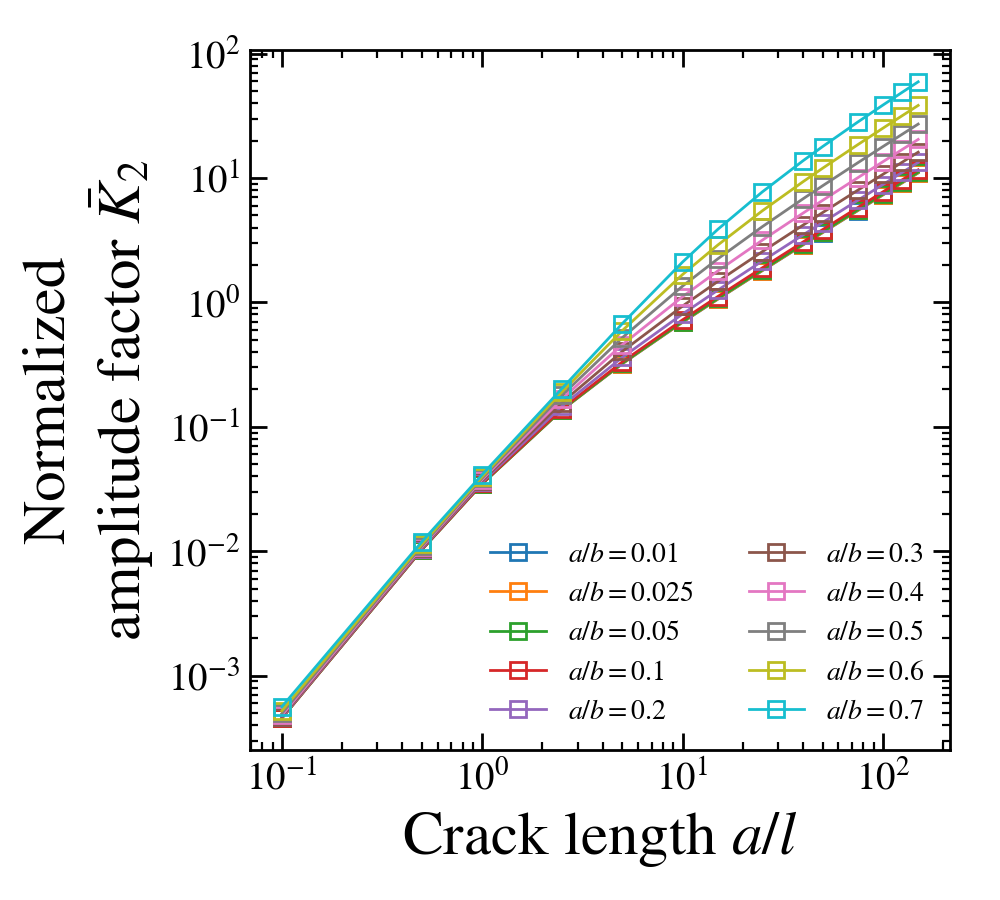}
\caption{Numerical SGE solution for SENT test: (a) J-integral, (b) amplitude factor $K_1$, (c) amplitude factor $K_2$}
\label{fig:jk_center_2}
\end{figure}

\clearpage

\begin{figure}[h!]
\centering
(a)\includegraphics[width=0.38\textwidth]{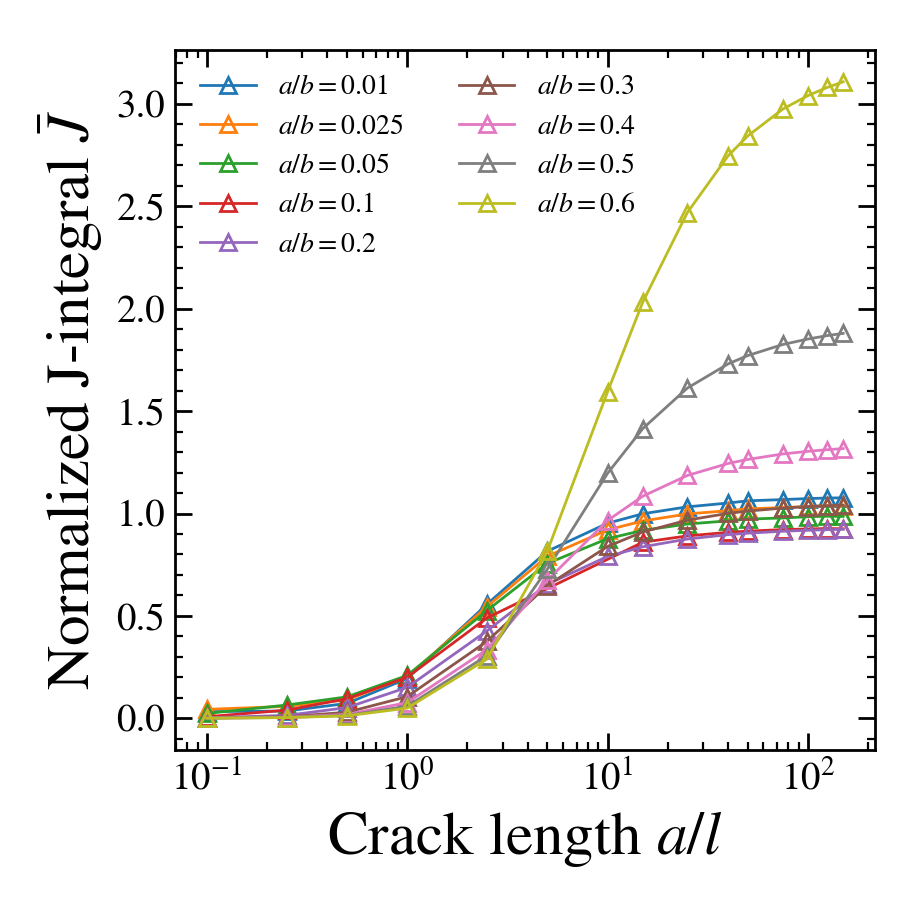}
(b)\includegraphics[width=0.38\textwidth]{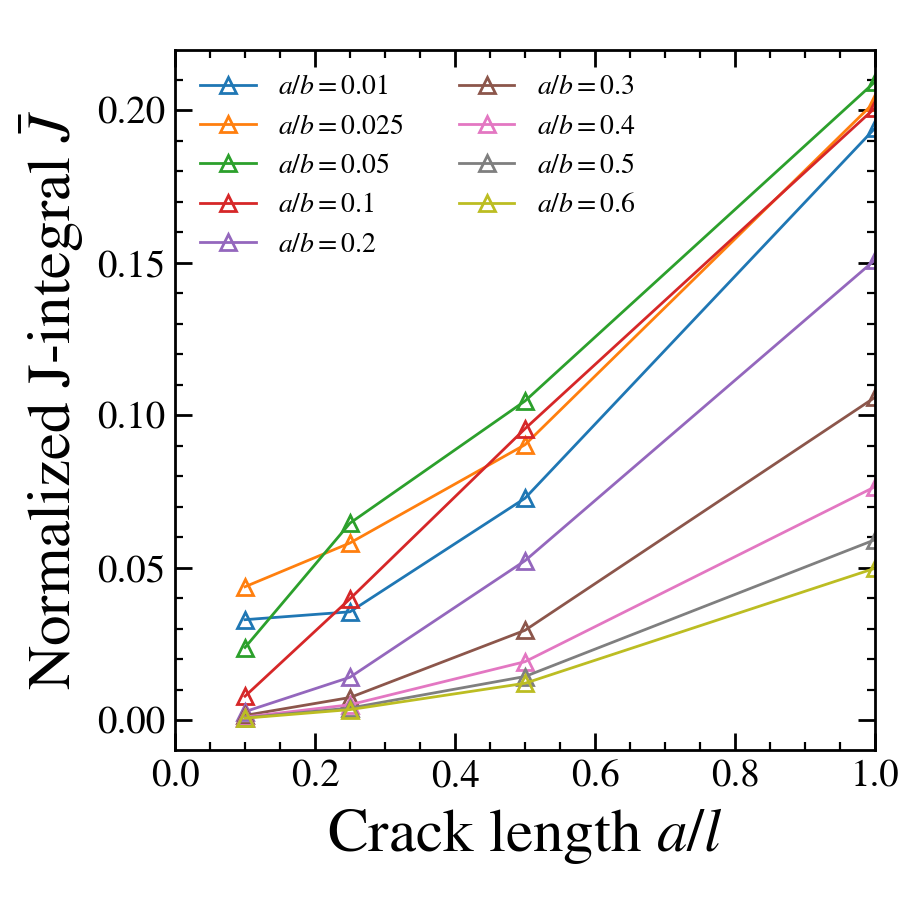}\\
(c)\includegraphics[width=0.4\textwidth]{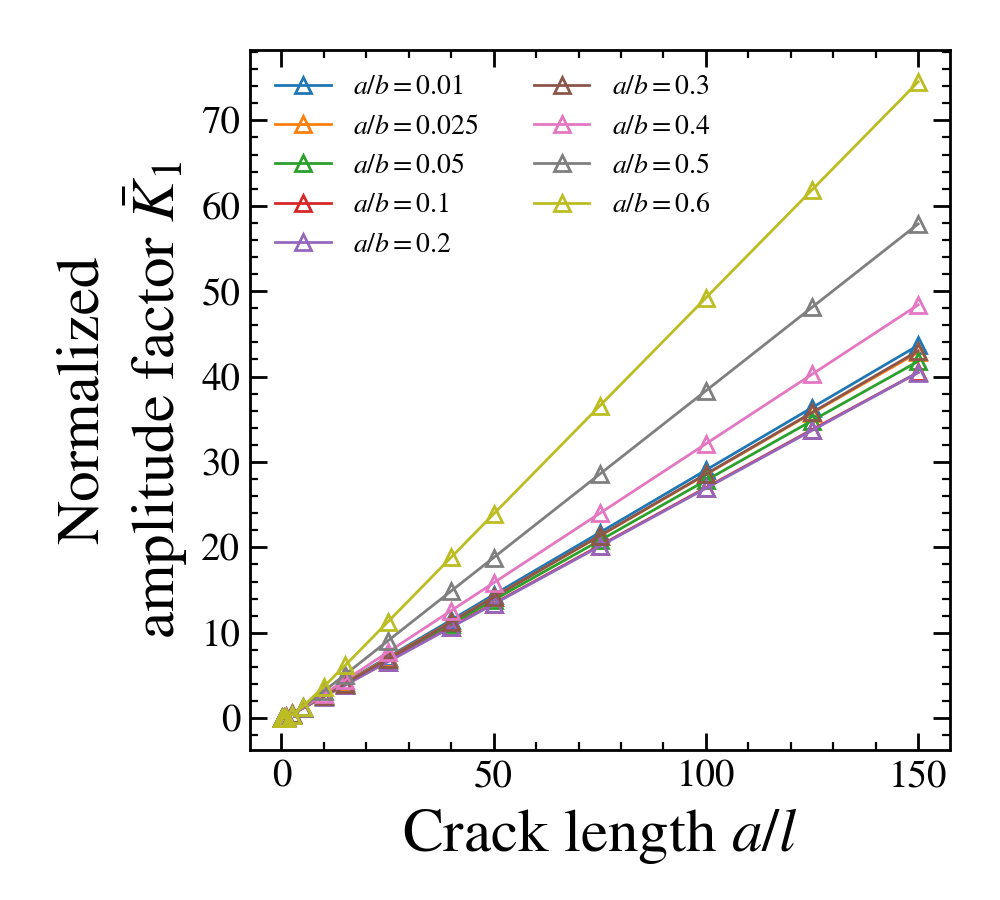}
(d)\includegraphics[width=0.4\textwidth]{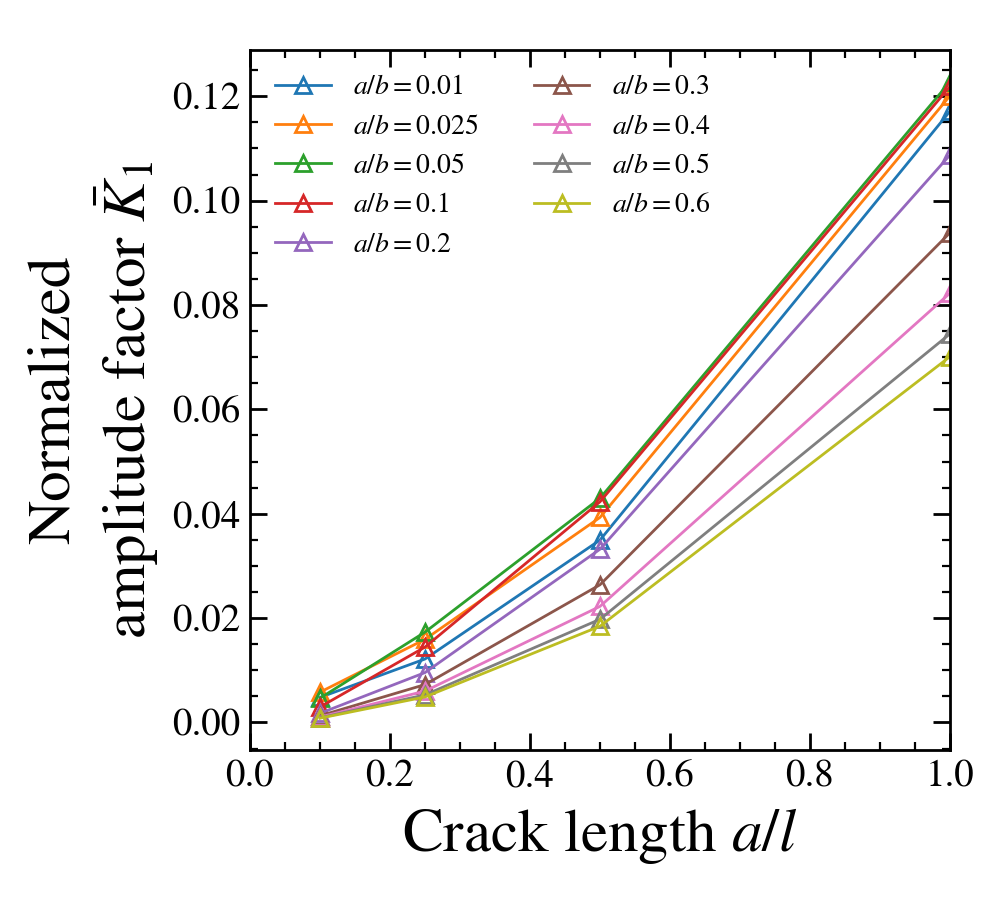}\\
(e)\includegraphics[width=0.4\textwidth]{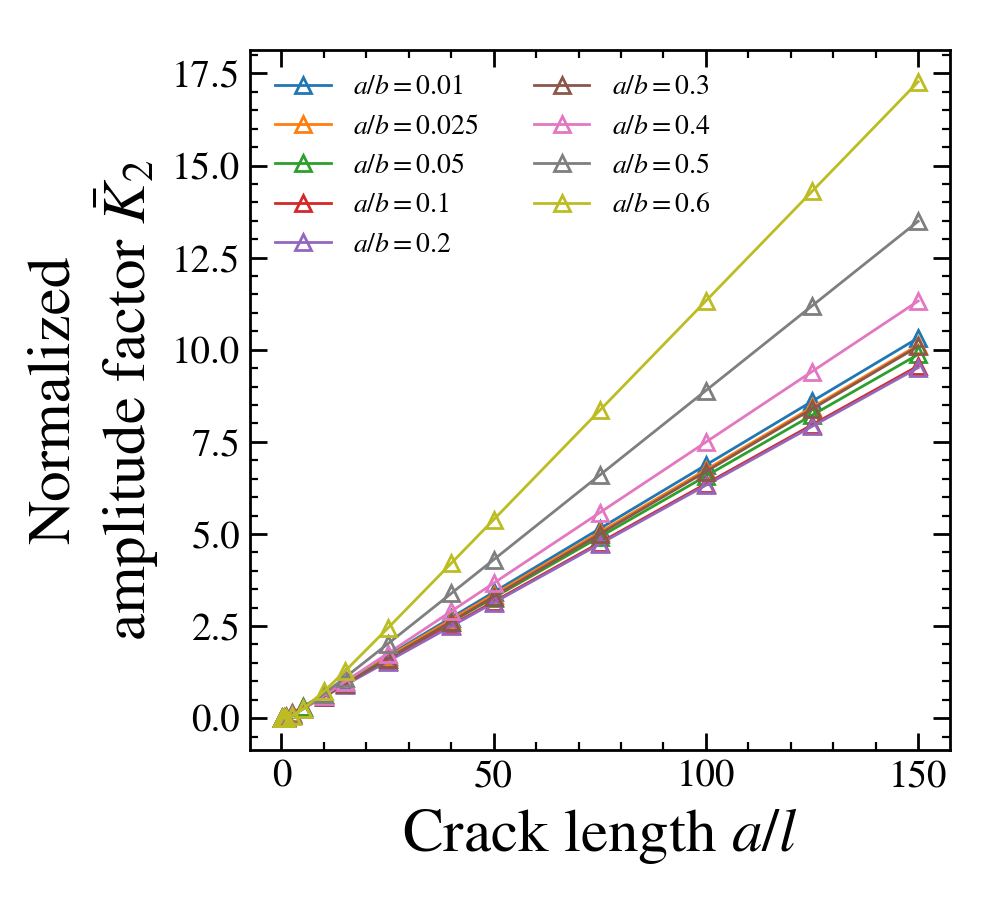}
(f)\includegraphics[width=0.4\textwidth]{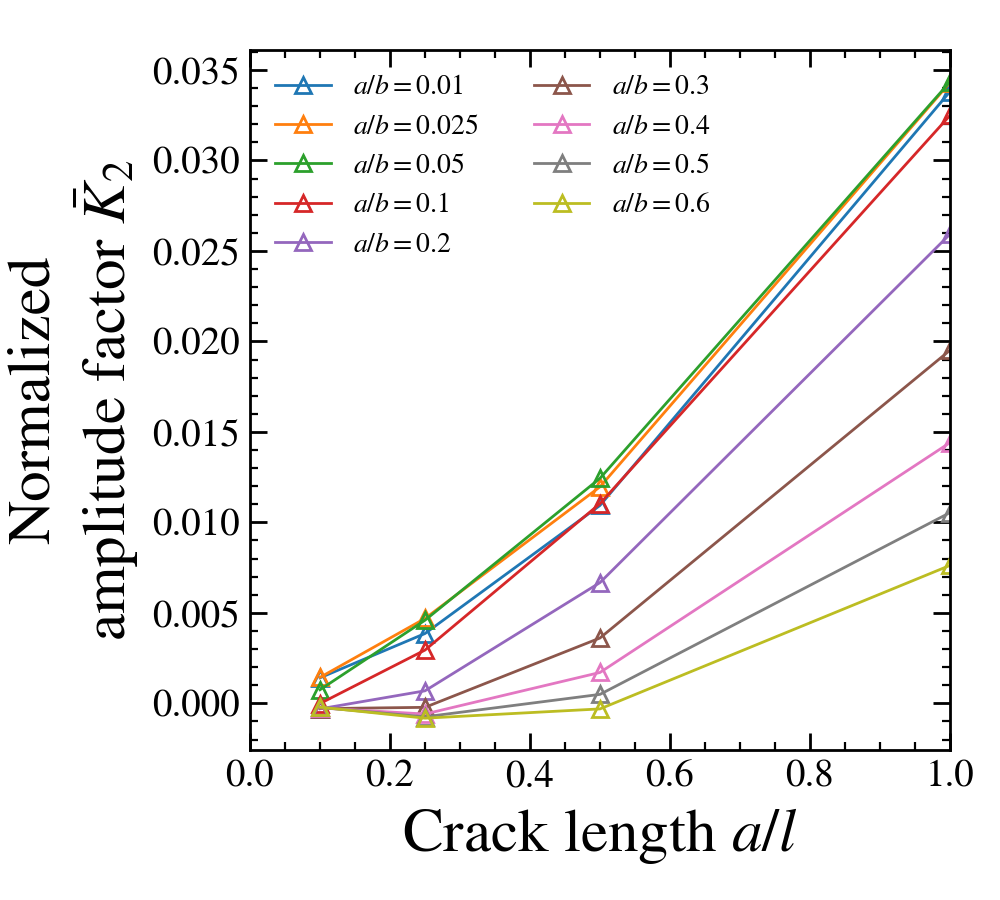}
\caption{Numerical SGE solution for SENB test: (a, b) J-integral, (c, d) amplitude factor $K_1$, (e, f) amplitude factor $K_2$}
\label{fig:jk_center_3}
\end{figure}

\clearpage

\footnotesize
%\section*{References}
%\bibliographystyle{unsrt} 
\bibliography{refs.bib}

\end{document}